\documentclass[aps,prx,twocolumn,superscriptaddress,noeprint,longbibliography, nofootinbib]{revtex4-1}

\usepackage[colorlinks=true]{hyperref}
\hypersetup{citecolor = blue}
\usepackage{graphicx}% Include figure files
\usepackage{bm}% bold math
\usepackage{physics}
\usepackage{xcolor}
\usepackage{enumitem}
\usepackage{amsmath, amssymb}
\usepackage[normalem]{ulem}
\usepackage{natbib}

\newcommand{\be}{\begin{equation} }
\newcommand{\ee}{\end{equation} }
\newcommand{\ba}{\begin{eqnarray} }
\newcommand{\ea}{\end{eqnarray} }
\newcommand{\n}{\nonumber \\ }

\newcommand{\Hc}{H_{\rm sym}}

\newcommand{\qdg}{Q^{+}}

\newcommand{\bit}{\begin{itemize}}
\newcommand{\eit}{\end{itemize}}

\newcommand{\balpha}{ \bm{\alpha}}
\newcommand{\bbeta}{ \bm{\beta}}
\newcommand{\bw}{ \mathbf{w}}
\newcommand{\Hz}{ H_{\text{SG}}}
\newcommand{\Hs}{H_{\rm SG}}

\begin{document}

\title{From tunnels to towers: quantum scars from Lie algebras \\and $q$-deformed Lie algebras}

\author{Nicholas O'Dea}
\affiliation{Department of Physics, Stanford University, Stanford, CA 94305, USA}
\author{Fiona Burnell}
\affiliation{Department of Physics, University of Minnesota Twin Cities, MN 55455, USA}
\author{Anushya Chandran}
\affiliation{Department of Physics, Boston University, MA 02215, USA}
\author{Vedika Khemani}
\affiliation{Department of Physics, Stanford University, Stanford, CA 94305, USA}

\begin{abstract}

We present a general symmetry-based framework for obtaining many-body Hamiltonians with scarred eigenstates that do not obey the eigenstate thermalization hypothesis. Our models are derived from parent Hamiltonians with a non-Abelian (or $q$-deformed) symmetry, whose eigenspectra are organized as degenerate multiplets that transform as irreducible representations of the symmetry (`tunnels'). We show that large classes of perturbations break the symmetry, but in a manner that preserves a particular low-entanglement multiplet of states -- thereby giving generic, thermal spectra with a \emph{shadow} of the broken symmetry in the form of scars. The generators of the Lie algebra furnish operators with `spectrum generating algebras' that can be used to lift the degeneracy of the scar states and promote them to equally spaced `towers'. Our framework applies to several known models with scars, but we also introduce new models with scars that transform as irreducible representations of symmetries such as SU(3) and $q$-deformed SU(2), significantly generalizing the types of systems known to harbor this phenomenon.
Additionally, we present new examples of generalized AKLT models with scar states that do \emph{not} transform in an irreducible representation of the relevant symmetry. These are derived from parent Hamiltonians with enhanced symmetries, and bring AKLT-like models into our framework. 

\end{abstract}

\maketitle

\section{Introduction and General Framework}

A central question in non-equilibrium quantum dynamics is whether reversible unitary dynamics in a closed quantum system can establish local thermal equilibrium. Much insight into quantum thermalization follows from the eigenstate thermalization hypothesis (ETH)~\cite{Jensen:1985aa, Deutsch1, Sred1, Rigol:2008bh, DAlessio:2016aa}, a strong version of which posits that \emph{every} finite temperature eigenstate of a thermalizing system reproduces thermal expectation values locally~\cite{KimIkedaHuse}. In contrast, there are classes of interacting, typically disordered, ``many-body localized" (MBL) systems that violate the ETH and never thermalize~\cite{Basko2006, Nandkishore2015, AbaninRMP2019}. 

More recently, attention has focused on weak ETH violating systems with so-called `many-body quantum scars'~\cite{ShiraishiMori, MoudDisc,  Bernien2017, TurnerWEB2018}.  Scars are non-thermal eigenstates embedded within an otherwise thermal eigenspectrum. These typically have sub-thermal entanglement entropy ( $\sim O(\log(|A|))$ or $\sim O(|\partial A|)$ for a subsystem $A$) and coexist at the same energy density as thermal volume-law entangled eigenstates. Scars constitute a vanishing fraction of the eigenspectrum -- and hence these systems still obey a \emph{weak} version of the ETH~\cite{LauchliWeakETH}; nonetheless, their presence can lead to measurable non-thermal dynamical signatures in quenches from atypical but experimentally amenable initial states~\cite{Bernien2017, Lev2020}. Indeed, the recent literature on scars followed from an interesting experimental observation of non-thermal (oscillatory) quench dynamics in a Rydberg atom chain that realizes a constrained `PXP' spin Hamiltonian~\cite{Bernien2017}. 

The mechanism leading to scars in the PXP model is still a largely open question, and most of the scarred eigenstates relevant to the quench dynamics are only approximately known~\cite{TurnerWEB2018, TurnerRyd2018, SigInt2019, AbaninPerfectScars, Ho2019, LM2019,  ISDWC, IadecolaMagnon2019,James2019, PhysRevX.10.011055,Surace2020,  Papic2020}. In contrast, by now, there are many lattice models with exactly known scar states, ranging from the celebrated AKLT model to a spin-1 XY model to deformed topological models~\cite{ShiraishiMori,  MoudDisc, LM2019,MoudEntang,  ISXY, ISDWC, Chattopadhyay,Sala2020,Shattering2020,MLM, MoudMPS, MoudHubb, MM, Ok2019, Pancotti2020}. Many such examples with \emph{exact} scar states can be understood via one (or both) of two complementary approaches: the first due to Shiraishi and Mori (SM) \cite{ShiraishiMori} relies on local projectors, and the second due to Refs.~\cite{MLM, MoudHubb} relies on the existence of a spectrum generating algebra (SGA) on the scarred subspace.

The SM prescription~\cite{ShiraishiMori} relies on two ingredients: (1) a set of local projectors, $\{P_i\}$ centered around site $i$, that generically do not commute with one other, and (2)  one or more states $|\psi_s\rangle$ that are simultaneously annihilated by all the $P_i$ and span a subspace $\mathcal{S}$. The  $|\psi_s\rangle$ are then scarred eigenstates, annihilated by Hamiltonians of the form 
\begin{equation}
H_A^{\rm SM} = \sum_i P_i h_i P_i, 
\label{eq:SM}
\end{equation} 
where the $h_i$ are generic local operators of finite range. The $h_i$ operators ensure that the rest of the spectrum is thermalizing and non-integrable.
If, additionally, there exist special Hamiltonians $H'$ that commute with all the $\{P_i\}$, then these can be added to $H_A^{\rm SM}$ to impart different energies to the states in $\mathcal{S}$. Note that $H=H_A^{\rm SM} + H'$ does not have explicit symmetries, but the Hilbert space nevertheless \emph{dynamically} splits into disconnected `Krylov sectors': the subspaces $\mathcal{S}$ and its complement do not mix because  $\mathcal{S}$ is annihilated by $H_A^{\rm SM}$. 

Separately, Refs.~\cite{MLM, MoudHubb} furnished a complimentary framework that unified the existence of `towers' of scar states in three different models: the AKLT spin chain, a spin-1 XY model and a domain-wall conserving model~\cite{MoudDisc, ISXY, ISDWC}. In these models, scars $\{|\psi_n\rangle\}$ were generated by repeatedly acting with a particular operator $Q^+$ on a particular low entanglement eigenstate $|\psi_0\rangle$ of $H$  so that $|\psi_n\rangle = (Q^+)^n|\psi_0\rangle$. 
It was shown that, in all these cases, $Q^+$ acts as a spectrum generating `ladder' operator when restricted to the scarred subspace~\cite{MLM, MoudHubb}:
\begin{equation}
\left([H, \qdg] - \omega \qdg\right)|\psi_n\rangle =0,
\label{eq:MLM}
\end{equation}
which implies that the $|\psi_n\rangle$ are equally spaced energy eigenstates of $H$  with $E_n= \omega n + E_0$. Furthermore, the particular form of the $\qdg$ operator  is such that the states $|\psi_n\rangle$ have low entanglement. Ref.~\cite{MLM} discussed various example Hamiltonians obeying Eq.~\eqref{eq:MLM} which had the form $H = \Hs + H_A$, such that $\Hs$ has a `spectrum generating' algebra (SGA):
\begin{align}
    [\Hs, \qdg] = \omega \qdg,
    \label{eq:SGA}
\end{align}
and $H_A$ annihilates the scars, $H_A|\psi_n\rangle = 0$. Similar to SM, these contain a piece that annihilates the scars and one that gives them energy.   

While such constructions have been very useful for explicitly deriving and unifying the presence of scars in specific `one-off' models, \emph{qualitative} pictures of when and how scars may arise more generally are still largely missing. For example, it is still largely unclear where in the general space of operators and states we may expect to find a set $\{ H, \qdg, |\psi_0\rangle \}$ such that the conditions in Eq.~\eqref{eq:MLM} leading to (weak) ETH violations are met. In contrast, we have many phenomenological notions for how (strong) ETH violation arises in MBL systems: this generally requires strong disorder and weak short-ranged interactions, and MBL systems are understood as having an extensive set of emergent local integrals of motion~\cite{Abanin2013CL, Huse:2014ac}.  

In this work, we attempt to bridge this gap by presenting a very general symmetry based framework for obtaining scar towers. We start with parent Hamiltonians, $\Hc$, with a continuous non-Abelian symmetry $G$ (or a $q$-deformed version thereof, $G_q$). The generators of the symmetry furnish a natural set of spectrum generating `raising' operators $Q^+$, that connect multiplets of degenerate eigenstates in $\Hc$. We show that there are general ways to perturb $\Hc$ that break the symmetry in a manner that preserves a \emph{shadow} of the symmetry in the form of scars. For example, eigenstates of $\Hc$ in superselection sectors with at most $O(\text{poly}(L))$ basis states will have at most $O(\log(L))$ entanglement in a system of size $L$. Families of perturbations can be chosen that preserve certain such low entanglement subspaces of $\Hc$, but generically mix between all other sectors - thereby leading to the embedding of scarred eigenstates in otherwise thermal spectra\footnote{We note that $\Hc$ by itself is not considered to be scarred because features such as the presence of low entanglement eigenstates result from symmetries of $\Hc$; indeed thermalization (and ETH) is always discussed with reference to symmetry appropriate equilibrium ensembles.}. Our perspective is reminiscent of KAM theorems that concern the fate of integrable models with extensively many symmetries to the addition of small perturbations, and specifically whether remnants of the integrability can be preserved under the action of the perturbation.

\begin{figure}
\centering
\includegraphics[width=\columnwidth]{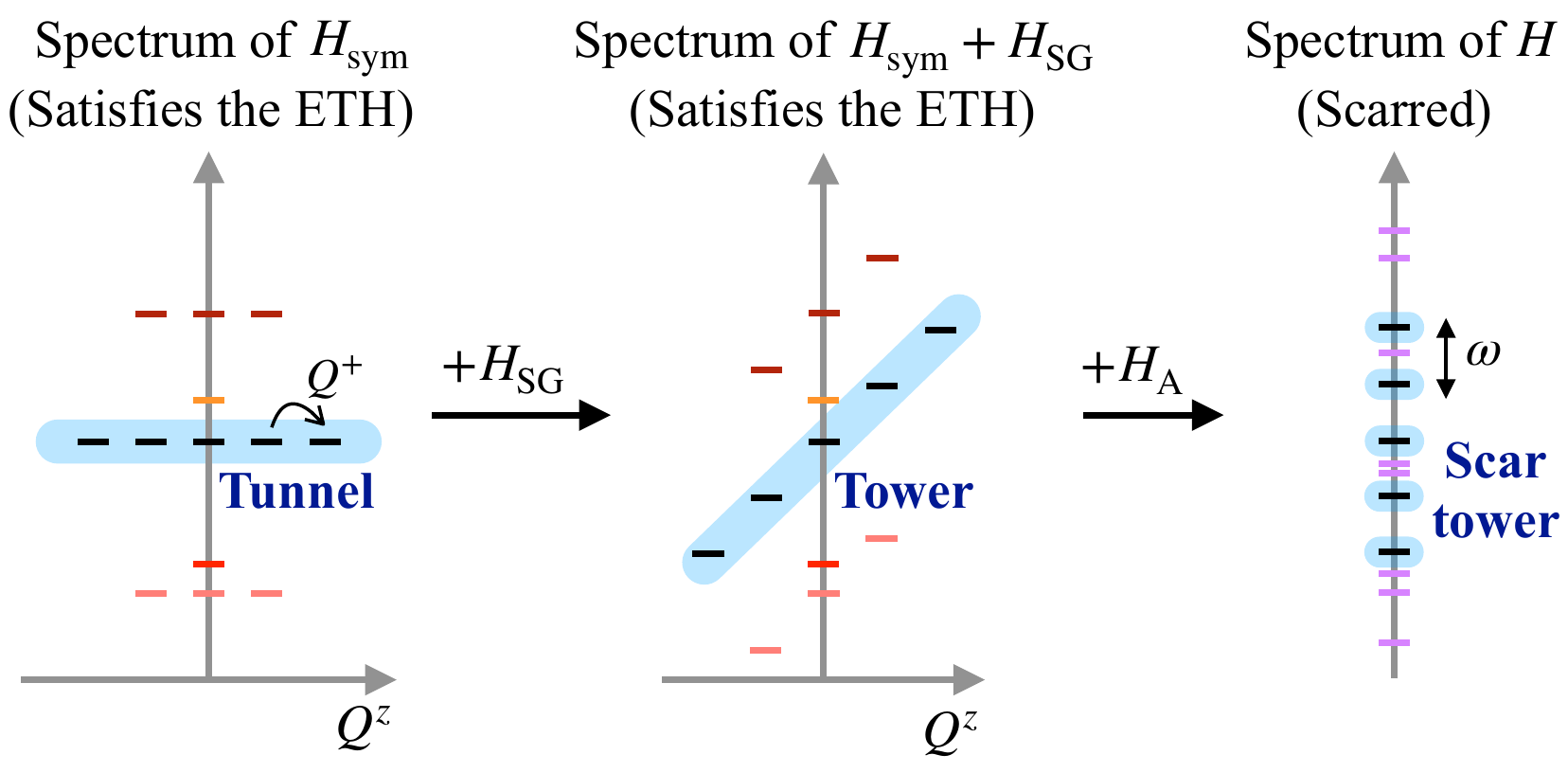}~
\caption{Schematic sketch of the tunnels-to-towers framework for obtaining scars. For simplicity, we consider the case of symmetric scars obtained by perturbing an SU(2) symmetric model $\Hc$. The generators of the symmetry furnish operators $\{Q^+, Q^-, Q^z\}$ associated with the SU(2) algebra.  
(a) $\Hc$ has `tunnels' of degenerate eigenstates with the same eigenvalue for the Casimir $Q^2$ but different eigenvalues for $Q^z$. Each tunnel is denoted by a different color. One can move between states in a tunnel using $Q^\pm$. (b) Adding $\Hz \propto Q^z$ preserves the eigenstates, but breaks the degeneracy of the tunnels. Instead, states in each tunnel get promoted to `towers' and acquire an evenly spaced harmonic spectrum because of the SGA $[Q^z, Q^\pm] = \pm Q^\pm$. (c) An $H_A$ can be chosen to annihilate a specific tower of states (highlighted) but generically break all symmetries and mix between the other states so as to make the rest of the spectrum thermal. The chosen tower of states are scars in $H=\Hc+\Hz+H_A$. \label{fig:schematic}}
\end{figure}

Our framework makes extensive use of the generators of the Lie algebra of the symmetry group $G$, which furnish a natural set of spectrum generating operators (SGOs) with `raising/lowering action'. (For example,  operators $\{ Q^+, Q^-, Q^z\}$  associated with an SU(2) symmetry have the SGA: $[Q^z, Q^\pm] = \pm Q^\pm$.)  We obtain scarred models via a three step process:
\begin{itemize}
    \item First, $\Hc$ contains multiplets of degenerate eigenstates --- tunnels --- that transform as irreducible representations (irreps) of the symmetry $G$. Each multiplet is labeled by its eigenvalues under the Casimir operators of $G$, and states within the multiplet are distinguished by their eigenvalues under the generators of $G$ in the Cartan subalgebra. Raising operators connect between the states in a multiplet. As an example, an eigenstate $|\psi_0\rangle$ of an SU(2) symmetric $\Hc$ is labeled also by its eigenvalues under the Casimir $Q^2$, and the Cartan generator $Q^z$. Then, $|\psi_n\rangle = (Q^+)^n|\psi_0\rangle$ will be a degenerate eigenstate with the same $Q^2$ but different $Q^z$ eigenvalue since $[\Hc, Q^+]= [Q^2, Q^+] = 0$, and $[Q^z, Q^+] = Q^+$. 
    
    \item Next, tunnels in $\Hc$ can be promoted to equally spaced `towers' of non-degenerate eigenstates in the Hamiltonian $H_s= \Hc + \Hs$. Here $\Hs$ is typically chosen to be a linear combination of the generators in the Cartan subalgebra, which commute with and share the eigenstates of $\Hc$, but have a SGA with the raising operators.   For example, if $G=SU(2)$, choosing  $\Hz = \omega Q^z$ gives the states $|\psi_n\rangle$ energy $E_n = E_0 + n \omega$ because $[\Hs, Q^+] = \omega Q^+$. 
    
    We emphasize that, even though the addition of $\Hs$ breaks the symmetry $G$, the eigenstates of $H_s$ and $\Hc$ are still the same and only their energy eigenvalues are different: in particular,  degenerate tunnels of states become non-degenerate towers\footnote{More generally, we will also consider larger non-Abelian symmetries (such as SU(3)) where the eigenspectra of the multiplets may have more complex `pyramidal' relations.}. 
    
    We will also discuss models where the scar tower does not transform as an irrep of the symmetry group G and/or where $\Hs$ is \emph{not} a generator of the symmetry but still has a SGA with the raising operators. This is possible when $\Hc$ has an expanded symmetry, which allows $\Hc$ to be simultaneously diagonalized with $\Hz$ and have tunnels of degenerate eigenstates that do not transform as an irrep.      
    
    \item Finally, to make $H_s$ a scarred model, we introduce symmetry breaking perturbations $H_A$ that annihilate a particular low entanglement tunnel of states $\{|\psi_n\rangle\}$ built upon a particular low entanglement `base state' $|\psi_0\rangle$. $H_A$ can typically be chosen to be generic enough to mix states across the various symmetry sectors of $\Hc$ so as to make the rest of the spectrum generic and thermal. In all, 
\begin{equation}
    H = \Hc + \Hs + H_A
\label{eq:tunneltotower}
\end{equation} 
obeys the condition in Eq.~\eqref{eq:MLM}, and has towers (or pyramids) of scar states generated by raising operators of the non-Abelian symmetry $G$ acting on a low entanglement base state $|\psi_0\rangle$ which is an eigenstate of each of the three terms in $H$.  
\end{itemize}

This three-step process is schematically illustrated in Fig.~\ref{fig:schematic}. Several scarred models in the literature can be understood as special cases of our general framework~\cite{Shibata, ISXY,Buca2020, MM, MoudHubb}. Importantly, our work also furnishes a natural way to get many new scarred Hamiltonians derived from various non-Abelian and $q$-deformed non-Abelian symmetries. In what follows, we flesh out the ingredients for our framework in more detail in Section~\ref{sec:ingredients}. We then discuss two qualitatively distinct families of scars. In the first, discussed in Section~\ref{sec:symmetricscars}, the scarred eigenstates inherit the parent symmetry and transform as a single irreducible representation of $G$ (or $G_q$). These represent generalizations of perturbed $\eta$-pairing models that have been discussed in the literature~\cite{MM, MoudHubb}. The second, discussed in Section~\ref{sec:aklt}, is a generalization of various AKLT like models where the scars do \emph{not} inherit the symmetry $G$. However, as we discuss, these can be viewed as arising from parent Hamiltonians with an enhanced symmetry group larger than $G$. We conclude in Section~\ref{sec:conclude}, and present various technical details in a series of appendices. 

\section{Ingredients of the framework}
\label{sec:ingredients}

We now discuss in more detail our framework for constructing families of Hamiltonians with towers of scarred states. For specificity, we will always consider a one-dimensional chain with $L$ sites, with a spin -$S$ degree of freedom (i.e. a $(2S+1)$-state Hilbert space) on each site~\footnote{It is easy to see that the general philosophy of our constructions apply \emph{mutatis mutandis} to systems in higher dimensions.}. We denote the {\it physical} spin operators on site $j$ as $S^\pm_j, S^z_j$, with 
\begin{equation}
    S^\pm = \sum_j S^\pm_j \ , \ \ S^z = \sum_j S^z_j,
\label{eq:spinops}
\end{equation}
where $S_j^\pm$ are the usual spin raising and lowering operators on site $j$ and $S_j^z$ measures the $z$ polarization of the spin. We refer to the resulting SU(2) algebra as the spin-SU(2) algebra, and to any associated symmetry in our model as a spin-SU(2) symmetry.  

\subsection{Lie Algebras, Raising Operators and $\Hz$}

To describe the possible Hamiltonians of the form in Eq.~\ref{eq:tunneltotower},
our first task is to characterize a suitable set of ladder operators $Q^+$ associated with the Lie algebra of a non-Abelian symmetry $G$ in $\Hc$. Note that $\Hc$ will \emph{not} generically have spin-SU(2) symmetry and the $Q$ operators will generally be distinct from the physical spin operators in Eq.~\eqref{eq:spinops}. 
Here we treat the case where $G$ acts as a product of onsite symmetries.  A different context in which operators with suitable commutation relations emerge naturally is in $q$-deformed Lie algebras. In that case the commutation relation between the $q$-deformed SU(2) (or SU(2)$_q$) raising and lowering operators requires an action that extends over all sites in the system; we discuss this example in detail in Sec. \ref{sec:qdef}.

We begin with the case $G=SU(2)$.  In this case, we define a single raising, lowering and Cartan  operator, each of which is a linear combination of the form
\begin{equation} \label{Eq:Q}
    Q^\pm = \sum_j e^{\pm  i k r_j}(Q^\pm_{j}), \,\,\, Q^z = \sum_j Q^z_j \ \ .
\end{equation} 
Here $k$ is a momentum index, and the operators $\{Q^\pm_{j}, Q^z_j\}$ are derived from the local SU(2) generators of the symmetry acting on site $j$. 
To ensure that the scar tower contains only $O(\text{poly}(L)) $ states, we also require that $(Q^+_{j})^{n_{\text{max}}} = 0$, and hence on a chain with $L$ sites, $(Q^\dag_{})^{n_{\text{max}}L } = 0$.

More generally, if $\Hc$ is invariant under a continuous non-Abelian symmetry group $G$, we can find a family of raising/lowering operators $Q_{\balpha^\pm}$ and $\Hz$ chosen from the associated Lie algebra $G$, such that these satisfy commutation relations corresponding to an SGA. 
To construct these, let $G$ be an $N$ dimensional semi-simple Lie algebra on a given site $j$ [$N = m^2 - 1$ for SU(m)].  We denote by $Q_{\mu,j}^z$ with $\mu = 1,\cdots R $ a maximal linearly independent set of commuting Hermitian generators that can be diagonalized simultaneously, called the Cartan subalgebra (CSA). $R$ is known as the rank of the algebra.

On each site, the $N-R$ generators that are not in the CSA can be rearranged into pairs of raising and lowering operators $\{Q_{\balpha^\pm,j} \}$ of different SU(2) subalgebras.  They satisfy the commutation relations
\begin{eqnarray} \label{Eq:QzQalphaCom}
[Q^z_{\mu,j}, Q_{\balpha,j}] &= &   \alpha_{\mu} Q_{\balpha,j} 
\end{eqnarray}
Here $\balpha$ are $R$-component vectors, known as {\it roots}. We will use the notation $\{\balpha\} = \{\balpha^+\} \cup \{\balpha^-\}$ to collectively refer to all roots, while  $\balpha^{\pm}$ denote positive/negative roots, such that the corresponding operators $Q_{\balpha^{\pm}}$ can be viewed as raising/lowering operators. A set of positive roots is a subset of all roots that contains only one of $\balpha^+$ and $\balpha^- \equiv -\balpha^+$, such that if $\balpha^+ + \bbeta^+$ is a root, then $\balpha^+ + \bbeta^+$ is also a positive root.  (Note that this allows for multiple choices of positive roots; the specific choice will not matter for our formalism\footnote{We have in mind the specific choice that a root $\balpha$ is positive if the first nonzero element of $\balpha$ is positive.}). We also define {\it simple roots}, $\balpha^{(S)}$, as the set of positive roots that cannot be written as a sum of other positive root vectors.  Simple root vectors have the nice property that any positive root vector can be written as a unique linear combination of simple root vectors with positive integer coefficients. In addition, all raising operators can be expressed as nested commutators of the raising operators associated with simple roots.  The raising and lowering operators on each site obey the commutation relations
\begin{equation} \label{Eq:alphabetaCom2}
[ Q_{ \balpha,j} , Q_{ \bbeta,j}]   \propto\begin{cases} 
\sum_{\mu} \frac{\alpha_{\mu}}{\alpha \cdot \alpha} Q^z_{\mu,j}   & \text{if }  \bbeta =-\balpha\\
\\
Q_{ \balpha + \bbeta,j} & \text{if } \balpha + \bbeta \text{ is a root, and } \\
&\bbeta\neq-\balpha \\
\\
0 &\text{otherwise.}
\end{cases}
\end{equation}

In a chain with $L$ sites, we can therefore define the global raising/lowering operators as
\begin{equation} \label{Eq:Qdag}
    Q_{\balpha^\pm}= \sum_j  e^{ i r_j k_{\balpha^\pm} } Q_{\balpha^\pm, j}. 
\end{equation}
Here $k_{\balpha^\pm}$ can be chosen independently for each simple root\footnote{If $\bbeta = \sum_i n_i \balpha_i $ where $\balpha_i$ are simple roots, then we require $k_{\bbeta} = \sum_i n_i k_{\balpha_i}$, to ensure that the operators $\{ Q_{\balpha^\pm},  \sum_j Q^z_{\mu,j} \}$ satisfy the commutation relations of $G$}.   
Next, choosing 
\begin{equation} \label{Eq:HzGen}
    \Hz = \sum_{\mu=1}^R h_\mu \sum_j Q^z_{\mu,j}
\end{equation}
 to be a linear combination of the generators in the CSA,
all the raising/lowering operators obey the desired SGA
\begin{equation} \label{Eq:Raising}
    [  \Hz, Q_{\balpha} ] = \omega_{\balpha} Q_{\balpha} 
\end{equation}
where 
$\omega_{\balpha} = \sum_\mu h_\mu \balpha_\mu$.   We thus find that for $R>1$ there are multiple linearly independent choices of the coefficients $h_\mu$, which in general exhibit different spectra for the scar states.  For a given choice of $\Hz$, the different raising operators $Q_{\balpha}$ also create multiple distinct branches of the scar tower, associated with different fundamental frequencies $\omega_{\balpha}$.

To summarize: a subset of the generators of a Lie algebra can always be combined to furnish one or more pairs of raising and lowering ladder operators, $Q_{\balpha^\pm}$, associated with embedded SU(2) subalgebras. The remaining generators $Q^z_\mu$ form the Cartan subalgebra and have spectrum-generating commutation relations with $Q_{\balpha}$ (cf. Eq.~\ref{Eq:QzQalphaCom}). When $\Hz$ is chosen to be as a linear combination of $Q^z_\mu$, as in Eq.~\eqref{Eq:HzGen}, then $\Hz$ can be simultaneously diagonalized with $\Hc$ and the Casimirs, and has spectrum-generating commutation relations with $Q_{\balpha}$ (cf. Eq.~\eqref{Eq:Raising}). This immediately implies that specific multiplets of eigenstates of $\Hc$ with the same eigenvalue for the Casimirs but different eigenvalues of $\Hz$ are degenerate. Each of these multiplets forms a ``tunnel" in the spectrum of $\Hc$ that transforms as a single irreducible representation of $G$, and acting with $\{ Q_{ \balpha^\pm} \}$ moves between different states in a given tunnel (Fig.~\ref{fig:schematic}a). 
When $\Hz$ is added to the Hamiltonian, the degeneracies are broken and the eigenstates in the tunnels acquire energy spacings that are integer superpositions of $\omega_{\balpha}$ (cf. Eq.~\eqref{Eq:Raising}), thereby getting promoted to `towers' (or pyramids) of states (Fig.~\ref{fig:schematic}(b)). The final step, discussed in the next two subsections, is to add a term $H_A$ to the Hamiltonian that annihilates a \emph{particular} tunnel of low entanglement states built upon a particular `base state'\footnote{Note that while the spectrum of $\Hc$ has many tunnels of degenerate eigenstates, not all of these will have low entanglement.}; $H_A$ generically breaks all symmetries and mixes between all other states so as to give a thermal spectrum with the chosen states embedded as low entanglement scars.

Interestingly,  in our discussion of generalized AKLT models in Sec.~\ref{sec:aklt}, we will encounter examples where $\Hz$ \emph{cannot} be expressed in terms of the generators of the CSA, but nevertheless has the desired `raising action' in its commutation relations with $Q_{\balpha}$. For example, the total spin-z operator $S^z$ obeys the commutation $[S^z, Q^+]= 2S Q^+$ for the `raise by $2S$' $Q^+$ operators in Eq.~\eqref{eq:qsu2}, but $S^z$ is linearly independent from $Q^z$. 
In these cases, the parent Hamiltonian $\Hc$ generally has a larger symmetry, so that its eigenstates can still be simultaneously diagonalized with $\Hz$ and the picture of tunnels to towers still applies -- however, the states in the tower of scars need not have a definite eigenvalue under the Casimir $Q^2$ and are not contained within a single irreducible representation of $G= SU(2)$.

\subsection{Base state $|\psi_0\rangle$}
  
In order to construct our candidate scar tower, the next ingredient we need is to select a specific multiplet of degenerate tunnel states in $\Hc$ that will get promoted to form a scar tower. In order for the scars to have low entanglement, the tower should be built by acting with the raising operators $Q_{\balpha^+}$ on a particular  low entanglement  ``base" state $|\psi_0\rangle$.  As discussed above, we will require $|\psi_0\rangle$ to be an eigenstate of $\Hc$ and $\Hz$.  In general, the  scar space consists of a discrete set of states of the form:
\begin{equation}  \label{Eq:psin}
    |\psi_{n_1, n_2, ... n_k}\rangle = Q_{\balpha^+_k}^{n_k}... Q_{\balpha^+_2}^{n_2} Q_{\balpha^+_1}^{n_1} |\psi_0 \rangle
\end{equation}
where $\balpha^+_j$ are positive roots and $0 \leq n_j \in \mathbb{Z}$. Importantly, we need the the number of states in the scar tower to grow at most polynomially in $L$. This is possible because,  as for SU(2), $Q_{\balpha,j}^{n_{\text{max}}} =0$ as an operator on each site: any state can be raised/lowered by at most a fixed amount in any direction. Hence the $n_j$ can grow at most linearly with $L$. Additionally, we can choose the $\{\balpha_{j}\}, \, j=1,2,...,k$ to be a redundant ordering on the simple roots such that $k$ is independent of the length of the chain.\footnote{We can find this choice for many cases, such as when $(Q_{\balpha})_i$ furnishes a fundamental representation of SU(N) on each site and $|\psi_0\rangle$ is the fully polarized state. We expect it to be generically true, but we do not have a general proof.}

The base states that we consider come in two types.  First, $|\psi_0 \rangle$ can be a low-entanglement eigenstate of $\Hc, \Hz$ and the relevant Casimir operators.  The resulting scar states, which we refer to as symmetric scars, transform in a single irreducible representation of the symmetry group $G$.  Base states of this form (with $G$=SU(2)) are relevant to the spin-$1$ $XY$-model~\cite{ISXY,MLM, MM, MoudHubb}, as well as the $\eta$-pairing states of the Hubbard model and other electronic models after appropriate mappings from spin lattices to electronic models~\cite{MM}. 
Second, we may choose $|\psi_0 \rangle$ to be an eigenstate of $\Hz$, but not of the relevant Casimirs.  In this case the scar pyramid is not contained within a single irreducible representation of $G$, and the associated parent Hamiltonian $\Hc$ must have additional degeneracies not explained by the symmetry $G$.  This scenario arises in various AKLT-like model Hamiltonians exhibiting exact quantum scars.

For symmetric scars, one simple choice of base state is a maximally polarized state.  For example, if $G$=SU(2),  we can take $|\psi_0\rangle$ to be the maximally spin-polarized state, which is the only state in the symmetry sector labeled by $(Q = Q_{\rm max}, Q_z = -Q_{\rm max})$. Acting with the raising operator on this state $n$ times generates  unique state in the symmetry sector labeled by  $(Q = Q_{\rm max}, Q_z = -Q_{\rm max}+n)$.

For general $G$, the maximally polarized state is obtained as follows. We will work in a basis $\{ | \bw \rangle \}$ of simultaneous eigenstates of all $Q^z_\mu$.  (This is analogous to working in the basis of $\sigma^z$ eigenstates in the SU(2) case.) Here $\bw$, known as the weight vector, describes the eigenvalues of the $Q^z_\mu$, via 
\be
Q^z_\mu |\bw \rangle = w_\mu|\bw \rangle \ .
\ee
The commutation relations (\ref{Eq:QzQalphaCom}) 
imply that 
\be
Q_{\balpha^\pm} | \bw \rangle \propto   | \bw \pm \balpha \rangle
\ee
i.e. acting with $Q_{ \balpha^\pm}$ on a state $|\bw \rangle$ changes the eigenvalue of $Q^z_\mu$ by an amount $\pm \balpha_\mu$ (which can be 0 for some choices of $\balpha, \mu$), while preserving the value of the Casimirs.  
Note that the coefficient of proportionality can be 0, in which case $ | \bw \pm \balpha \rangle$ is not a state in our Hilbert space.  There is always a unique ``lowest weight" state $|\bw_{\text{min}}\rangle$ such that $Q_{\balpha^-} |\bw_\text{min} \rangle = 0$. 
The general maximally polarized base state is thus a product of  lowest-weight states on each site in our system: $|\psi_0 \rangle = \prod_i |\bw_{\text{min},i}\rangle$.   By definition, this is an eigenstate of the many-body Casimirs and all $Q^z_\mu$.   
Acting with all non-vanishing products of the form (\ref{Eq:psin}) will then generate all states in the corresponding irrep of the Lie algebra $G$. 
We note that for a system with $L$ sites, the maximum number of states in any such representation grows only polynomially with $L$, guaranteeing that our scar subspace is sub-extensive.  
  
We now argue that the states $|\psi_{n_1, n_2, ... n_k}\rangle $ are low entanglement eigenstates of $(\Hc + \Hz)$ and hence good candidate scar states once $H_A$ is added.  First, they are eigenstates since 
\begin{align}
    (\Hc &+ \Hz)|\psi_{n_1, ... n_k}\rangle  \nonumber \\
    &= (\Hc+\Hz) Q_{\balpha_k}^{n_k}... Q_{\balpha_1}^{n_1} |\psi_0 \rangle \nonumber \\
    &=Q_{\balpha_k}^{n_k}... Q_{\balpha_1}^{n_1} \left(\Hc + \sum_{i=1}^k  n_i \omega_{\balpha_i}  +  \Hz \right) |\psi_0 \rangle \nonumber \\
    &= \left(E_0 + \sum_{i=1}^k  n_i \omega_{\balpha_i}\right) Q_{\balpha_k}^{n_k}... Q_{\balpha_1}^{n_1}|\psi_0\rangle \ \ ,
\end{align}
where the third line follows from Eq.~\ref{Eq:Raising} and the fact that $[\Hc, Q_{\balpha_k}]=0$, and the last line follows from the fact that we require $|\psi_0 \rangle$ to be an eigenstate of $\Hz$ and $\Hc$ with eigenvalue $E_0$.

Second, the states $|\psi_{n_1, ... n_k}\rangle$ all have entanglement that grows at most logarithmically in the subsystem size, provided that $|\psi_0 \rangle$ has low entanglement.  To see this, observe that if $|\psi_0 \rangle$ has finite (or $\log(L)$) entanglement, it can be approximated (up to exponentially small corrections) by a matrix product state with bond dimension $d$ for some $d$ that is finite (or ${\rm poly}(L)$).  
In fact, the choices of $|\psi_0 \rangle$ that we use here will all be exact matrix product states.
Further, for all choices of $Q^+$ operators considered in this work - for Lie algebras and $q$-deformed Lie algebras - the operator 
 $(Q^+)^n $ can be expressed as a matrix-product operator (MPO) of bond dimension $n+1$.  We show this in Appendix \ref{App:MPO}, by generalizing an argument due to Mougdalya et al in Ref.~\cite{MoudEntang}.  
Thus the state $Q_{\balpha_k}^{n_k}... Q_{\balpha_2}^{n_2} Q_{\balpha_1}^{n_1} |\psi_0 \rangle$ has entanglement entropy of \emph{at most} $S \sim \log(d) + \sum_k \log (n_k+1)$.  Since the maximum possible value of $n_i$ grows polynomially with  $ L $, we see that states in our scar tower have entanglement entropy that scales at most logarithmically, rather than linearly, with $L$.  This is a defining characteristic of a quantum scar eigenstate.

\subsection{Annihilation operators $H_A$}

Finally, our construction requires an operator $H_A$ that behaves like a generic, thermal Hamiltonian on the non-scarred eigenstates, but with the special property that
\begin{equation}
   H_A  |\psi_{n_1, ... n_k} \rangle = 0
\end{equation}
for any $\{ n_\mu \}$-- i.e.  it annihilates all states in the scar tower.

In general, we will consider two types of $H_A$ operators.  The first is of the Shiraishi and Mori form in Eq.~\eqref{eq:SM} which requires a set of local projectors $\{P_i \}$ which annihilate all scar states:
\begin{equation}
    P_i (Q_{\balpha_k}^{n_k}... Q_{\balpha_2}^{n_2} Q_{\balpha_1}^{n_1}) |\psi_0 \rangle = 0
\end{equation}
for all $i$ and any set of powers $n_\mu$.
In general, we will restrict ourselves to translation-invariant $h_{i}$ in Eq.~\eqref{eq:SM}, to ensure that eigenstates of $H_A^{(\text{SM})}$ are not many-body localized.  By choosing these $h_i$ operators sufficiently generically and with sufficiently large (but finite) range $r$, quite generally we expect that $H_A^{(\text{SM})}$ can be chosen to be ergodic on those states that it does not annihilate.

In many cases, appropriate projectors $P_i$ can be deduced from the properties of the group.  
For example, scars built atop a polarized state for an SU(2) spin-symmetric system will have maximum possible total spin for any pair of neighboring sites, so that bond-wise projectors onto states with total spin less than this maximal value must annihilate the scars.  Letting $\Pi^{\text{max}}_{i, i+1}$ be the projector onto the maximal total $Q$-spin state between sites $i$ and $i+1$, an appropriate set of bond-wise projectors is $P_{i,i+1} = 1 - \Pi^{\text{max}}_{i, i+1}$.
Likewise, for higher rank Lie groups, bond-wise projectors can be obtained by exploiting the symmetry of the state $|\bw_{\text{min},i},\bw_{\text{min},i+1} \rangle $ under interchanging indices $i$ and $i+1$.  For example,  if $|\bw_{\text{min},i}\rangle$ is a state in the fundamental representation of the group $G$, for $k=0$ the corresponding many-body state is in the completely symmetric representation (a single row, in terms of Young Tableaux).  We can therefore define the projector $\Pi_{i, i+1}$ onto completely symmetric states along the bond $(i, i+1)$. 

At this point, it is also worth commenting on the role of the momentum $k$ in defining SU(2) generators as in Eq.~\eqref{Eq:Q}.
First, the commutation relations are invariant under locally re-defining:
\be
Q^+_i \rightarrow e^{i \phi_i} Q^+_i \ , \ \ Q^-_i \rightarrow e^{-i \phi_i} Q^-_i  \ ,
\ee
and thus under changes in $k$.  The scar models that we discuss have particular values of $k$, for example $k=\pi$ for the spin-1 XY and AKLT models. This is because the states in the scar tower, and hence the choices of annihilating projectors $P_{i, i+1}$, will be $k$-dependent.  In general, for certain choices of the momentum $k$,  the $P_{i, i+1}$ may not have a simple, physical form in terms of the underlying spin operators.  

With a little more care, we can similarly locally redefine the raising operators $Q_{\alpha^+}$ for other Lie Groups $G$. In this case, we can locally and freely redefine the simple positive roots with phases, $\phi^{\balpha^{+(S)}}_i$; this then constrains the phase choice for the other positive roots and the negative roots so that, similar to the $SU(2)$ case, the commutation relations do not depend on $k$; rather, different $k$ correspond to different scar towers, annihilated by different bond-wise projectors $P_{i,i+1}$.

The second type of term that we include in $H_A$ are ``as a sum" annihilators.  These are operators of the form
\begin{equation} \label{Eq:Hasum}
    H_A^{\Sigma} = \sum_i \beta_i O_i
\end{equation}
where $O_i$ is a local operator centered at site $i$ which does not on its own annihilate the scar tower.  In this case there is no freedom to adjust the relative coefficients $\beta_i$ at different sites, since only specific superpositions annihilate all scar states. Including such operators is sometimes necessary for understanding the structure of scars in a given model; for example, Ref.~\cite{MLM} worked out a particular $H_A^{\Sigma}$ for the AKLT model. In other cases, including such terms can lead to physical and potentially experimentally realizable examples of Hamiltonian with scars, such as the one in Eq.~\eqref{Eq:MM_SM_Physical} presented in Ref.~\cite{MM}. 
Additionally, in some cases Hamiltonians of the SM form, Eq.~\eqref{eq:SM}, annihilate not only the desired scar tower, but also some of the states outside of the scar tower.  Thus in order to ensure that the only non-ergodic states in our spectrum are the scar states, it is also useful to include ``as a sum" annihilators in $H_A$. 

In order to identify the scarred models described here, we have carried out an exhaustive search for the possible contributions to $H_A$. Specifically, we present a general algorithm which, given a particular set of `target' states, constructs Hamiltonians for which the target states are eigenstates.   This is a generalized version of the covariance-matrix algorithm presented in Ref. \cite{XLQ2017}, and we recapitulate some of the main points of the algorithm for completeness.  (This method can also be useful for identifying $\Hz$ and $\Hc$.) 

Consider any $m$-dimensional linear space of Hermitian operators of interest $\mathcal{H}$ and construct a Hermitian basis $\{h_\alpha\}$ for this space. Then, given a target state $|\psi\rangle$, the null space of the $m$ by $m$ matrix 
\begin{equation}
    C_{\alpha \beta}^{|\psi\rangle} = \frac{1}{2}\langle\psi|h_\alpha h_\beta + h_\beta h_\alpha| \psi \rangle - \langle\psi|h_\alpha| \psi \rangle \langle\psi|h_\beta| \psi \rangle
\end{equation}
corresponds to the space of Hermitian operators in $\mathcal{H}$ for which the state $|\psi\rangle$ is an eigenstate. That is, from any vector $\vec{c}$ in the null space, we can construct a Hermitian operator $\sum_\alpha c_\alpha h_\alpha$  with $| \psi \rangle$ as an eigenstate. Because the covariance matrix has non-negative eigenvalues, the null space of a sum of covariance matrices $C_{\alpha \beta}^{|\psi_n\rangle}$ for multiple states $|\psi_n\rangle$ corresponds to the space of Hermitian operators in $\mathcal{H}$ that have \emph{all} the $|\psi_n\rangle$ as eigenstates. Finally, if one desires Hamiltonians that annihilate the target states, such as $H_A$, then dropping the $- \langle\psi|h_\alpha| \psi \rangle \langle\psi|h_\beta| \psi \rangle$ piece of the covariance matrix suffices.

The dimension $m$ of the covariance matrix depends on the size of the space of interest $\mathcal{H}$, but is often quite small. For example, the space of translationally invariant sums of at-most-range-2 operators is just $(2S+1)^2 ((2S+1)^2-1)$ dimensional, which is independent of $L$. Thus, the null space of the covariance matrix can be computed very quickly.  More computational effort is required to calculate the elements of the covariance matrix, and this calculation scales with the size of the eigenstates $|\psi\rangle$. However, in the case of translationally invariant (or periodic with fixed period) models and states, calculations on a small size chain can capture the null space of the infinite $L$ covariance matrix. Further, when $|\psi_n\rangle$ has a matrix product state (MPS) representation (as is the case for all of the scar states we discuss), MPS techniques are useful to calculate the elements of the covariance matrix. 

A complementary algorithm for obtaining as-a-sum annihilators was discussed in reference~\cite{MM}, which specialized to scar towers and translationally invariant operators and relied on matrix product methods. We emphasize that the covariance-matrix algorithm above does not need such specializations, and hence can be used for a wider class of target states and Hamiltonians where matrix product methods may not be readily amenable. Relaxing the restriction on translation invariance allowed us to discover a wider class of nearest-neighbor models with the spin-$1$ AKLT scar states as eigenstates than had been reported previously in the literature; we discuss this example and its generalizations in Appendix \ref{app:as-a-sum}. 
We note that this method can also be used to directly search within different classes of operators that may be of interest to different experimental setups. For example, $h_\alpha$ could be chosen to be a staggered field, $h_\alpha = \sum_i S_i^z (-1)^i$ or a particular kind of two or three body interaction. This method is also general enough to find examples of Hamiltonians that embed any specific set of target states of interest -- that may or may not be derived from symmetries -- and hence can be used to construct special `one-off' models with scars. 

At this point, we have discussed all the ingredients that enter our framework for constructing scars from symmetries, specifically: (i) $\Hc$ with a non-Abelian symmetry $G$,  (ii) the ladder operators $Q^\pm$ derived from embedded SU(2) sub-algebras of the Lie algebra, (iii) choices for $\Hz$ that may or may not be built from the CSA of the Lie algebra, (iv) choices for the base-state $|\psi_0\rangle$ that may or may not be an eigenstate of the Casimirs of $G$ and (v) choices for $H_A$ that annihilate the tower of scar states. We note that once a particular tower of states has been identified by the action of $Q^\pm$ on $|\psi_0\rangle$, then $H_A$ is the most important piece since it ensures that the Hamiltonian acts non-generically only on the scarred manifold but is well thermalizing on the rest of the spectrum. Indeed, in many cases, the simplest choice of $\Hc=0$ works. Likewise, while $\Hz$ is used to give different energies to the scar states, this is not required, and models with degenerate low entanglement are still scarred. 
In the next two sections, we present several examples of existing and new scarred Hamiltonians that lie within our framework.

\section{Symmetric scars}
\label{sec:symmetricscars}
In this section, we discuss symmetric scars, in which the scarred subspace transforms as an irreducible representation of  $G$ (or a $q$-deformed version thereof), even though the Hamiltonian as a whole is \emph{not} invariant under the symmetry.
In all such models, the scar tower is obtained by acting with raising operators 
on a low entanglement base state that has a definite eigenvalue under all the Casimir operators of $G$.
Section~\ref{sec:su2} presents various examples, several of which have been presented in the literature previously, where the scars are derived from an SU(2) or ``$\eta$-pairing SU(2)" symmetry. In Section~\ref{Sec:IntScars}, we review known examples \cite{Shibata,Buca2020} where integrability is the source of the underlying symmetry. In Section~\ref{sec:obscureroot}, we review a known example of scars \cite{ISDWC, MLM} for which the raising operators do not have a direct connection to root systems.  
Section~\ref{sec:higherspin} generalizes to higher rank Lie groups, while Section~\ref{sec:qdef} considers $q$-deformed SU(2) symmetry. For most of our examples, the base state will be a state with maximum eigenvalue under the Casimir, such as a spin-polarized state for SU(2) or its analog for general $G$, but we also present new examples of scar towers built on non-polarized states in Sec.~\ref{sec:0-}. 

\subsection{Symmetric scars from SU(2) symmetry}
\label{sec:su2}

\subsubsection{Spin-SU(2) symmetry}

We start with a particularly simple example where the symmetry group $G$ is the spin-SU(2) symmetry. Here $Q^\pm = S^\pm$, the only generator in the Cartan subalgebra is $Q^z=S^z$, and the Casimir is $Q^2 = S^2 =\frac{1}{2}(S^+S^-+S^-S^+)+(S^z)^2$. The corresponding parent Hamiltonian $\Hc$ is spin-SU(2) symmetric, and states in the scar tower share a common value of $S^2$.
Any Hamiltonian $\Hc$ with spin-SU(2) symmetry has a tunnel of $(L+1)$ degenerate states built upon a polarized base state. For example, when $S=1/2$, we can build a tunnel by acting with $S^+$ on $|\psi_0\rangle = |\downarrow\downarrow\downarrow\cdots \downarrow\rangle$. Each of these has maximal $S^2$ eigenvalue but different $S^z$ eigenvalues, and take the form $$|\psi_n\rangle = (Q^{+})^n|\psi_0\rangle \propto \left|Q^2 = \frac{L}{2} \left(\frac{L}{2}+1\right), Q^z = -\frac{L}{2} + n\right \rangle.$$ Each $|\psi_n\rangle$ is the unique eigenstate in a particular symmetry sector characterized by $(Q^2 = Q_{\rm max}(Q_{\rm max}+1), Q^z)$ eigenvalues. As discussed above, the form of $Q^+$ ensures that that these states have at most logarithmic entanglement.

The degeneracy of these states can be lifted by adding a term $\Hz = \Omega S^z$ to $\Hc$, which promotes the tunnels to towers. Finally, we can consider a Shiraishi-Mori type $H_A$, as in Eq.~\ref{eq:SM}, with projectors onto two-site singlets on neighboring sites $P_{i,i+1} = (1/4-\vec{S}_i\cdot \vec{S}_{i+1})$. Because the $|\psi_n\rangle$ have maximal total spin, they are annihilated by each of these singlet projectors. Indeed, Ref.~\cite{AbaninPerfectScars} constructed a model of `perfect scars' of exactly this form: 
\begin{equation}
    H = \Omega \sum_{i}S^z_i + \sum_i V_{i-1, i+2}P_{i,i+1}
\label{eq:dimaperfect}
\end{equation}
where $V_{i,j} = \sum_{\mu, \nu}  J^{\mu, \nu}_{i,j}S_i^\mu S_j^\nu$ is an arbitrary operator that is used to break the spin-SU(2) symmetry. In this ``perfect scar" model, $Q^+=S^+$,  $\Hc=0$, $\Hz = \Omega S^z$, and $H_{A} = \sum_i V_{i-1, i+2}P_{i,i+1}$. Note that even though $\Hc=0$, the action of $H_A$ makes the model well thermalizing outside the scarred subspace, and the scarred states still inherit the SU(2) algebra.

A different example with the same maximal spin scar states is given in reference \cite{MM}: 
\begin{equation}\label{Eq:MM_SM_Physical}
    H = \sum_{i}  J_1\vec{S}_{i} \cdot \vec{S}_{i+1} + J_2 \vec{S}_{i} \cdot \vec{S}_{i+2} +
    D \hat{z}\cdot (\vec{S}_{i} \cross \vec{S}_{i+1})
\end{equation}
Unlike the previous model, Eq.~\ref{eq:dimaperfect}, this model has a non-trivial $\Hc= J_1\vec{S}_{i} \cdot \vec{S}_{i+1} + J_2 \vec{S}_{i} \cdot \vec{S}_{i+2}$ with spin-SU(2) symmetry, but has $\Hz=0$ so that all the scars are degenerate (one could, of course, equally well add a term of the form $\Hz = S^z$). The final term, $H_A = D \sum_i \hat{z}\cdot (\vec{S}_{i} \cross \vec{S}_{i+1})$ breaks the SU(2) symmetry and annihilates the scars, but it is not of the SM form since it only annihilates the scar states as a complete sum, whereas previously each local projector individually annihilated the scar states. 

\subsubsection{$Q$-SU(2) symmetry}

Next, we consider a model where the operators $\{Q^\pm, Q^z\}$ satisfy SU(2) commutation relations, but are distinct from the spin-SU(2) operators. In particular, we can choose $Q^\pm, Q^z$ according to
\be
Q^{\pm} = \frac{1}{(2S)!}\sum_i e^{ikr_i}(S^\pm_i)^{2S} \ , \ \ Q^z = \frac{1}{2} [ Q^+, Q^- ]. 
\label{eq:qsu2}
\ee
This choice with spin $S=1$ and $k=\pi$ produces the SGO of two well-known scar models in the literature: the spin-1 AKLT model~\cite{MoudDisc, MoudEntang, MLM, MoudMPS} and the spin-1 XY model~\cite{ISXY, Chattopadhyay, MLM, MoudMPS}. Here $Q^\pm_i = (S^\pm_i)^{2S}$  raises (lowers) the spin $-S$ (spin $S$)  state to a spin $S$ (spin $-S)$ state, and annihilates all other states. Thus the $(2S+1)$ states on each site are divided into the $Q$-spin-$1/2$ doublet $\{|-S\rangle, |S \rangle \}$, and $2S-1$ $Q$-spin singlets\footnote{This is an example of an embedded SU(2)-sub-algebra of SU(2S+1) on each site.  For $S=1$, there are three independent SU(2) sub-algebras in SU(3), and the Gell-Mann matrices provide a natural basis for these embedded sub-algebras of which the choice described in Eq.~\ref{eq:qsu2} represents one.}.
Correspondingly, the operators $Q^+, Q^-$ and $Q^z$ obeys the Lie algebra SU(2), but the resulting $Q$-SU(2) symmetry is distinct from the spin-SU(2) symmetry. 

It is natural to also consider other onsite raising operators of the form $Q^+_i = (S^+_i)^n$ for $1 \leq n \leq 2S$. However, for $n<2S-1$ and $S>3/2$, these operators do not describe an SU(2) algebra but rather form higher-rank Lie group symmetries, as discussed in detail in Sec.~\ref{sec:higherspin}. 

As before, we use the operators $Q^+$ to construct scar states built upon a base state that is an eigenstate of both $Q^z$ and $Q^2$ so that all of the scars share the same eigenvalue of $Q^2$, but are distinguished by their eigenvalues under $Q^z$. 
A particular example of this kind in furnished in the spin-$1$ XY model ~\cite{ISXY,MLM, MM, MoudHubb}:
\begin{align}
    H &= \sum_{i} J(S^x_i S^x_{i+1} + S^y_i S^y_{i+1}) + J_3(S^x_i S^x_{i+3} + S^y_i S^y_{i+3})  \nonumber \\
       &+ h S^z_i +  D (S^z_i)^2.
\label{Eq: 1XY}
\end{align}
The scars are built by the action of $Q^+$ on the fully polarized down state $|\psi_0 \rangle = |---\cdots-\rangle$. Note that the first term $\propto J$ breaks $Q$-SU(2) symmetry and annihilates the scars, the term $h S^z$ acts as $\Hz$ and gives energy to the scars, while the term $\propto D$ commutes with $Q^z$ and $Q^{+}$. 
The third neighbor term is added to break a non-local SU(2) symmetry for which the scar states are the only states in their symmetry sector.   Specifically, the raising operator associated with this nonlocal SU(2) symmetry is obtained by replacing $(-1)^i \rightarrow e^{i\pi \sum_j=1^i S_j^z}$ in the expression (\ref{eq:qsu2}) for $Q^+$. The resulting ladder operator acting on our fully polarized base state generates the same scar tower.  Thus this non-local SU(2) symmetry and its corresponding Casimir must broken in order for our scar tower to violate ETH.

Similar physics is also at play in the $\eta$-pairing states of the Hubbard model on bipartite lattices~\cite{Yangeta}. The Hubbard model has both a spin SU(2) symmetry \emph{and} an independent ``$\eta$-pairing" SU(2) symmetry (which plays the role of the $Q$-SU(2) symmetry). The $\eta$-pairing states have low-entanglement~\cite{Vafek2017}, and are the unique states in the symmetry sector of maximal ``$\eta$-pairing" total spin (i.e. states with maximal eigenvalues under $Q^2$). Analogous to the examples above, the Hubbard model can be perturbed by a suitable $H_A$ to break the $\eta$-pairing SU(2) symmetry while preserving the $\eta$-pairing states as scarred eigenstates in the perturbed model~\cite{MoudHubb, MM}; the Hirsch model furnishes a notable example~\cite{MM}. Strikingly, there exists a simple mapping from spin-1 models above to electronic models that allows for translation between the scar states of the spin-1 XY model and the eta-pairing scars of the Hirsch model and some related electronic models~\cite{MM}.

\subsubsection{Scar towers from base states of non-maximal spin}
\label{sec:0-}
The above examples, drawn from previous literature, contain scar towers generated from a fully polarized state for the base state. In each case, this meant that the scar tower transformed in an irreducible representation of $Q$-SU(2) with maximal spin.  
We emphasize, however, that maximal spin (or, more generally, extremal Casimir eigenvalues) are not necessary for scar states, though they are useful for enumerating the bond-wise annihilators. 

\begin{figure}
\centering
\includegraphics[width=\columnwidth]{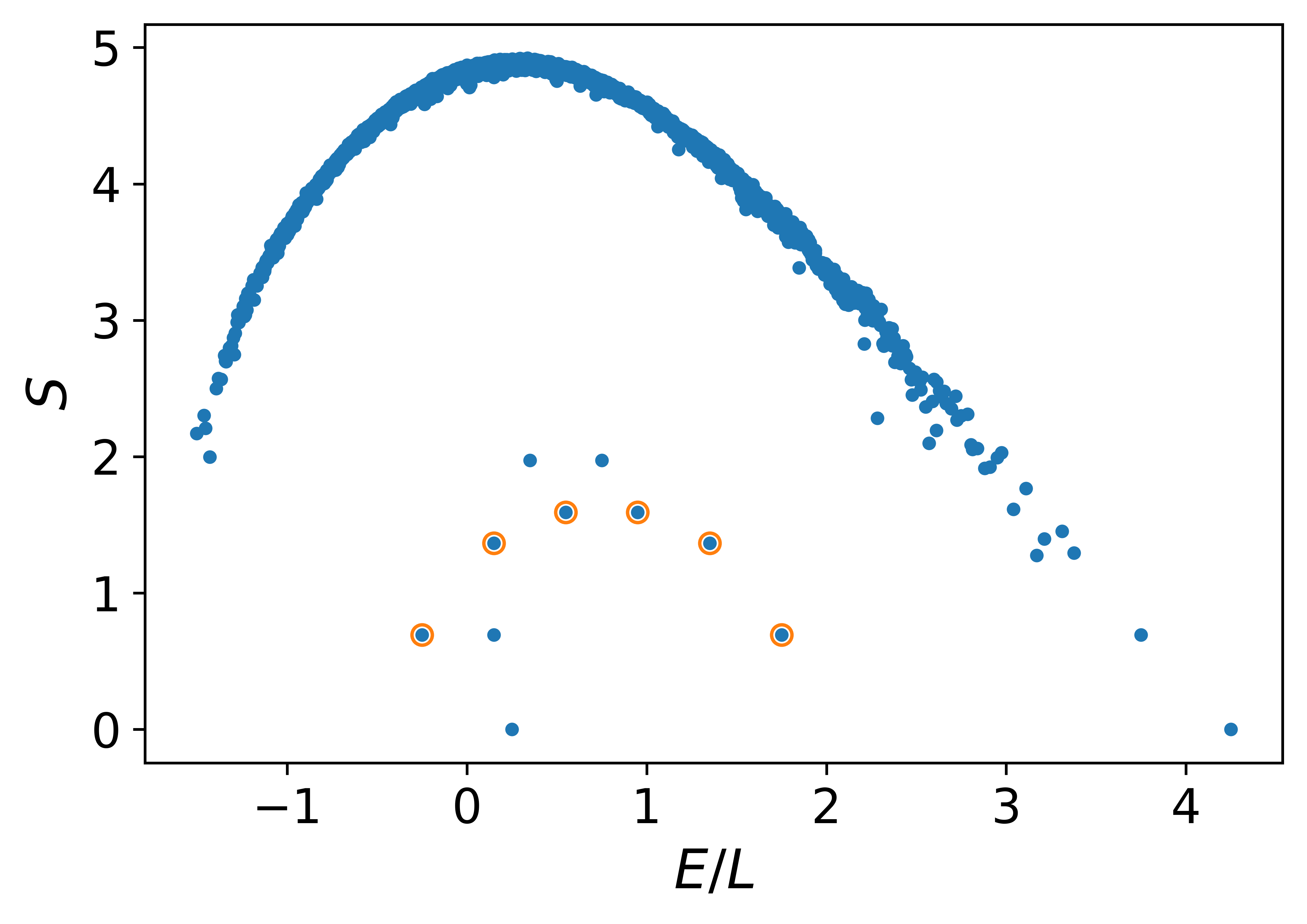}~
\caption{Entanglement entropy vs. energy density in the $k=0$ momentum sector of the periodic chain of length $L=10$ with Hamiltonian Eq.~\eqref{eq: 0-} and parameters $h=2$, $D=B_1=B_2 = J_z=1$, $J_1 = -3/4$, and $J_2 = 1/2$. The circled states are a scar tower that transforms in an irrep of $Q$-SU(2) with less than maximal spin. 
\label{fig:0-}}
\end{figure}

To demonstrate this, we offer a simple example.  Consider a spin-1 chain 
described by the Hamiltonian $H = \Hc+\Hz+H_A$ with:
\begin{align}
\label{eq: 0-}
\Hc &= D (S_i^z)^2 \nonumber\\
\Hz &= h S_i^z  \nonumber\\
H_A &=  \sum_i J_1 (\vec{S}_i \cdot \vec{S}_{i+1})^2 + J_2( (\vec{S}_i \cdot \vec{S}_{i+2})^2 + \vec{S}_i \cdot \vec{S}_{i+2}) \nonumber \\ 
&+ B_1 ((S^z_{i})^2 (S^x_{i+1}+S^y_{i+1})- (S^x_{i+1} + S^y_{i+1}) (S^z_{i+2})^2)   \nonumber \\ 
& + J_z S_i^z S_{i+1}^z  + B_2 (S^z_{i} (S^z_{i+1})^2 - (S^z_{i+1})^2 S^z_{i})
\end{align}
This model has a scar tower,  generated by acting with $Q^+ = \sum_{i} \frac{1}{2} (S^+_i)^2$, which is of the form in Eq.~\ref{eq:qsu2} with $k=0$ and $S=1$ on the base state $|\psi_0 \rangle =\frac{1}{\sqrt{2}}|0-0-...0-\rangle + \frac{1}{\sqrt{2}}|-0-0...-0\rangle$. As promised, $|\psi_0 \rangle$ is an eigenstate of the Casimir $Q^2$ with eigenvalue $L/4(L/4+1)$ which is less than the maximal Casimir $L/2(L/2+1)$.

To see that the terms act as labeled, observe that the terms with proportionality constants $B_1$, $B_2$, and $J_z$ all annihilate all states in this scar tower bond-wise, because every state in the scar tower has $|0\rangle$ on every other site. Similarly, $J_1$ is equal to the identity plus three times the projector onto the singlet state and is hence also a bond-wise annihilator on subtracting out the identity. $J_2$ is another bond-wise annihilator up to a factor of the identity, as it is equal to a linear combination of the identity and a projector onto the antisymmetric spin-$1$ states.

We emphasize that the $J_2$ term is sensitive to the momentum of $Q^+$; further, without this term, any states of the form $| a_1 0 a_2 0 a_3 0 ... \rangle$ where $a_i = \pm 1$ would be eigenstates. Finally, the terms $B_1$ and $B_2$ help us to break symmetries and make the model thermal. 
Collectively, the terms in $H_A$ are sufficient to render all but a few of the states outside the scar space thermal, as seen in Figure~\ref{fig:0-}; the fully polarized up and down states remain as eigenstates despite $S^z$ being broken, so those states are also scars. We also note that Eq.~\eqref{eq: 0-} contains only a subset of the operators that could be added to $H_{A}$ to make the model thermal; others can be found using the covariance-matrix algorithm.

More generally, base states with other eigenvalues under $Q^2$ and $Q^z$, such as those with eigenvalue $(Q_{\rm max}-p)(Q_{\rm max}-p+1)$ under $Q^2$, and $(-Q_{\rm max}+p)$ under $Q^z$ for some finite $p$, will also have low entanglement and can be used to build scar tunnels.

\subsection{Scars from integrability} \label{Sec:IntScars}

It is also possible to write scarred models that fit into our general framework, Eq.~\ref{eq:tunneltotower}, using commutation relations derived from algebraic structures following from integrability. Ref.~\cite{Shibata} furnished such an example using integrable clock models~\cite{Vernier2019}. In this case, $\Hc$ is an integrable model with extensively many conserved quantities associated with an infinite-dimensional Onsager algebra. The conserved quantities take the form of sums of local operators; some of the conserved quantities raise the total z-magnetization by fixed amounts, and can be chosen as the analog of our $Q^+$.  Then, picking $H_{SG} = S^z$, Ref.~\cite{Shibata} derived a scarred model by adding an $H_{A}$ that broke integrability and annihilated scars built on a polarized down state. In a similar spirit, Ref.~\cite{Buca2020} used the fact that \emph{quasilocal} conserved operators in certain parameter regimes of a spin-1/2 XXZ model obey a SGA with $S^z$~\cite{ProsenXXZ2016} to derive the presence of persistent oscillations in these integrable models. However, the perturbations to integrability considered in Ref.~\cite{Buca2020} were general and not aimed at preserving a scar tower. In general, the quasilocality may make finding a choice of $H_{A}$ more challenging in this case. 

\subsection{Raising operators without connections to root structures}
\label{sec:obscureroot}
Before we discuss higher rank and $q$-deformed groups, we emphasize that Eq.~\ref{eq:tunneltotower} does not require direct connections to the root structures of named Lie algebras. It is enough to have a dynamical symmetry indicated by the presence of an SGA between the raising operator $Q^+$ and $\Hs$, along with an $H_A$ that annihilates the scars and can make the rest of the spectrum thermal. For example, Ref.~\cite{ISDWC} investigated an exact set of scar states in the spin-1/2 domain-wall-conserving model with a raising operator lacking a clear connection to roots. The scars were generated by $Q^+_{\rm DWC} = \sum_i P^-_{i-1} \sigma^+_i P^-_{i+1}$, with projectors $P^-$ onto spin down, acting repeatedly on the fully spin down product state. As argued in Ref.~\cite{ISDWC}, the ``lowering" operator to undo the action of $Q^+_{\rm DWC}$ on the scar states is necessarily non-local, meaning that the scar states do not transform in a single representation of any symmetries associated with $Q^+_{\rm DWC}$, similar to the AKLT and generalized AKLT scars discussed below. However, \textit{unlike} the AKLT scars for which $Q^+_{\rm AKLT}$ is a generator of $\rm SU(2)$, $Q^+_{\rm DWC}$ is not clearly related to the root structures of named Lie algebras: the projectors spoil the $\rm SU(2)$ commutation relations with $Q^+_{\rm DWC}$'s Hermitian conjugate, and the commutators of the resulting operators do not close readily, obscuring the relation to other Lie algebras. Nevertheless, the Hamiltonian and scar states still follow Eq.~\ref{eq:tunneltotower} despite being generated by operators $Q^+$ without direct connections to the root systems of Lie algebras.

\subsection{Higher rank Lie group symmetric scars}
\label{sec:higherspin}

Another new class of examples that our symmetry-based perspective on scars makes natural is scar states associated with continuous symmetry groups $G$ other than SU(2).  As discussed in Section \ref{sec:ingredients}, these differ from the SU(2) case in a few important ways.  
First, in general there are multiple choices of raising operators.  Second, there are multiple choices of $\Hz$, which in general satisfy commutation relations of the form (\ref{Eq:Raising}).  Depending on the choice of $\Hz$, we can therefore engineer scar states with multiple distinct frequencies, or with exact degeneracies in their spectra that reflect the more complex Lie group symmetry.

The general idea of the construction closely parallels the SU(2) case.  Choosing $|\psi_0 \rangle = \prod_i |\bw_{\text{min},i}\rangle$ to be a product of the lowest weight state at each site, 
we have $Q_{\balpha^-,i} |\bw_{\text{min},i}\rangle =0$. This is the analog of the polarized state, and has maximal eigenvalue under the Casimir. 
The scar tower then consists of all states of the form (\ref{Eq:psin}).

As a simple example, we consider a spin-$1$ chain. The three states a given site can be viewed as transforming in the 3-dimensional fundamental representation of SU(3), which has $N=8$ generators and rank $R=2$. The six generators that are not in the CSA furnish three raising and three lowering operators. In the $S^z_i$ basis $|+_i \rangle = (1,0,0)^T, \ | 0_i \rangle = (0,1,0)^T, \ |-_i \rangle = (0,0,1)^T$, the three raising operators of SU(3) are 
\begin{align}
    Q_{\balpha_1^+,i}&=\left[
\begin{array}{ccc}
0 & 0& 0 \\ 
0& 0 & 1 \\ 
0& 0& 0  
\end{array}\right], \, \nonumber \\
Q_{\balpha_2^+,i} &=\left[
\begin{array}{ccc}
0 & 1 & 0 \\ 
0 & 0 & 0 \\ 
0 & 0 & 0   
\end{array}\right], \, \nonumber \\
Q_{\balpha_3^+,i} &=\left[
\begin{array}{ccc}
0 & 0 & 1 \\ 
0 & 0 & 0 \\ 
0 & 0 & 0   
\end{array}\right]
\end{align}
Using the following basis for the 2 generators of the Cartan subalgebra on site $i$:
\begin{equation}
    Q^z_{1,i} = \left[ \begin{array}{ccc}
1 & 0 & 0 \\ 
0 &- 1 & 0 \\ 
0 & 0 & 0   
\end{array}\right], \, Q^z_{2,i} = \frac{1}{\sqrt{3}} \left[ \begin{array}{ccc}
1 & 0 & 0 \\ 
0 & 1 & 0 \\ 
0 & 0 & -2  
\end{array}\right], \,
\end{equation}
the roots are
\begin{equation} \label{Eq:TheRoots}
    \balpha_1^+ = (-1, \sqrt{3}) \ , \ \ \balpha_2^+ = (2,0) \ , \ \ \balpha_3^+ = (1, \sqrt{3} ) \ .
\end{equation}
Note that $Q_{\balpha_3^+,i} = [ Q_{\balpha_2^+,i}, Q_{\balpha_1^+,i}] = Q_{\balpha_2^+,i} Q_{\balpha_1^+,i}$; hence $Q_{\balpha_1^+,i}$ and $Q_{\balpha_2^+,i}$ generate the complete set of states on site $i$ from the lowest weight state $|-\rangle_i$. 

We now consider the global raising operators $Q_{\balpha^+_j} = \sum_{i=1}^L Q_{\balpha^+_j,i}$. 
With a base state $|\psi_0\rangle = \prod_i |-_i\rangle$, the scar space is spanned by the $\frac{1}{2}(L+1)(L+2)$ states
\begin{equation}
\label{Eq: steq}
    Q_{\balpha^+_2}^m Q_{\balpha^+_1}^n|\psi_0\rangle \textrm{ for } 0 \leq m \leq n \leq L,
\end{equation}
\emph{i.e.} by all states in the $\frac{1}{2}(L+1)(L+2)$-dimensional irreducible representation of SU(3) containing the lowest weight state $|\psi_0 \rangle$.
These states will have at most $\log(m+1)+\log(n+1)$ entanglement entropy, since both $Q_{\balpha^+_1}^n$ and $Q_{\balpha^+_2}^n$ have MPO representations with bond-dimension $n+1$; see Appendix \ref{App:MPO}. 

There are two natural physical operators for $\Hz$: $S^z = \sum_i S^z_i$ and $\sum_i (S^z_i)^2$. 
These can be expressed as 
\begin{align}
    S^z_i &=\frac{1}{2} Q^z_{1,i} +\frac{ \sqrt{3}}{2} Q^z_{2,i} \\
    (S^z_i)^2 &= \frac{2}{3} \mathbf{1} + \frac{1}{2}Q^z_{1,i} - \frac{\sqrt{3}}{6} Q^z_{2,i}
\end{align}
Using Eqs.~\eqref{Eq:HzGen}, \eqref{Eq:Raising}, \eqref{Eq:TheRoots},  both $Q_{\balpha^+_1}$ and $ Q_{\balpha^+_2}$ raise $S^z_i$ by one, while 
$Q_{\balpha^+_1}$ ($Q_{\balpha^-_2}$) decreases (increases) the eigenvalue of  $(S^z_i)^2$ by 1. Correspondingly, the eigenvalue of $S^z$ on the state in Eq.\eqref{Eq: steq} is $-L+m+n$, while the eigenvalue of $\sum_i (S^z_i)^2$ is $L-n+m$.

We now turn to $H_A$. At the two-site level, the states in the tower will only contain the six symmetric states $|--\rangle,\, |-0\rangle + |0-\rangle,\, |00\rangle,\, |+-\rangle + |-+\rangle,\, |+0\rangle + |0+\rangle\, \text{and } |++\rangle$. The three antisymmetric states $|1,-1\rangle = \frac{1}{\sqrt{2}}(|-0\rangle - |0-\rangle)$, $|1,0\rangle =\frac{1}{\sqrt{2}}(|+-\rangle - |-+\rangle)$, and $|1,1\rangle =\frac{1}{\sqrt{2}}(|+0\rangle - |0+\rangle)$ do not appear, and so we can use projectors onto these states as Shiraishi-Mori-type projectors. Correspondingly, there are 9 bond-wise annihilators on a given pair of sites $(i, i+1)$ that we can use in $H_A$: 
\begin{align} \label{Eq:P1P9}
&P_{1,i} = |1,-1\rangle \langle 1,-1|, \ \ \ 
P_{2,i} = |1,0\rangle \langle 1,0|,  \ \ \ \nonumber\\
&P_{3,i} = |1,1\rangle \langle 1,1|, \ \ \ P_{4,i} = |1,-1\rangle \langle 1,0|+h.c., \nonumber \\  
&P_{5,i} = i|1,-1\rangle \langle 1,0| + h.c., \ \ \   P_{6,i} = |1,-1\rangle \langle 1,1| + h.c., \nonumber \\ 
&P_{7,i} = i|1,-1\rangle \langle 1,1| + h.c., \ \ \  P_{8,i} = |1,0\rangle \langle 1,1| + h.c.,  \nonumber\\
&P_{9,i} = i|1,0\rangle \langle 1,1| + h.c. 
\end{align}
 We also found eight ``as-a-sum" annihilators through the covariance-matrix algorithm discussed above, but we will not discuss these annihilators further here.

\begin{figure}[h]
\includegraphics[width=\columnwidth]{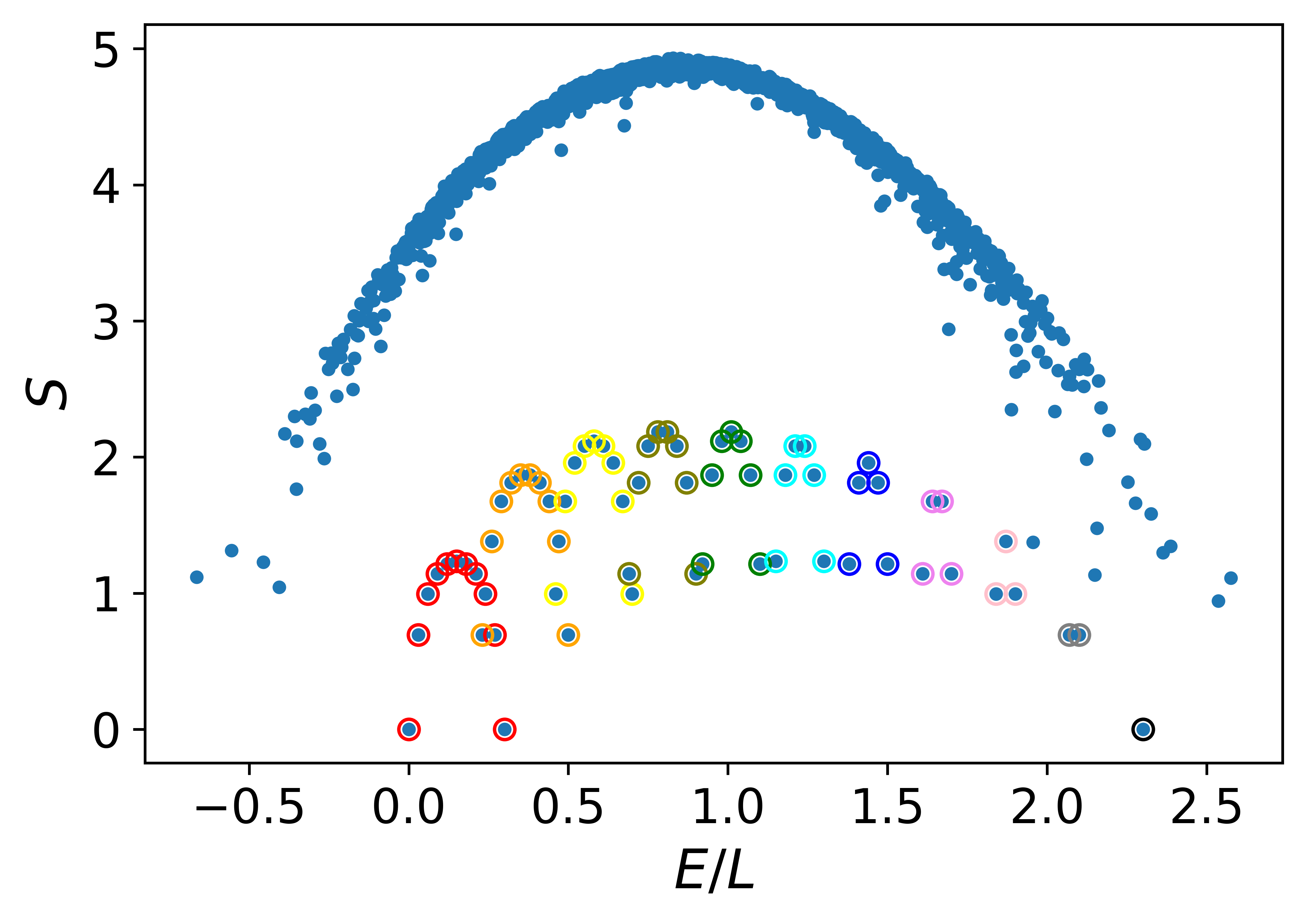}
\caption{\label{fig: su3ee} Entanglement entropy in the momentum $k=0$ and spatial inversion-symmetric sector of the SU(3)-scarred Hamiltonian in Eq. \eqref{eq: hsu3} for a periodic chain of length $L=10$. The scar states are colored according to their values of $m$ in Eq.~\eqref{Eq: steq}. }
\end{figure}

The terms in Eq.~\eqref{Eq:P1P9} break SU(3) symmetry and keep the Hamiltonian from commuting with the two SU(3) Casimirs; hence taking $H_A$ to be a linear combination of these bond-wise annihilators at each site is sufficient to eliminate the symmetry, and indeed can lead to a spectrum that is ergodic in the non-scarred Hilbert space.
This is seen in Figure \ref{fig: su3ee}, which shows the entanglement entropy in the $k=0$ and inversion-symmetric sector of the Hamiltonian
\begin{align}
\label{eq: hsu3}
    H &= \sum_j  S^z_j + 1.3 (S^z_j)^2  -.5 P_{4,j} + 1.2 P_{5,j} \nonumber \\
    &+ .9 P_{6,j}  - 1.7 P_{7,j}+ 1.7 P_{8,j} - 1.5 P_{9,j}
\end{align}
We've colored the scar states according to their values of $m$ from Eq.~\ref{Eq: steq}; we have an evenly spaced tower of states for each value of $m$, with $n$ ranging from $m$ to $L$. As we increase $m$ by one, the resulting tower has one fewer state than the previous.  For the parameters chosen here, increasing $m$ corresponds to increasing the energy by $2.3$, while increasing $n$ corresponds to decreasing the energy by $.3$. 

These scar states also allow us to construct families of simple product states that exhibit periodic or quasiperiodic revivals. For any two complex parameters $c_1$ and $c_2$ the product states
\begin{equation}
\label{Eq: stsu3}
    |c_1, c_2\rangle = \otimes_i \left( \frac{|+\rangle_i + c_1|0\rangle_i + c_2 |-\rangle_i}{\sqrt{1+|c_1|^2+|c_2|^2}}\right)
\end{equation}
are superpositions of the scar states with no overlap with states outside the scar sector. To see this, first define $n_{+}$, $n_{0}$, and $n_{-}$ as the number of $|+\rangle$, $|0\rangle$, $|-\rangle$ in a given $S^z$ product state, $S_L$ as the set of permutation operators on a length $L$ chain, and the normalization $\mathcal{N} = \left(\frac{1}{\sqrt{1+|c_1|^2+|c_2|^2}}\right)^L$. Then expanding the product above yields
\begin{align}
\begin{split}
\label{Eq: stsu32}
    |c_1, c_2\rangle = & \mathcal{N} \sum_{ \substack{n_+, n_0, n_- \geq 0; \\n_+ + n_0+n_-=L}} \frac{c_1^{n_0} c_2^{n_-}}{n_+! n_0! n_-!}  \\
    &\sum_{\mathcal{P} \in S_L}  \mathcal{P} \left( \otimes_{i=1}^{n_+}|+\rangle_i   \otimes_{j=n_++1}^{n_+ + n_0}|0\rangle_j     \otimes_{k=n_++n_0+1}^L|-\rangle_k \right) \\
    = & \mathcal{N} \sum_{0\leq m \leq n \leq L}  \frac{c_1^{n-m} c_2^{L-n}}{m!n!} Q_{\balpha^+_2}^m Q_{\balpha^+_1}^n |\psi_0\rangle
\end{split}
\end{align}
where $|\psi_0\rangle = |--\cdots--\rangle$ is the polarized state as before.

Let $h$ be the coefficient of $\sum_i S_i^z$ and $D$ be the coefficient of $\sum_i (S_i^z)^2$ in the Hamiltonian; we noted above that the scar states had energy $L(D-h)+(h-D)n + (h+D)m$. Thus, for $(h-D)$ and $(h+D)$ commensurate, the above family of product states undergoes exactly periodic revivals, while for $(h-D)$ and $(h+D)$ incommensurate we have quasiperiodic revivals for $c_1, c_2 \neq 0$. The periodic revivals will have the usual period associated with two different nonzero angular frequencies; i.e., writing $\frac{h+D}{h-D} = \frac{p}{q}$ for integer $p$ and $q$, we will have a period of $\frac{p}{h+D} 2\pi$. There are also a couple interesting limits of the coefficients. Setting $c_1$ to zero, $c_2$ to zero, or $c_1$ and $c_2$ to infinity will give perfectly periodic revivals with periods $\frac{\pi}{h}$, $\frac{2\pi}{h-D}$ and $\frac{2\pi}{h+D}$ respectively.

\begin{figure}[h]
\includegraphics[width=\columnwidth]{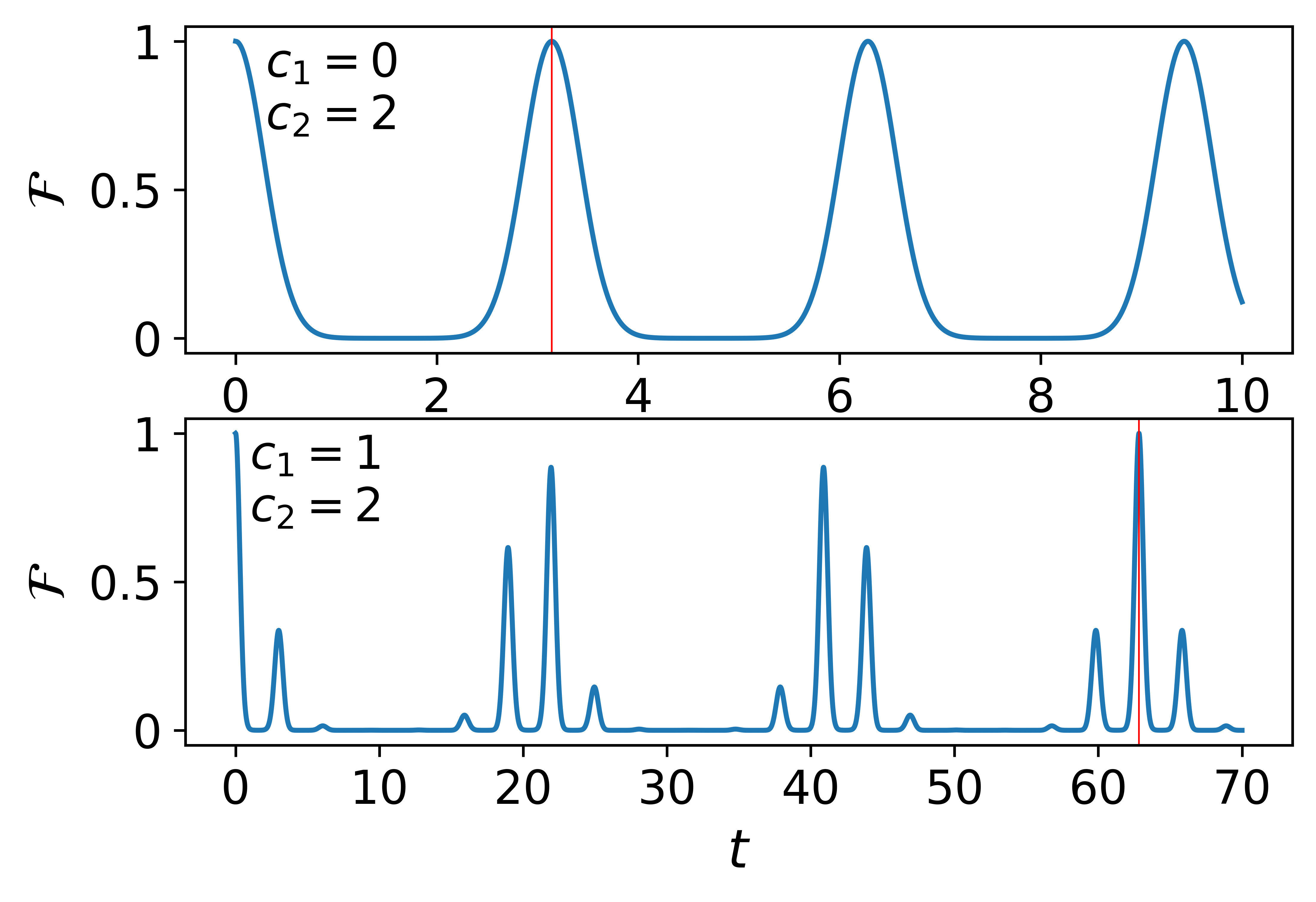}
\caption{\label{fig: stack} Fidelity $\mathcal{F} = |\langle \psi(t)|\psi(0)\rangle|^2$ of the state in Eq.~\ref{Eq: stsu3} with $c_1=0,\, c_2=2$ (top) and $c_1=1,\, c_2=2$ (bottom) under the time evolution generated by the Hamiltonian in Eq.~\ref{eq: hsu3} for $L=10$. The periods of $\pi$ and $20\pi$ are marked with vertical red lines.}
\end{figure}

We demonstrate periodic revivals from initial states $|c_1, c_2\rangle$ with $c_1=0,\, c_2=2$ with the Hamiltonian in Eq.~\ref{eq: hsu3} in Fig.~\ref{fig: stack} (top). The choice of $c_1=0$ means $|c_1, c_2\rangle$ will only have overlap with the scar states that have only $|+\rangle$'s and $|-\rangle$'s, and so the scar states will be separated by an energy spacing of $2h =2$, giving a period of $\pi$. We also demonstrate periodic revivals with $c_1=1, c_2=2$ with the Hamiltonian in Eq.~\ref{eq: hsu3} after a period of $20\pi$. The structure of these revivals is more complicated; there are partial revivals at intermediate times, arising from the two different sets of energy spacings.

\subsubsection{Higher rank Lie symmetries from spin operators}

A priori, it is not obvious under what conditions higher rank Lie group symmetries would arise in real solid-state systems, such as spin chains. 
In fact, however, we are naturally led to these if we consider raising operators of the form
\be \label{Eq:Sn}
Q^+_i = \frac{1}{\mathcal{N}} (S^+_i)^n \ , Q^-_i = (Q^+_i)^\dag \ , Q^z_i = [Q^+_i, Q^-_i] \ ,
\ee 
for $1 \leq n \leq 2S$, where $\mathcal{N}$ is a normalization constant. For $n = 1$ or $n= 2S$, with a suitable choice of $\mathcal{N}$ the operators $Q^+_i, Q^-_i$, and $Q^z_i$ form an SU(2) algebra.
For $n<2S-1$ and $S>3/2$, however, in general Eq. (\ref{Eq:Sn}) does not describe an SU(2) algebra.  Indeed, the set $Q^+_i, Q^-_i, Q^z_i$ is not closed under commutation. Closing these operators under commutation leads to a set of raising operators associated with a larger Lie group symmetry.  

To illustrate this, we begin with $S=5/2$, and $n=2$ in Eq. (\ref{Eq:Sn}).  We define
\begin{align}
    &P^+ = [Q^z, Q^+] \ , \ \ P^- = -[Q^z, Q^-] \ , \ \ P^z = [P^+, P^-] \nonumber \\ 
 &   R^+ = [ P^+,Q^+] \ , \ \ R^- = [Q^-, P^-]
\end{align}
Here $P^+$ ($P^-$) is a second, linearly independent operator that raises (lowers) $S^z$ by 2, and $R^+$ ($R^-$) is a third raising operator, which raises (lowers) $S^z$ by 4.  It is convenient to change basis, defining the raising operators:
\begin{align} \label{Eq:Lambdas}
  &  Q_{\balpha_1^+} =\frac{13}{12 \sqrt{5}} Q^{+} + \frac{1}{12 \sqrt{5}} P^{+} \ , \ \ Q_1^z = \frac{1}{1512}(- Q^z + P^z )   \nonumber \\
&   Q_{\balpha_2^+} = - \frac{1}{36} (Q^{+} +  P^{+}) \ , \ \ Q_2^z = \frac{3037}{7560 \sqrt{3}} Q^z -\frac{13}{7560 \sqrt{3}} P^z\nonumber \\
 &   Q_{\balpha_3^+} =\frac{1}{36 \sqrt{5}} R^{+}
\end{align} 
where we have taken $\mathcal{N} = \frac{1}{2 \sqrt{2}}$, and as usual, the lowering operators are given by $Q_{- \balpha_j} = Q_{\balpha_j}^\dag$.  As above, the roots $\balpha_j$ are given by Eq. (\ref{Eq:TheRoots})
The 3 raising operators $Q_{\balpha_j}$ act on our states according to:
 
\begin{eqnarray}
    Q_{\balpha_1^+}|5/2,-5/2 \rangle=|5/2,-1/2\rangle \ , \ \  Q_{\balpha_1^+}|5/2,1/2\rangle = |5/2,5/2\rangle \n
     Q_{\balpha_2^+}|5/2,-1/2 \rangle =  |5/2,3/2\rangle \ , \ \  Q_{\balpha_2^+}|5/2,-3/2\rangle = |5/2,1/2\rangle \n
     Q_{\balpha_3^+}|5/2,-5/2\rangle = |5/2,3/2\rangle \ , \ \  Q_{\balpha_3^+}|5/2, -3/2\rangle = |5/2,5/2\rangle \nonumber 
\end{eqnarray}
where states not shown are annihilated by the raising operator in question.  

It is straightforward to check that the raising operators $\{  Q_{\balpha^+_j} \}$, together with the corresponding lowering operators $\{  Q_{\balpha^-_j} \}$, and two diagonal generators $Q^z_j$, obey the commutation relations of the 8 generators of the Lie group SU(3).  
Under the action of these 8 matrices, the 6 states in $s=5/2$ split into two sets of three, which are not connected by any raising operator.  Thus the representation on each site consists of 
one copy of the fundamental (triplet) representation of SU(3), containing the states $|5/2,-5/2\rangle,|5/2, -1/2\rangle,$ and $|5/2,3/2\rangle$, and a copy of the conjugate (anti-fundamental) representation, containg the states  $|5/2,-3/2\rangle, |5/2,1/2\rangle, |5/2,5/2\rangle$.

For general spin $S$ and $n=2$, we show the following in Appendix \ref{App:SpinLie}.  For half-integer $S $, the relevant Lie algebra is SU($S+1/2$), with the $2S+1$ states on each site dividing into a copy of the $(S+1/2)$-dimensional fundamental representation, and a copy of its conjugate.    For integer $S$, the algebra can be divided into two sets of operators, which act only on even and odd integer spins, respectively.  This leads to a Lie algebra structure SO(S +1) $\times$ Sp($S$) for even $S$, and SO($S$) $\times$ Sp($S+1$) for odd $S$.  For even $S$, the Hilbert space at each site corresponds to a copy of the $S+1$-dimensional vector representation of SO($S+1$), containing the even-integer spins, and a copy the $S$ dimensional fundamental representation of Sp($S$), containing the odd-integer spins.  For odd $S$ the $S+1$ odd-integer spins transform in the $(S+1)$ dimensional fundamental representation of Sp($S+1$), while the $S$ even-integer spins transform in the $S$-dimensional vector representation of SO($S$).

For these examples, though the Cartan generators $Q^z_{\mu,i}$ all commute with $S^z_i$, it is not in general the case that $S^z_i$ can be expressed as a linear combination of the $Q^z_{\mu,i}$, since it is not necessarily traceless when acting on each irreducible representation of the relevant Lie group in the Hilbert space. 
 Thus a natural alternative to an $\Hz$ of the form (\ref{Eq:HzGen}) is to take $\Hz = S^z = h \sum_i S^z_i$, which satisfies an SGA commutation relation of the form (\ref{Eq:Raising}) for all raising operators $Q_{\balpha^+_j}$.  In this case the value of $\omega$ is fixed by how much $Q_{\balpha_j^+}$ raises $S^z$.  Thus all frequencies are integer multiple of the elementary frequency $2 n h$, and in general multiple $Q^+_i$ operators will be associated with the same frequency.

With this choice, we find degeneracies in the scar tower characteristic of the underlying larger Lie group symmetry. For example, consider the spin-5/2 system described above, with an SU(3) symmetric scar tower.  
The two operators  $Q_{\balpha^+_1}, Q_{\balpha^+_2}$ both raise $S^z$ by $2$, while $Q_{\balpha^+_3} = Q_{\balpha^+_2}Q_{\balpha^+_1}$ raises $S^z$ by $4$.  Taking $|\psi_0\rangle = |-5/2,-5/2,-5/2, ... \rangle$ to have energy $0$, we see that  $(Q_{\balpha^+_1})^2 |\psi_0\rangle $ and $Q_{\balpha^+_2} Q_{\balpha^+_1}  |\psi_0\rangle$ are linearly independent states with the same energy of $4 h$. 
In contrast, in the SU(2) case, all states in the scar tower have distinct energies, since each power of $Q^+$ applied to $|\psi_0 \rangle$ necessarily raised the eigenvalue of $S^z$ by the same amount.

\subsection{$q$-deformed towers}
\label{sec:qdef}
In the preceding sections, we considered scar states that transformed in a single irreducible representation of some group. However, we can also consider scar states transforming in representations of ``$q$-deformed groups". $q$-deformed groups have found many applications, including solving the quantum Yang-Baxter equation \cite{Majid1990}, describing anyons \cite{lerda1993anyons}, and phenomenologically describing perturbations to otherwise symmetric models \cite{bonatsos1999quantum}. Ref.~\cite{qgroupbook} gives a thorough introduction to the mathematics of quantum groups, and we use that reference's notation for the two kinds of q-deformations $[n]_q$ and $[[n]]_q$ mentioned below. For the purposes of this work, we restrict our attention to SU$_q$(2), though we expect that our key results generalize to other $q$-deformed groups. 

The characteristic feature of $q$-deformed groups is a parameter $q$ that modifies the generator algebra. For example, SU$_q$(2) has the following algebra:
\begin{equation}\label{Eq: qcomm}
    [\Tilde{S}^z, \Tilde{S}^\pm] = \pm \Tilde{S}^\pm\text{ and }[\Tilde{S}^+, \Tilde{S}^-] = [2\Tilde{S}^z]_q
\end{equation}
where 
\begin{equation}
    [x]_q = \frac{q^x - q^{-x}}{q-q^{-1}}
\end{equation}
The deformation is such that $q \rightarrow 1$ returns the algebra to the usual SU(2) algebra. 

For real, positive $q$, the representations of $q$-deformed SU(2) that satisfy the algebra share many similarities with the usual representations. The irreducible representations are $2S+1$-dimensional with $S^z$ independent of $q$ 
\begin{equation}\label{eq: qdefz}
    (\Tilde{S}^z) = S^z
\end{equation} 
and with
\begin{equation}\label{eq: qdefplus}
    \langle m' | (\Tilde{S}^\pm) |m\rangle = \sqrt{[S\mp m]_q [S\pm m+1]_q }\delta_{m', m\pm 1}.
\end{equation} 
The $\Tilde{S}^\pm$ operators are the same as ${S}^\pm$ for spin $S< 1$. 

The Casimir operator that commutes with the generators and labels the multiplets is $\Tilde{S}^2= \Tilde{S}^-\Tilde{S}^+ + [S^z]_q[S^z+1]_q$ with eigenvalues $[S]_q[S+1]_q$; that such an operator commutes with the generators can be checked by explicit computation. We will also define
\begin{equation}
    \Tilde{S}^x = \frac{\Tilde{S}^+ + \Tilde{S}^-}{2}, \,\,\,
    \Tilde{S}^y = \frac{\Tilde{S}^+ - \Tilde{S}^-}{2i}
\end{equation}
for use below.

However, because of the deformation, some of the usual properties of representations of Lie algebras no longer hold. In the regular SU(2) algebra, if we had a representation $\{S^+, S^-, S^z\}$ we could form a direct product representation, 
\begin{equation}
    \{S^+ \otimes \mathbb{I} + \mathbb{I} \otimes S^+,\, S^- \otimes \mathbb{I} + \mathbb{I} \otimes S^-,\, S^z \otimes \mathbb{I} + \mathbb{I} \otimes S^z\},
\end{equation}
which would also satisfy the algebra. This is how we would describe the action of SU(2) on, say, two spin-$S$ particles. Such a set of would-be generators generally fail to satisfy the $q$-deformed algebra  - instead, for SU$_q$(2), we have that the operators
\begin{equation}
    \{\Tilde{S}^+ \otimes q^{S^z} + q^{-S^z} \otimes \Tilde{S}^+,\, \Tilde{S}^- \otimes q^{S^z} + q^{-S^z} \otimes \Tilde{S}^-, \,S^z \otimes \mathbb{I} + \mathbb{I} \otimes S^z \}
\end{equation}
satisfy the deformed algebra if $\{\Tilde{S}^+, \Tilde{S}^-, S^z\}$ do. Similarly, $\Tilde{S}^\pm$ acting on a chain of length $L$ picks up `tails' of diagonal operators to the left and right for each site: 
\begin{equation}\label{Eq: qsplus}
    \Tilde{S}^\pm = \sum_{i=1}^L (\otimes_{j=1}^{i-1} q^{-S^z_j}) \otimes \Tilde{S}^\pm_i \otimes (\otimes_{j=i+1}^{L} q^{S^z_j})
\end{equation}
while $\tilde{S}^z$ is the same as $S^z$. 

For generating scar towers, we'll consider a single-site representation of SU$_q$(2):
\begin{equation}\label{Eq: qqcomm}
    [Q_i^z, \Tilde{Q}_i^\pm] = \pm \Tilde{Q}_i^\pm\text{ and }[\Tilde{Q}_i^+, \Tilde{Q}_i^-] = [2Q_i^z]_q
\end{equation}
From the single-site representation, we can construct a chain-wide representation through
\begin{equation}\label{Eq: qqsplus}
    \Tilde{Q}^\pm = \sum_{i=1}^L e^{\pm i \phi_i} (\otimes_{j=1}^{i-1} q^{-Q^z_j}) \otimes \Tilde{Q}^\pm_i \otimes (\otimes_{j=i+1}^{L} q^{Q^z_j}),
\end{equation}
for arbitrary phase-factor $\phi_i$. The freedom to choose an arbitrary phase factor while maintaining the commutation relations may seem surprising, as $\Tilde{Q}^\pm$ is a sum of tailed operators. Nevertheless, $(\otimes_{j=1}^{i-1} q^{-Q^z_j}) \otimes \Tilde{Q}^+_i \otimes (\otimes_{j=i+1}^{L} q^{Q^z_j})$ and $(\otimes_{j=1}^{m-1} q^{-Q^z_j}) \otimes \Tilde{Q}^-_m \otimes (\otimes_{j=m+1}^{L} q^{Q^z_j})$ commute for $i \neq m$, and the phase factors cancel for $i = m$, so the phases don't affect the commutation relations.

For $\phi_i = k r_i$, powers of these operators, $(\Tilde{Q}^\pm)^n$, have a simple MPO representation with bond-dimension $n+1$; see Appendix \ref{App:MPO}. It is striking that the MPO representation is linear in $n$ rather than exponential in $n$. This means that $(Q^\pm)^n$ will only increase entanglement entropy by a factor of at most $O(\log(n))$, rather than by $n$. Thus, $\Tilde{Q}^+$ with an associated $q$-deformed symmetry and symmetric base states are good candidates for scar states with additional $q$-deformed symmetry relative to the Hamiltonian.

To illustrate some of these ideas, consider the following $q$-deformations of previously discovered models. The simplest is a $q$-deformation of the model with SU(2) scars in Eq.~\eqref{eq:dimaperfect}: 
\begin{equation}\label{Eq: qdefDima}
    H= \Omega \sum_{i}S^z_i + \sum_{i=1}^{L-1} V_{i-1, i+2}\tilde{P}_{i,i+1}
\end{equation}
where
$\tilde{P}_{i,i+1}$ is a projector onto the two-site $q$-deformed spin-singlet. Explicitly,

\begin{equation}\label{Eq: projector}
    \tilde{P}_{i,i+1} = \frac{1}{[2]_q}(\sqrt{q}|\uparrow\downarrow\rangle-\frac{1}{\sqrt{q}}|\downarrow\uparrow\rangle)(\sqrt{q}\langle\uparrow\downarrow|-\frac{1}{\sqrt{q}}\langle\downarrow\uparrow|)
\end{equation}
The new scar states are those simultaneous eigenstates of $S^z$ and $\tilde{S}^2$ generated by $\Tilde{S}^+$
on the $S^z = -L/2$ state.

We can also modify more complicated models and extract $q$-deformed scar states. We introduce here a deformed version of the spin-1 XY model \cite{ISXY,MLM, MM, MoudHubb}  in Eq.~\ref{Eq: 1XY},
\begin{equation}\label{Eq: qXY}
    H = \sum_{i=1}^{L-1} J\Tilde{h}^{(q)}_{i,i+1} + \sum_{i=1}^{L-3}J_3\Tilde{h}^{(q^3)}_{i,i+3}  + \sum_{i=1}^{L}h S^z_i +  D (S^z_i)^2
\end{equation}
where 
\begin{equation}\label{Eq: qhXY}
    \tilde{h}^{(q)}_{i,i+1} = (\Tilde{S}^x_i q^{S^z_{i+1}}) (q^{-S^z_{i}}\Tilde{S}^x_{i+1}) + (\Tilde{S}^y_i q^{S^z_{i+1}}) (q^{-S^z_{i}}\Tilde{S}^y_{i+1})
\end{equation}
and 
\begin{equation}\label{Eq: q3hXY}
    \tilde{h}^{(q^3)}_{i,i+3} = (\Tilde{S}^x_iq^{3 S^z_{i+3}}) (q^{-3S^z_{i}}\Tilde{S}^x_{i+3}) + (\Tilde{S}^y_iq^{3 S^z_{i+3}}) (q^{-3 S^z_{i}}\Tilde{S}^y_{i+3})
\end{equation}
Here, $\Tilde{S}^x$ and $\Tilde{S}^y$ are deformed using the value of the deformation parameter given in the superscript of $\tilde{h}$.  
This model has scar states generated by a deformed version of the raising operator in Eq.~\eqref{eq:qsu2} with $k=\pi$, $S=1$ and with a deformation parameter of $\frac{1}{q^2}$:
\begin{equation}\label{Eq: defQp}
    \Tilde{Q}^\pm = \sum_{i=1}^L(-1)^i (\otimes_{j=1}^{i-1} q^{2Q^z_j}) \otimes \Tilde{Q}_i^\pm
     \otimes (\otimes_{j=i+1}^{L} q^{-2Q^z_j})
\end{equation}
with $\Tilde{Q}_i^\pm = \frac{1}{2}(S_i^{\pm})^2$. Notice that $\Tilde{Q}_i^\pm$ is \emph{not} $\frac{1}{2}(\Tilde{S}_i^{\pm})^2$: in the undeformed model, $Q_i^+ = \frac{1}{2}(S^+_i)^2$ furnishes a reducible representation which is a direct sum of a singlet plus a doublet, and we want the $q$-deformed $\Tilde{Q}_i$ to furnish a reducible representation which is a direct sum of a deformed singlet and deformed doublet; however the $q$-deformations of singlets and doublets are independent of $q$, c.f. Eq.~\eqref{eq: qdefplus}. One can check directly that this $\Tilde{Q}^+$ is a ladder operator for $SU_{1/q^2}(2)$.
The scar states are $(\Tilde{Q}^+)^n|\psi_0\rangle$, where $|\psi_0\rangle$ is the fully-polarized down state of $|-...-\rangle$. There exists a separate ladder operator for which $(-1)^i \rightarrow e^{i \pi \sum_{j=1}^i S^z_j}$ that also generates these \emph{same} scar states. This second operator is a ladder operator for a separate SU$_{1/q^2}$(2) symmetry, so we must be careful to break the two different SU$_{1/q^2}$(2) Casimirs associated with the different ladder operators. The tower of scar states is annihilated by the $J$ and $J_3$ terms in Eq. (\ref{Eq: qXY}), and these terms keep the Casimirs from commuting with the Hamiltonian:
the nearest neighbor term $J$ is sufficient to violate conservation of the Casimir of the first ladder operator, while the $J_3$ term violates conservation of the Casimir corresponding to the first and second ladder operators. The similarities to the discussion of the original spin-1 XY model in Eq.~\ref{Eq: 1XY} should be clear. 

\begin{figure}[h]
\includegraphics[width=\columnwidth]{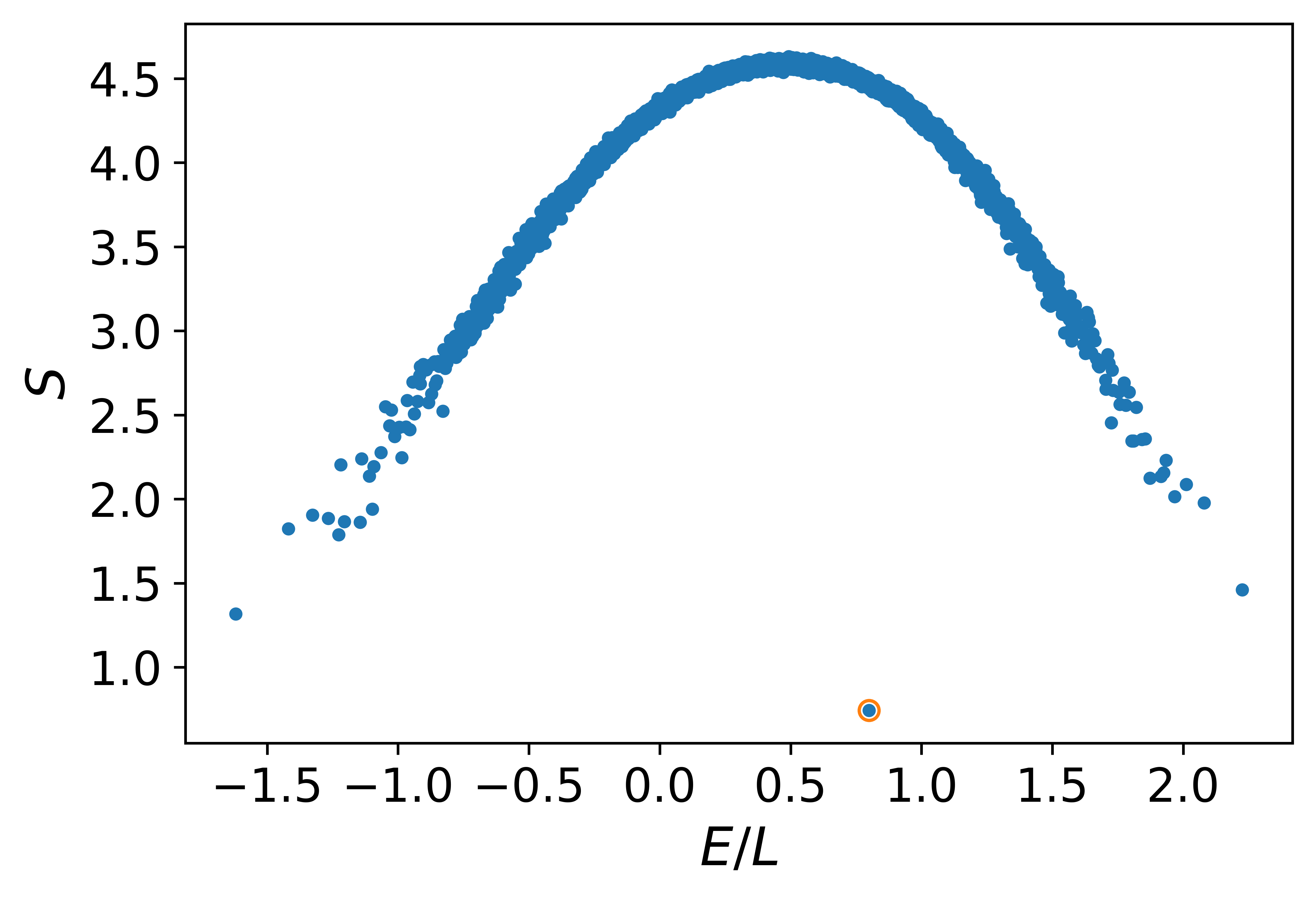}
\caption{\label{fig: qxy} Entanglement entropy in the $S^z = -2$ sector of the $q$-deformed spin-1 XY Hamiltonian for $q=1.2$ and $J=J_3=h=D=1$. The chain has open boundary conditions and a length $L=10$. The scar state within this sector is circled.}
\end{figure}

In figure \ref{fig: qxy}, we plot the entanglement entropy in the $S^z = -2$ sector of the $q$-deformed XY Hamiltonian in equation \ref{Eq: qXY}, for $q=1.2$ and $J=J_3=h=D=1$. This symmetry sector contains only a single scar state circled in orange. The scar state is $(\Tilde{Q}^+)^4$ acting on the fully polarized $|---...----\rangle$ state with energy $E/L=.8$

We turn next to the issue of revivals from simple initial states. 
For the undeformed $q=1$ model, Ref.~\cite{ISXY} noted a simple product state that would undergo perfect revivals. This product state was an eigenstate of $Q^x = (Q^+ + Q^-)/2$ with maximal eigenvalue:
\begin{equation}
    |c=1, q=1\rangle = \otimes_i \left( \frac{|+\rangle - (-1)^i |-\rangle}{\sqrt{2}}\right)
\end{equation}
Such a state has maximal $q=1$ $Q^2$ by virtue of having maximal $Q^x$ and hence is a superposition of scar states. The scar states are evenly spaced in energy with spacing $2h$, yielding perfect revivals of the state under the time evolution of Eq.~\ref{Eq: qXY} with $q=1$. Furthermore, grouping the product states within the state above according to their total $z$-magnetization, it follows that 
\begin{equation}
\label{Eq:cq1}
    |c, q=1\rangle = \otimes_i \left( \frac{|+\rangle - (-1)^i c |-\rangle}{\sqrt{1+|c|^2}}\right)
\end{equation}
is also a superposition of the scar states, for arbitrary complex parameter $c$. We can also write this state more explicitly in terms of the $q=1$ $Q^+$ operator by noting that
\begin{align}
\begin{split}
    |c, q=1\rangle =& \mathcal{N} \otimes_i \left(|-\rangle - (-1)^i \frac{1}{c} |+\rangle\right) \\
    =& \mathcal{N} \otimes_i \left(1-\frac{1}{c}Q^+_i\right)|\psi_0\rangle \\
    =& \mathcal{N} \otimes_i \left(e^{-\frac{1}{c}Q^+_i}\right)|\psi_0\rangle \\
    =& \mathcal{N} e^{-\frac{1}{c}Q^+}|\psi_0\rangle
\end{split}
\end{align}
where we have defined $\mathcal{N} = (-1)^{\frac{1}{2}L(L+3)} \left(\frac{c}{\sqrt{1+|c|^2}}\right)^L$.

For our deformed model, we can write a similar family of product states:
\begin{equation}
\label{Eq:cq}
    |c, q\rangle = \otimes_i \left( \frac{|+\rangle - (-q^2)^i c|-\rangle}{\sqrt{1+q^{4i}|c|^2}}\right)
\end{equation}
Here, $c$ is an arbitrary complex parameter. Note that the $q$-dependence changes across the chain. This product state is again a superposition of the (now $q$-deformed) scarred states. To write the superposition explicitly, we'll introduce the notation for $m$ and $n$ positive integers,
\begin{equation}
    [[n]]_q  = \frac{1-q^n}{1-q}, \, [[n]]_q !  = \prod_{m=1}^n [[m]]_q.
\end{equation} 
Then, comparing $(\Tilde{Q}^+)^n|\psi_0\rangle$ to the terms in $|c,q\rangle$ with $n$ sites $|+\rangle$, we have the explicit superposition
\begin{equation}
    |c, q\rangle = -\mathcal{N} \sum_{n=0}^L c^{L-n} \frac{q^{L(L+1)-n(n+L)}}{[[n]]_{\frac{1}{q^4}}!} (\Tilde{Q}^+)^n|\psi_0\rangle 
\end{equation}
where
\begin{equation}
    \mathcal{N} = \left(\prod_{i=1}^L \frac{1}{\sqrt{1+|c|^2q^{4i}}}\right)
\end{equation}

\begin{figure}[h]
\includegraphics[width=\columnwidth]{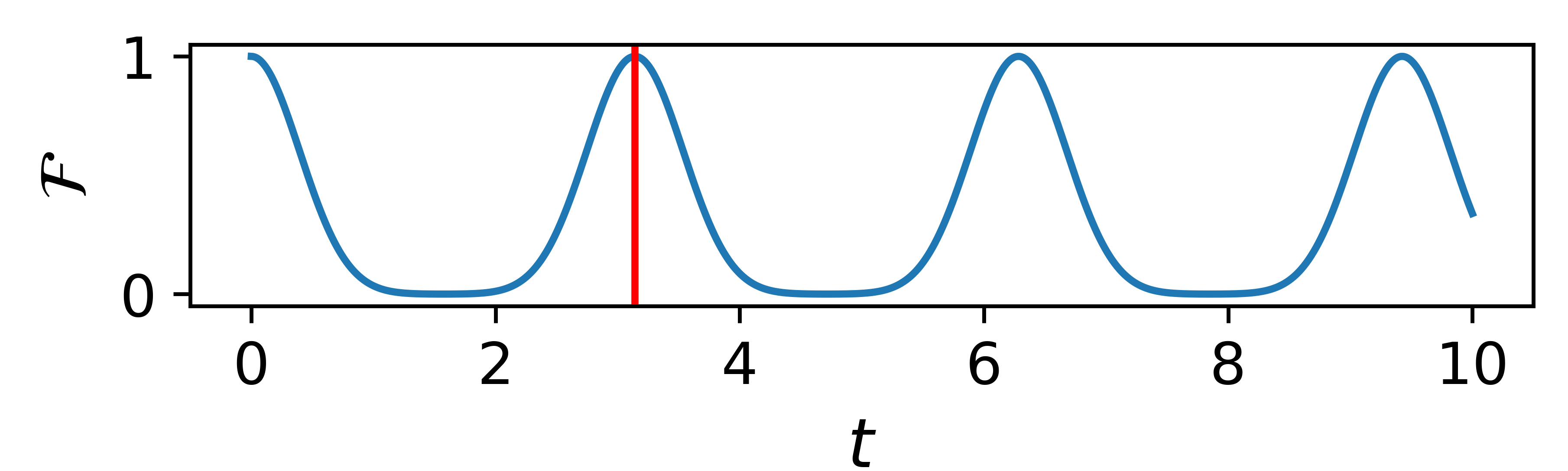}
\caption{\label{fig: qxyfid} Fidelity $\mathcal{F} = |\langle \psi(t)|\psi(0)\rangle|^2$ of the state in Eq.~\ref{Eq:cq} with $c = 1$ under the time evolution generated by the Hamiltonian in Eq.~\ref{Eq: qXY} for $q=1.2$ and $L=10$. The period of $\pi$ is marked with a vertical red line.}
\end{figure}

Furthermore, the scar states are again evenly spaced in energy with spacing $2h$, yielding perfect revivals of this product state under the time evolution of Eq.~\ref{Eq: qXY} for the corresponding value of $q$. As an example, we demonstrate in Fig.~\ref{fig: qxyfid} periodic revivals of Eq.~\ref{Eq:cq} with $c=1$ with the Hamiltonian in Eq.~\ref{Eq: qXY} with $q=1.2$  after a period of $\pi$.

The above discussion holds for any values of $q>0$. We note as an aside that for $q=1+\epsilon$ for small $\epsilon$, the spin-1 XY model is only weakly deformed from the regular spin-1 XY model. Further, the scar states are also close to their undeformed counterparts (although when $L$ is large,  their overlap is only nonnegligible when $\epsilon$ is small relative to $\frac{1}{L}$). This implies that the simple product state $|c, q=1\rangle$ which shows perfect revivals in the undeformed spin-1 XY model will show imperfect revivals in the deformed model with a magnitude of oscillation decaying with growing $\epsilon$ and $L$.

\section{Generalized AKLT Scars}
\label{sec:aklt}
Thus far, we have constructed scarred Hamiltonians in which the scarred eigenstates transform as a single irreducible representation of a (possibly $q$-deformed) symmetry group $G$, and have a unique eigenvalue for the Casimir operator(s) $C$.  Here, we describe a qualitatively different family, in which the scarred states are \emph{not} eigenstates of $C$ and do \emph{not} transform in a single irrep of the symmetry. These will be built by considering $\Hc$ with an \emph{enhanced} symmetry, which allows us to pick base states that do not have a definite eigenvalue for the Casimir(s) but are nevertheless eigenstates of $\Hc$.

For specificity, we focus on one-dimensional spin-$S$ generalized AKLT chains for which the Hamiltonians can be written as a sum of projectors. In Sec.~\ref{Sec:GenAKLTModels} we present two new models, the $q$-deformed and SO(2S+1) generalizations of the spin-$S$ AKLT model, and show that they have towers of scarred eigenstates generated by the action of the ladder operator:
\be  \label{Eq:QpAKLT}
Q^+_{\textrm{AKLT}} = \sum_j \frac{1}{(2S)!}(-1)^j (S^+_j)^{2S}
\ee
on their respective ground states. 
Note that $Q^+_{\textrm{AKLT}}$ is the same as the raising operator associated with $Q$-SU(2) discussed earlier, Eq.~\eqref{eq:qsu2} with $k=\pi$. However, because the base state is not an eigenstate of $Q^2$, the projector onto the resulting \emph{asymmetric} scarred manifold does \emph{not} commute with the $Q$-SU(2) symmetry. Previous work showed that the spin-$S$ AKLT model also has an asymmetric tower of scarred eigenstates generated by the same $Q^+_\textrm{AKLT}$ \cite{MoudDisc, MLM}. 

To understand how this fits with our broader symmetry-based picture, we note that Ref.~\cite{MLM} showed that the spin-$1$ AKLT model can be deformed to an $\Hc$ with $Q$-SU(2) symmetry while preserving the scar states as eigenstates. 
To make the AKLT ground state an eigenstate of $\Hc$ --- despite not being an eigenstate of $Q^2$ --- also requires degeneracies between tunnels of states with different values of the Casimir $Q^2$, allowing superpositions with indefinite $Q^2$ to be eigenstates of $\Hc$. This points to an expanded symmetry in $\Hc$ leading to a much larger set of degeneracies. By taking advantage of these degeneracies, one can prepare base states that are eigenstates of $\Hc$ and $\Hz$, even if they are not eigenstates of $Q^2$ and $Q^z$. We discuss in Section~\ref{sec:parent} a nontrivial set of $\Hc$ for a very general set of base states with indefinite $Q^2$. These $\Hc$ allow us to decompose the generalized AKLT Hamiltonians into $\Hc$, $\Hs$, and $H_A$ and connect to our broader symmetry-based framework. 

Finally, in Section \ref{Sec:DisjointGR}, we study deformations between scarred Hamiltonians with different sets of scar towers. In particular, we discuss generalizations of a deformation in Ref.~\cite{MoudMPS} between the spin-$S$ AKLT model and an integrable point with $Q$-SU(2) symmetry. 

\subsection{Structure of scars in generalized AKLT models}

The scars in the AKLT model and its generalizations are a consequence of the following general structure, first discussed by Refs.~\cite{MLM, MoudMPS}. Consider a Hamiltonian $H= \sum_j h_{j,j+1}$, a base state $\ket{\psi_0}$ with zero energy, and a ladder operator $Q^+ = \sum_j e^{ikj} Q^+_j$. We assume periodic boundary conditions with $kL$ being a multiple of $2\pi$; we discuss the generalization to open boundary conditions briefly in the next subsection and in more detail in Appendix \ref{appendix:OBC}. We also assume that we can group the two-site Hilbert space into the three \emph{disjoint} subspaces, $\mathcal{G}$, $\mathcal{R}$, and $\mathcal{M}$. The subspace $\mathcal{G}$ (not to be confused with the non-Abelian symmetry group $G$ of $H_\textrm{sym}$) contains all 2-site configurations (which we will refer to as bonds) that are present in the base state. The subspace $\mathcal{R}$ contains the image of all bonds in $\mathcal{G}$ under the action of 
$q^+_{j,j+1} = Q^+_j + e^{ik} Q^+_{j+1}$, while the subspace $\mathcal{M}$ is the complement of $\mathcal{G} \cup \mathcal{R}$. If $h_{j,j+1}$ and $q^+_{j,j+1}$ have the following general forms,
\begin{eqnarray}
    h_{j,j+1}&=&\left[
\begin{array}{c|c|c}
h_{\mathcal{M}\mathcal{M}} & 0& 0 \\ \hline
0& \omega I& 0 \\ \hline
0& 0& 0   
\end{array}\right] \label{Eq:hform} \\
q^+_{j,j+1} &=&\left[
\begin{array}{c|c|c}
q^+_{\mathcal{M}\mathcal{M}} & 0 & 0 \\ \hline
q^+_{\mathcal{R}\mathcal{M}}& 0& q^+_{\mathcal{R}\mathcal{G}} \\ \hline
q^+_{\mathcal{G}\mathcal{M}}& 0& 0   \label{Eq:qform}
\end{array}\right],
\end{eqnarray}
then the model has a scar tower generated by $Q^+$ with energy spacing $2\omega$. The above result follows from explicitly calculating the commutator of $H$ and $Q^+$, as shown in Eq.~\eqref{eq: Hpbc}. Here, the matrices $h_{\mathcal{MM}}$, $q^+_{\mathcal{MM}}$, $q^+_\mathcal{{RM}}$, $q^+_\mathcal{{GM}}$ are completely free and can be zero or nonzero. Additionally, the choice of $h_\mathcal{{MM}}$ can depend on the site $j$. However, the blocks labeled $0$ must be zero. 

In other words, if $\mathcal{G}$ and $\mathcal{R}$ are disjoint, $(q^+_{j,j+1})^2=0$ when acting on $\mathcal{G}$, and the Hamiltonian takes the form above, then the model is scarred. We show that the generalized AKLT models in Sec.~\ref{Sec:GenAKLTModels} satisfy these conditions (with $\omega=1$), and their ground states are not eigenstates of $Q^2$. Hence, they each have asymmetric scar towers.

Remarkably, we can also construct a large, continuously connected class of matrix product states (MPS) such that $\mathcal{G}$ and $\mathcal{R}$ are disjoint for $Q^+ = Q^+_{\textrm{AKLT}}$. These are spin-$S$, bond-dimension $S+1$ matrix product states of the form:
\begin{align}\label{Eq:MPSScarBaseSt}
    \ket{\psi_0, A} &= \sum_{\textbf{m}} \textrm{Tr}[A^{[m_1]}A^{[m_2]}...A^{[m_L]}]|m_1...m_L\rangle \\
    A^{[m]}_{ij} &= 0 \text{ for } j-i \neq m \nonumber
\end{align}
with $m_i = {-S, -S + 1, \cdots S}$. The condition $A^{[m]}_{ij} = 0 \text{ for } j-i \neq m$ means $A^{[m]}$ is nonzero only on the $m$th diagonal.\footnote{We use the usual convention for $m$th diagonal: $m=0$ is the main diagonal, $m=S$ is the upper right corner, and $m=-S$ is the lower left corner.}  Each such state acts as a base state for a different tower generated by $Q^+_{\rm AKLT}$, and we can enumerate the states in $\mathcal{G}$, $\mathcal{R}$, $\mathcal{M}$ to construct new Hamiltonians of the form in Eq.~\ref{Eq:hform} scarred by the corresponding tower. This class of states includes the $q$-deformed and regular spin-$S$ AKLT ground states~\cite{Motegi2010, Santos2012}, though the SO(2S+1) AKLT models' ground states have a different structure for $S>1$~\cite{Tu2008}.

To bring these myriad scar towers into our symmetry-based framework, we note in Section \ref{sec:parent} that all states of the form in Eq.~\ref{Eq:MPSScarBaseSt}, as well as all of the SO(2S+1) AKLT models' ground states, enjoy a canonical $Q_{\rm AKLT}$-SU(2) symmetric parent Hamiltonian for which $(Q^+)^n|\psi_0\rangle$ are all eigenstates of the same energy. This parent Hamiltonian differs for different base states $|\psi_0\rangle$, but its form is the same in terms of the corresponding base state's $\mathcal{G}$, $\mathcal{R}$, and $\mathcal{M}$. This parent Hamiltonian further allows us to write all of the generalized AKLT models' Hamiltonians and all the Hamiltonians of the form in Eq.~\ref{Eq:hform} in terms of $\Hc$, $\Hs$, and $H_A$.

Because the MPS are continuously connected, we can give continuous deformations between scarred Hamiltonians along which the asymmetric scarred states persist and are continuously deformed. To demonstrate the power of this large class of states, in \ref{Sec:DisjointGR} we revisit Ref.~\cite{MoudMPS}'s deformation of the spin-$1$ AKLT model to an integrable point along a path with (continuously varying) scarred eigenstates. We show that there are many such deformations between the spin-$S$ AKLT model and corresponding high-symmetry integrable points.

\subsection{The generalized AKLT models}
\label{Sec:GenAKLTModels}
Affleck, Kennedy, Lieb and Tasaki (AKLT) introduced the spin-$1$ AKLT model to analytically describe the Haldane gap in integer spin chains~\cite{AKLT}. Subsequent work discovered that the AKLT chain has fractionalized edge spins in open chains, and is a symmetry-protected topological phase with non-local string order~\cite{KennedyTasaki,PollmannSPT2012}. The spin-1 AKLT chain's interesting properties prompted many generalizations, including generalizations to spin-$S$, $q$-deformed spin-$S$ \cite{Batchelorqdef, Klumperqdef, Totsuka1994, Motegi2010, Santos2012,Quella}, and other symmetry groups like SO(2S+1)~\cite{Tu2008}. These generalizations are all examples of the Haldane phase with exact matrix product ground states. We demonstrate that these models have a second curious property in common, not directly related to the Haldane phase: they all have scar towers generated by $Q^+_{\text{AKLT}}$ on appropriate ground states.

Each term $h^\alpha_{j,j+1}$ in the Hamiltonian $H^\alpha=\sum_j h^\alpha_{j,j+1}$ of a generalized AKLT model of type $\alpha$ can be expressed as a sum of projectors. For the spin-$S$, $q$-deformed spin-$S$, and the SO(2S+1) AKLT models, we have: 
\begin{align}
\begin{split}
& h^{\rm S}_{j, j+1} = \sum_{t=S+1}^{2S} P^{(t)}_{j,j+1} \\
&h^{\rm S_q}_{j, j+1} =  \sum_{t=S+1}^{2S} \Tilde{P}^{(t)}_{j,j+1} \\
& h^{\rm SO(2S+1)}_{j, j+1} = \sum_{k=1}^{S} P^{(2k)}_{j,j+1} 
\end{split}
\end{align}
Here, the two-site operators $P^{(t)}_{j,j+1}$ and $\Tilde{P}^{(t)}_{j,j+1}$ project onto total spin $t$ and $q$-deformed total spin $t$ respectively. We give their explicit forms in terms of spin operators and $q$-deformed spin operators in Appendix~\ref{appendix: Projectors}. 
The projectors $P^{(t)}_{j,j+1}$ are SU(2) invariant, and thus $H^S$ and $H^{SO(2S+1)}$ are SU(2) invariant\footnote{The larger SO(2S+1) symmetry of $H^{\rm SO(2S+1)}$ comes from identifying $\sum_{k=1}^{S} P^{(2k)}_{j,j+1}$ as a projector onto the $2S^2 + 3S$-dimensional irreducible representation of SO(2S+1) within the direct product of two fundamental representations~\cite{Tu2008}.}. In comparison, $\Tilde{P}^{(t)}_{j,j+1}$ is SU$_q$(2)-symmetric for all $j$ except for $j=L$, implying that $H^{S_q}$ is SU$_q$(2)-symmetric with open boundary conditions but not periodic boundary conditions. Notice that $H^S$ and $H^{SO(2S+1)}$ agree for $S=1$. We also emphasize here that although $H^{S_q}$ has $q$-deformed SU(2) symmetry, its scar tower is generated by the ``usual" raising operator $Q^+_\textrm{AKLT}$ defined in Eq.~\ref{Eq:QpAKLT}, and \emph{not} by $q$-deformed raising operators of the form Eq.~\eqref{Eq: qqsplus}.

The three models have exactly known matrix product ground states. With periodic boundary conditions, the ground states are frustration-free and unique. With open boundary conditions, the regular and $q$-deformed spin-$S$ AKLT models have $(S+1)^2$ frustration-free ground states, and the SO(2S+1) AKLT Hamiltonian has $4^S$ frustration-free ground states.

We briefly comment on the difference between the scar towers in periodic and open chains (see Appendix \ref{appendix:OBC} for more details). Assume that we have disjoint $\mathcal{G}$ and $\mathcal{R}$ with $h$ and $q$ as given in Eq.~\eqref{Eq:hform} and \eqref{Eq:qform}. In periodic chains, we have 
\begin{equation}
\label{eq: Hpbc}
  [H_{\textrm{pbc}}, Q^+] = 2\omega Q^+ + \sum_{j=1}^{L}e^{ikj} A_{j,j+1}  
\end{equation}
where $A_{j,j+1} = [h_{j,j+1}, q^+_{j,j+1}] - \omega q^+_{j,j+1}$, $H_\textrm{pbc} = \sum_{j=1}^L h_{j,j+1} $. Computing the operator $A_{j,j+1}$ using Eqs.~$\eqref{Eq:hform}$ and $\eqref{Eq:qform}$, we see that $A_{j,j+1}$ annihilates all the 2-site configurations on sites $(j,j+1)$ that appear in the states of the scar tower. The form in Eq.~\eqref{eq: Hpbc} matches the condition in Eq.~\eqref{eq:MLM}.
In open chains, the commutator is modified to be:
\begin{equation}
\label{eq: Hobc}
    [H_{\textrm{obc}}, Q^+] =2\omega Q^+ +\sum_{j=1}^{L-1}e^{ikj}A_{j,j+1} - e^{ik}\omega Q^+_1 - e^{ikL} \omega Q^+_L
\end{equation}
The critical change is the presence of $Q^+_1$ and $Q^+_L$ acting on the physical edge spins, which restricts which of the ground states in open boundary conditions will be a good base state for the tower of states. We argue in Appendix~\ref{appendix:OBC} that $Q^+_1$ and $Q^+_L$ must individually annihilate the physical edge spins in these models, which we show occurs for $S^2$ out of $(S+1)^2$ ground states of the regular and $q$-deformed spin-$S$ AKLT models and $4^{S-1}$ out of $4^S$ ground states in the SO(2S+1) AKLT model. This discussion of open boundary conditions is especially important for the $q$-deformed AKLT models, as the models lose their interesting SU$_q$(2) symmetry with periodic boundary conditions.

In Appendices \ref{appendix:qplus1} and \ref{appendix:qplus2}, we prove that $\mathcal{G}$ and $\mathcal{R}$ are disjoint, and that Eqs.~\eqref{Eq:hform} and ~\eqref{Eq:qform} hold with $\omega=1$ and $h_\mathcal{{MM}} = I$ for all three models. Appendix \ref{appendix:qplus1} additionally shows that $\mathcal{G}$ and $\mathcal{R}$ are disjoint under $Q^+_{\rm AKLT}$ for all base states of the form given in Eq.~\ref{Eq:MPSScarBaseSt}. This furnishes the proof that the spin-S, $q$-deformed spin-S, and the SO(2S+1) AKLT models have asymmetric scar towers generated by $Q^+_\textrm{AKLT}$ on their respective ground states with periodic boundary conditions. The discussion of open boundary conditions follows additionally from Appendix~\ref{appendix:OBC}. 

\begin{figure}[htb]
\includegraphics[width=\columnwidth]{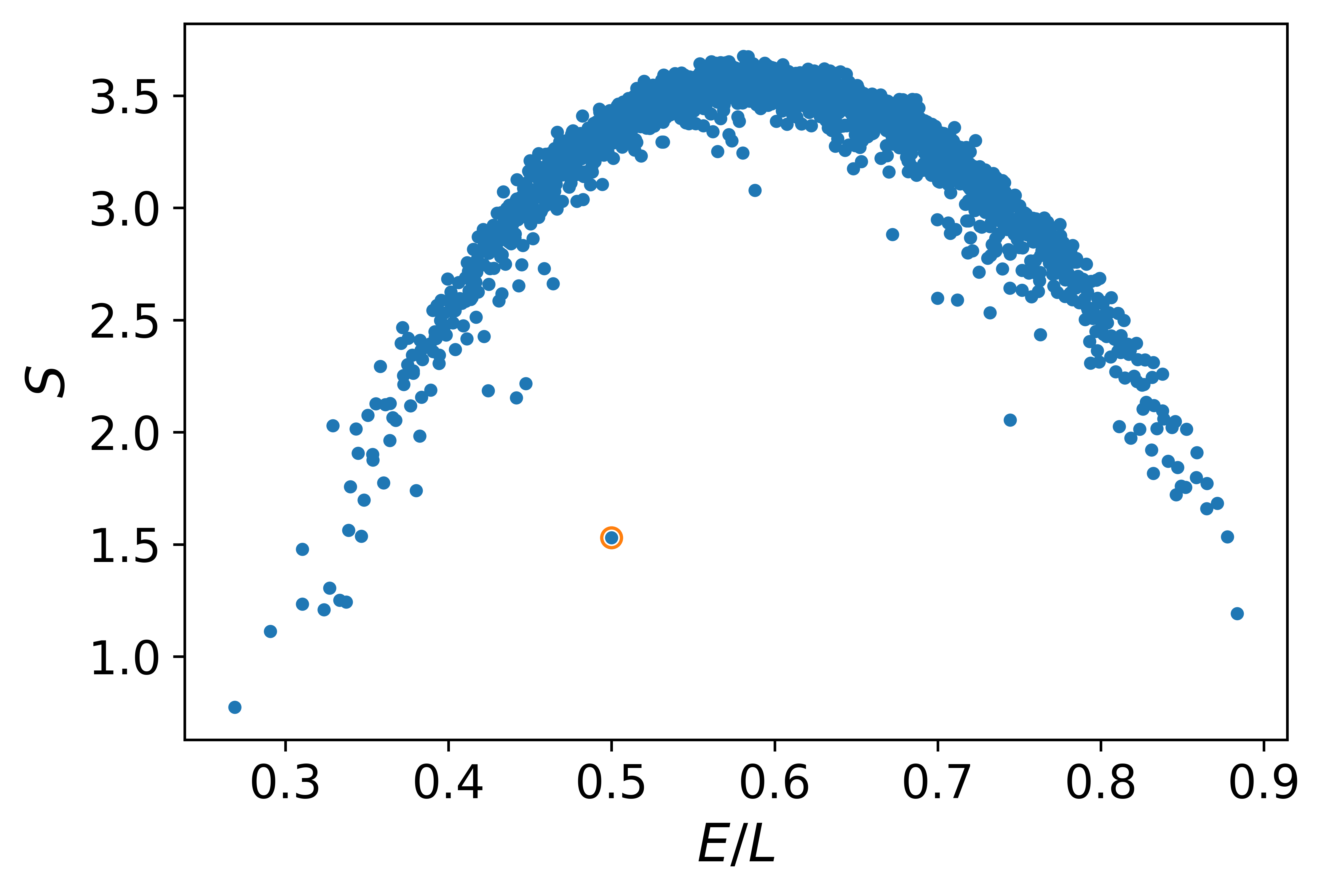}
\caption{\label{fig:qdefaklt2}
 Entanglement entropy in the $S^z=7$ and $\Tilde{S}^2 = [7]_q[8]_q$ sector of the $q$-deformed spin-$1$ AKLT Hamiltonian for $q=1.2$, $L=12$, and open boundary conditions. The circled scar state at energy $E/L=0.5$ is $(Q^+)^3$ on the $S^z=1$ ground state. }
\end{figure}

Fig.~\ref{fig:qdefaklt2} shows the eigenstate entanglement entropy of the open $q$-deformed spin-$1$ AKLT model for $q=1.2$ in a fixed $S^z$ sector vs the energy density. In every $S^z=2m+1, \Tilde{S}^2 = [2m+1]_q [2m+2]_q$ sector with integer $m\geq 0$, we expect a unique scar state at energy density $2m/L$ generated by the action of $(Q^+_\textrm{AKLT})^m$ on the $S^z=1$ ground state~\footnote{As follows from Appendix \ref{appendix:OBC}, for $S=1$ in open boundary conditions, the $S^z=1$ $q$-deformed ground state is the only ground state hosting a tower of eigenstates generated by the action of $Q^+$.}. The fact that $\Tilde{S} = 2m+1$ follows from an argument analogous to the $q=1$ case in \cite{MoudDisc}. The circled state at $E/L = 0.5$ is thus the predicted scar state in the $S^z=7$ and $\Tilde{S}^2 = [7]_q [8]_q$ sector. 

\subsection{Parent Hamiltonians for the generalized AKLT models} \label{sec:parent}

Even though our generalized AKLT models have scar states that do not transform in an irreducible representation of the SU(2) symmetry associated with $Q^+$, we can bring these models into our symmetry-based framework by decomposing their Hamiltonians into $\Hc$, $\Hs$, and $H_A$. More specifically, we detail how to write the spin-$S$ AKLT models, the $q$-deformed spin-$S$ AKLT models, the SO(2S+1) AKLT models, and all the models of the form Eq.~\ref{Eq:hform} with the base states of Eq.~\ref{Eq:MPSScarBaseSt} in terms of a canonical Q-SU(2) symmetric parent Hamiltonian $\Hc^{(C)}$ and an appropriate $\Hs$ and $H_A$. Our proposed canonical parent Hamiltonian is different for each model, but it can be written universally in the language of $\mathcal{G}$, $\mathcal{R}$, and $\mathcal{M}$ discussed above; for this reason, we call this parent Hamiltonian ``canonical". For ease, we will assume periodic boundary conditions in this section.

It will be useful in the following to decompose $\mathcal{M}$ further. As before, define $\mathcal{G}$ as those bonds in the base state of the tower and define $\mathcal{R}$ as the span of the bonds to which $\mathcal{G}$ is mapped under the action of $q^+$. However, additionally define $\mathcal{L}$ as the span of the bonds to which $\mathcal{G}$ is mapped under the action of $q^-$, the Hermitian conjugate of $q^+$. Finally, define $\Tilde{M}$ as the complement of $\mathcal{G} \cup R \cup \mathcal{L}$. It follows that $\mathcal{M} = \mathcal{L} \cup \Tilde{\mathcal{M}}$.

$\mathcal{L}$ and $\Tilde{\mathcal{M}}$ have a simple physical meaning in terms of scar towers. In particular, it is straightforward to verify that for the base states discussed above, choosing a Hamiltonian of the form (cf. Eq.~\ref{Eq:hform})
\begin{eqnarray}
    h_{j,j+1}&=&\left[
\begin{array}{c|c|c|c}
h_{\Tilde{\mathcal{M}} \Tilde{\mathcal{M}}} & 0& 0 & 0 \\ \hline
0& \omega_{\mathcal{R}} I& 0 &0 \\ \hline
0& 0& 0 & 0 \\ \hline
0& 0& 0 & \omega_{\mathcal{L}} I
\end{array}\right] \label{Eq:hform2}
\end{eqnarray}
will have an evenly spaced scar tower generated by $Q^+_{\rm AKLT}$ with spacing $2\omega_{\mathcal{R}}$, and an evenly spaced scar tower generated by $Q^-_{\rm AKLT} = (Q^+_{\rm AKLT})^\dagger$ with spacing $2\omega_{\mathcal{L}}$ on the ground state in periodic boundary conditions.\footnote{An analogous discussion of the form of $q^-$ and disjoint $\mathcal{G}$, $\mathcal{L}$, and $\mathcal{R}$ follows immediately from the arguments in Appendices \ref{appendix:qplus1} and \ref{appendix:qplus2}.} The above makes clear the physical meaning of this decomposition: a two-site local perturbation with support only in $\Tilde{M}$ will not affect the existence of the scar states, a perturbation with support in $\mathcal{L}$ but not $\mathcal{R}$ will typically break the tower generated by $Q^-_{\rm AKLT}$ but cannot affect the $Q^+_{\rm AKLT}$ tower, and a perturbation with support in $\mathcal{R}$ but not $\mathcal{L}$ will generically break the tower generated by $Q^+_{\rm AKLT}$ but cannot affect the tower generated by $Q^-_{\rm AKLT}$. To clarify the role of $Q^-_{\rm AKLT}$, we emphasize that $Q^-_{\rm AKLT}$ will not undo the action of $Q^+_{\rm AKLT}$ on the states in the tower, since the scar states are superpositions of states with different eigenvalues under the Casimir $Q^2$. Furthermore, for the models we describe, $(Q^+)^n (Q^-)^m|\psi_0\rangle$ for $n,m >0$ will generically NOT be an eigenstate of the Hamiltonian.

Additionally, for the base states we're considering, $q^+$ has a special form in terms of its action on $\Tilde{M}$, $\mathcal{R}$, $\mathcal{G}$, and $\mathcal{L}$. 
\begin{eqnarray}
    q^+_{j,j+1}&=&\left[
\begin{array}{c|c|c|c}
0 & 0& 0 & q^+_{\Tilde{\mathcal{M}}\mathcal{L}} \\ \hline
q^+_{\mathcal{R}\Tilde{\mathcal{M}}}& 0& q^+_{\mathcal{R}\mathcal{G}} &0 \\ \hline
0& 0& 0 & q^+_{\mathcal{G}\mathcal{L}} \\ \hline
0& 0& 0 & 0
\end{array}\right] \label{Eq:qform2}
\end{eqnarray}
As before, the $q^+_{R\Tilde{M}}$ and the like can be either zero or nonzero, but the blocks labeled with zeroes must be zero. This form is stricter than that given in Eq.~\ref{Eq:qform}, and we prove this form for all the base states we're considering in Appendix \ref{appendix:qplus3}.

We introduced $q^-$ and $\mathcal{L}$ in order to motivate the form of a useful contribution to $\Hc$. First, consider the operator
\begin{equation}\label{Eq:HZ2}
 H_{Z2} = \sum_j \left( P_{j,j+1}^{(\mathcal{R})} - P_{j,j+1}^{(\mathcal{L})} \right)
\end{equation}
where $P_{j,j+1}^{(\mathcal{R})}$ projects onto the bonds in $R$ and $P_{j,j+1}^{(\mathcal{L})}$ projects onto the bonds in $\mathcal{L}$; i.e. 
\begin{eqnarray}
    P_{j,j+1}^{(\mathcal{R})} - P_{j,j+1}^{(\mathcal{L})} &=&\left[
\begin{array}{c|c|c|c}
0 & 0& 0 & 0 \\ \hline
0& I& 0 &0 \\ \hline
0& 0& 0 & 0 \\ \hline
0& 0& 0 & -I
\end{array}\right]. \label{Eq:HZ22}
\end{eqnarray}
Using $[P_{j,j+1}^{(R)} - P_{j,j+1}^{(\mathcal{L})}, q^+_{\rm AKLT}] = q^+_{\rm AKLT}$, we can show 
\begin{equation}\label{Eq:Z2Qcomm}
    [H_{Z2}, Q^+_{\rm AKLT}] = 2Q^+_{\rm AKLT}
\end{equation}
i.e., $H_{Z2}$ is a candidate operator for $\Hs$. Indeed, $H_{Z2}$ has the base state of interest as an eigenstate as it annihilates all the bonds in $\mathcal{G}$, so it correctly acts like $\Hs$. Note that we also have $S^z$ as another operator that acts like $\Hs$: since $Q^+_{\rm AKLT}$ raises the total z-magnetization by $2S$, $[S^z, Q^+_{\rm AKLT}] = 2SQ^+_{\rm AKLT}$. It is useful here to then select only one of these operators for $\Hs$ and to take an appropriate linear combination of the two for $\Hc$. We take
\begin{equation}\label{Eq:canonical}
    \Hs = \frac{1}{S} S^{z}, \, \Hc^{(C)} = H_{Z2} - \frac{1}{S} S^{z}.
\end{equation}
Here $\Hc^{(C)}$ is the canonical parent  Hamiltonian. It commutes with $Q^+_{\rm AKLT}$, and hence is $Q$-SU(2) symmetric. As $S^z$ and $H_{Z2}$ annihilate the base state, $\Hc^{(C)}$ also annihilates the base state and hence, under this Hamiltonian, all the scar states have zero energy due to the $Q$-SU(2) symmetry.
Interestingly, $H^{(C)}_{\rm sym}$ is generically an ``as-a-sum" annihilator of the scar states; these  can  be difficult to predict a priori.

$\Hc^{(C)}$ allows us to write a unified version of the regular and generalized spin-$S$ AKLT models in terms of $\Hs$, $\Hc$, and $H_A$.  All of these models are described by the Hamiltonian
\begin{equation}\label{Eq:AKLTcanon}
    H_{\rm unify} = \Hc^{(C)} + \frac{1}{S}S^z + \sum_j P_{j,j+1}^{(\mathcal{L})} + \sum_j P_{j,j+1}^{(\mathcal{M})},
\end{equation}
where $P_{j,j+1}^{(\mathcal{M})}$ projects onto the bonds in $\mathcal{M}$. This Hamiltonian is a sum of projectors with unit coefficients onto all bonds except those in the base state.
 
In Eq. \ref{Eq:AKLTcanon}, we identify $\Hc = \Hc^{(C)}$, $\Hs = \frac{1}{S}S^z$, and $H_A = \sum_j P_{j,j+1}^{\mathcal{L}} + \sum_j P_{j,j+1}^{\mathcal{M}}$.~\footnote{Within this $H_A$ will generically be some operators that will also commute with $Q^+$, such as $|2S,0\rangle\langle 2S,0|$ for the regular spin-$S$ AKLT model. For simplicity, we will group them in $H_A$ as they can all be understood as bond-wise annihilators within $h_\mathcal{{MM}}$. Quite generally, there will be multiple operators comprising a scarred Hamiltonian that are $Q$-symmetric and annihilate the tower of states, and while it is possible to identify all of them as contributing to $\Hc$, it can be useful to make a different delineation to emphasize an interesting symmetric parent Hamiltonian. See, for example, the discussion around Eq.~\ref{Eq:H2s}}. To fully specify the Hamiltonian in Eq. (\ref{Eq:AKLTcanon}), we must fix the form of the projectors -- i.e. we must specify the base state 
$|\psi_0 \rangle$. Picking $|\psi_0 \rangle$ as the ground state of any of the AKLT models discussed in Eq.~\ref{Eq:AKLTcanon} will return the corresponding generalized AKLT Hamiltonian; the resulting scar tower is obtained by acting with $Q^+_{\rm AKLT}$ on  $|\psi_0 \rangle$. 

A slight modification of the coefficients yields 
a more general Hamiltonian, of the form given in Eq.~\ref{Eq:hform}:
\begin{equation}\label{Eq:gencanon}
    H_{\rm gen} = \omega(\Hc^{(C)} + \frac{1}{S}S^z) + \omega \sum_j P_{j,j+1}^{(\mathcal{L})} + \sum_j {h_\mathcal{{MM}}}_{j,j+1} \ .
\end{equation}
We see that these models have the same structure as the generalized spin-$S$ AKLT models described by Eq. \ref{Eq:AKLTcanon}, 
with $\Hc = \omega\Hc^{(C)}$, $\Hs = \frac{\omega}{S}S^z$, and $H_A = \omega \sum_j P_{j,j+1}^{\mathcal{L}} + \sum_j {h_\mathcal{{MM}}}_{j,j+1}$.  By choosing base states of the form in Eq.~\ref{Eq:MPSScarBaseSt} to specify the projectors, we expose a unifying structure relating generalized AKLT scar towers to this more general family of scar states.  Moreover, a more general class of Hamiltonians with the same scar eigenstates is obtained by  changing the coefficient of $S^z$, as well as adding the alternating as-a-sum annihilator discussed in Appendix \ref{app:as-a-sum}.  All of these share a common form for the parent Hamiltonian $\Hc^{(C)}$ and $\Hs$.  

From Eqs.~\ref{Eq:AKLTcanon} and \ref{Eq:gencanon}, we see that $H^{(C)}_{\rm sym}$ is a natural part of the generalized AKLT Hamiltonians,  expressible in terms of a simple, universal form in terms of the spaces $\mathcal{R}$ and $\mathcal{L}$ of the base state.
It is interesting, however, to consider $\Hc^{(C)}$ as a model in its own right.   
For the spin-1 AKLT model, $H^{(C)}_{\rm sym}$ reduces to the integrable parent Hamiltonian $H_0$ discovered in Ref.~\ref{eq:MLM}. Though we do not know whether $H^{(C)}_{\rm sym}$ is integrable in the more general case, its spectrum is much more degenerate than required by $Q$-SU(2) symmetry alone, suggesting that additional non-abelian symmetries beyond $Q$-SU(2) are generically present in these models.
This enhanced degeneracy allows multiple multiplets with different $Q^2$ eigenvalues to have degenerate energy eigenvalues, which is necessary to allow the asymmetric scar states to be eigenstates of a $Q$-symmetric Hamiltonian.

We can provide a loose lower bound on the degeneracies for $S>1$ by noting first that $\mathcal{R}$ and $\mathcal{L}$ for any base state $|\psi_0\rangle$ will only contain bonds that have at least one $|S\rangle$ and one $|-S\rangle$ respectively. Then $H_{Z2}$ must have at least $(2S-1)^L$ states with zero energy, corresponding to those product states that lack both $|-S\rangle$ and $|S\rangle$, since those states will be annihilated by $P^{(\mathcal{L})}_{j,j+1}$ and $P^{(\mathcal{R})}_{j,j+1}$.  We have that $\Hc^{(C)} = H_{Z2} - \frac{1}{S}S^z$, but note that $S^z$ and $H_{Z2}$ commute. Since $S^z$ has only $2LS+1$ unique eigenvalues, the addition of $S^z$ to $H_{Z2}$ is not enough to change the asymptotically exponential-in-$L$ degeneracy of $\Hc^{(C)}$. 

It is also interesting to study parent Hamiltonians that have multiple sets of scar towers as eigenstates, such that the choice of $H_A$ will dictate which towers survive as eigenstates. As a simple example, consider the operator 
\begin{equation} \label{Eq:H2s}
    \Hc^{(2S)} = \sum_j|t=2S,t^z=0\rangle \langle t=2S,t^z=0|_{j,j+1} \ .
\end{equation} 
$Q^+_{AKLT}$ and $Q^-_{AKLT}$ will annihilate bonds with total spin $t$ even and total z-magnetization $t^z=0$, which means that $\sum_j |t=2S,t^z=0\rangle \langle t=2S,t^z=0|_{j,j+1}$ is $Q$-SU(2) symmetric. This operator annihilates the ground states of the regular spin-$S$ and SO(2S+1) AKLT models, since $|t=2S,t^z=0\rangle \langle t=2S,t^z=0|_{j,j+1}$ projects onto a bond that is not in said ground states. It follows by the $Q$-SU(2) symmetry that the states in the towers are all at zero energy. Similarly to $\Hc^{(C)}$, this example Hamiltonian has a very large degeneracy. We can provide a lower bound on the degeneracy by noting that bonds with $t^z \neq 0$ are annihilated by this Hamiltonian, which implies that at least $(2S+1)(2S-1)(2S)^{L-2} ~ (2S)^L$ product states are annihilated. Consider a given $S^z$ product state from left to right. We have $(2S+1)$ choices for the first site, but to ensure that the first bond is not $t^z=0$, we have $2S$ choices for the second site. This follows similarly for the third through $L-1$th sites, but the last site on the chain has two constraints from sites $L-1$ and $1$ and so there are $2S-1$ choices for that site. In the next section, we discuss deformations between scarred Hamiltonians and illustrate some of the connections between the Hamiltonians discussed above by way of an explicit example involving the spin-$S$ AKLT scar tower and a tower of eigenstates of an integrable point.

\subsection{Paths between scarred Hamiltonians}\label{Sec:DisjointGR}

In this section, we give examples of deformations between scarred Hamiltonians with \textit{different} scar towers along which the scar states deform continously. In particular, we use the above-discussed continuously connected class of spin-$S$, bond-dimension $S+1$ matrix product states of the form 
\begin{align}\label{Eq:MPSScarBaseSt2}
    \ket{\psi_0, A} &= \sum_{\textbf{m}} \textrm{Tr}[A^{[m_1]}A^{[m_2]}...A^{[m_L]}]|m_1...m_L\rangle \\
    A^{[m]}_{ij} &= 0 \text{ for } j-i \neq m \nonumber
\end{align}
to deform between the regular spin-$S$ AKLT model and an integrable point. 
We noted above that the tower of states $(Q^+_{\rm AKLT})^n \ket{\psi_0, A}$ are eigenstates of Hamiltonians of the form in Eq.~\ref{Eq:gencanon} for the corresponding $\mathcal{G}$, $\mathcal{L}$ $\mathcal{R}$, $\Tilde{\mathcal{M}}$ determined by $\ket{\psi_0, A}$. We also noted that this class of states includes the $q$-deformed and regular spin-$S$ AKLT models' ground states.

The above results allow us to generalize Ref.~\cite{MoudMPS}'s deformation of the spin-$1$ AKLT model's ground state to a ground state of an integrable point. Ref.~\cite{MoudMPS} described a deformation of the matrices in the spin-1 AKLT ground state to that of an eigenstate of an  integrable pure-biquadratic model (discussed below). That is, they considered
\begin{equation}
\label{Eq: MoudDef}
A^{[+]} = c_{+} \sigma^{+}, A^{[0]} = c_{0} \sigma^{z}, A^{[-]} = c_{-} \sigma^{-} 
\end{equation}
for varying $c_{\pm, 0}$. The spin-1 AKLT ground state has coefficients $c_0 =-1$, $c_{-} = -\sqrt{2}$, $c_{+} = \sqrt{2}$, while $c_0 = -1$, $c_{-} = -i$, $c_{+} = i$ corresponds to an eigenstate of the integrable pure-biquadratic model. The authors used ``numerical brute force" to verify their version of the conditions that $\mathcal{G}$ and $\mathcal{R}$ are disjoint and that Eq.~\ref{Eq:qform} holds for every choice of the $c_{\pm, 0}$ coefficients. They thus constructed a family of Hamiltonians with the form in Eq.~\ref{Eq:hform} that connected the spin-1 AKLT model to the pure-biquadratic model. However, we see that numerical brute force is not needed; the conditions on $\mathcal{G}$, $\mathcal{R}$ and $q^+$ follow as an immediate corollary of our results in Appendix~\ref{appendix:qplus1} on MPS of the form in Eq.~\ref{Eq:MPSScarBaseSt2}, as the matrices in Eq.~\ref{Eq: MoudDef} are only nonzero on the correct diagonals.

We will now generalize Ref.~\cite{MoudMPS}'s deformation of the spin-1 AKLT ground state to a ground state of the pure-biquadratic model to spin-$S$. We note that the Hamiltonian of the spin-1 integrable pure biquadratic point is equivalent to a sum of projectors onto two-site spin-singlets.
A spin-$S$ chain with Hamiltonian given by a sum of projectors onto spin-singlets (the singlet-projector model SP)
\begin{align}
    H_{\rm S, SP} = \sum_j P^{(0)}_{j,j+1},
\end{align}
is similarly integrable\footnote{It is Temperley-Lieb equivalent to the Bethe-ansatz solvable XXZ model \cite{Batchelor_1990}} \cite{Batchelor_1990, Klumper_1990}. Furthermore, we note that there is a simple matrix product eigenstate of $H_{\rm S, SP}$, which can be written in terms of the matrices $A_\textrm{S,AKLT}^{[m]}$ that define the ground state of the spin-$S$ AKLT model:
\begin{equation}
A^{[m]}_{\rm S, SP} = \sqrt{(-1)^m \binom{2S}{S+m}} A^{[m]}_{\rm S,AKLT}
\end{equation}
It can be shown that this resulting state is annihilated by every two-site spin-singlet projector. Just like $A^{[m]}_{\rm S, AKLT}$, $A^{[m]}_{\rm S, SP}$ is only nonzero on the $m$th diagonal, and hence serves as a nice endpoint for a deformation between the spin-$S$ AKLT model and the integrable spin-$S$ singlet-projector model. This construction reduces to the form of the MPS in Ref.~\cite{MoudMPS} for $S=1$. 

We note that the spin-$S$ singlet-projector model is spin-SU(2) and $Q$-SU(2) symmetric\footnote{It is in fact SU(2S+1) symmetric if one uses different representations of SU(2S+1) on alternating sites~\cite{Klumper_1990}}. 
The $Q$-SU(2) invariance arises because the spin singlet (and more generally any bond with even total spin and z-magnetization 0) is annihilated by $Q^+$ and $Q^-$, as mentioned in Appendix~\ref{appendix:qplus2}. Because the model is $Q$-SU(2) invariant, the scar states all have the same energy. The model in fact corresponds to $\omega = 0$, and $h_\mathcal{{MM}}$ zero except for the projector onto the spin-singlet. However, at the cost of breaking the spin-SU(2) and $Q$-SU(2) invariance, we can assign energies to the scar states by setting $\omega>0$. We can furthermore make the model thermalizing outside the scar manifold by introducing generic $h_\mathcal{{MM}}$, e.g. a sum of projectors with unit coefficients.

Between the endpoints of the spin-$S$ AKLT model and singlet-projector model are many different paths along which the scar states deform continuously. For example, one could take the path 
\begin{equation}
\label{Eq:defex}
    A^{[m]} = c_{m} A^{[m]}_{\rm S,AKLT}
\end{equation}
and interpolate $c_{m}$ between $1$ and $\sqrt{(-1)^m \binom{2S}{S+m}}$ as some function of some parameter $\lambda$, where $h_\mathcal{{MM}}$ and $\omega$ are functions of $\lambda$. 

\begin{figure}
\centering
\includegraphics[width=\columnwidth]{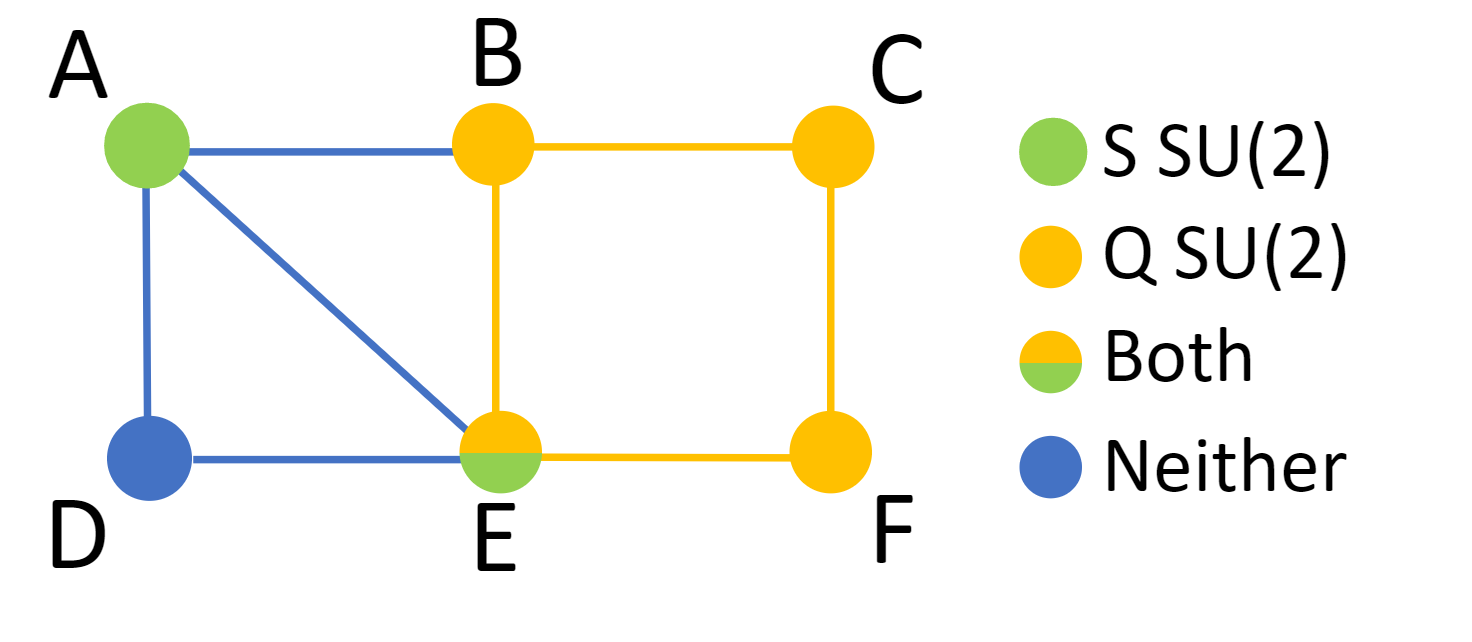}~
\caption{Schematic sketch of paths in the space of Hamiltonians. The paths and points are colored according to whether they have spin SU(2) ($S$-SU(2)) and/or $Q$-SU(2) symmetry. The label of each point corresponds to one of the Hamiltonians in Eq. ~\ref{eq:abcdef}.  Horizontal lines represent deformations of Hamiltonians that preserve a specific scar tower.  Vertical lines represent deformations that change the scar states according to Eq.~\ref{Eq:defex} while keeping the general structure of the Hamiltonian fixed. 
\label{fig:paths}}
\end{figure}

We give a schematic sketch of such a path and other interesting deformations in Fig.~\ref{fig:paths}. We look at deformations between six Hamiltonians labeled $A,B,C,D,E,F$. Here, 
\begin{align}
\begin{split}
\label{eq:abcdef}
   &A = H_{\rm S, AKLT} \\
   &B =  \Hc^{(2S)}    \\
   &C = H^{(C)}_{\rm sym}\left(A_{\rm S,AKLT}\right) \\
   &D = H_{\rm unify}\left(A_{\rm S, SP}\right) \\
   &E = H_{\rm S,SP} \\
   &F = H^{(C)}_{\rm sym}\left(A_{\rm S, SP}\right).
\end{split}
\end{align}

Note that we have made the dependence of $\Hc^{(C)}$ and $H_{\rm unify}$ on the base state explicit; $\Hc^{(C)}(A)$ and $H_{\rm unify}(A)$ refer to the $\Hc^{(C)}$ and $H_{\rm unify}$ in Eqs.~\ref{Eq:canonical} and ~\ref{Eq:AKLTcanon} constructed for the base state $|\psi_0, A\rangle$ in Eq.~\ref{Eq:MPSScarBaseSt}. Additionally, $H_{\rm S, AKLT}$ refers the regular spin-$S$ AKLT model.
The schematic is designed so that every horizontal line corresponds to a linear space of Hamiltonians that has a specific tower of states as eigenstates. That is, moving horizontally does not change the scar states, while moving vertically does: i.e. the Hamiltonians $A,\,B,\,C$ all have the spin-$S$ AKLT tower of states as eigenstates, while the Hamiltonians $D,\,E,\,F$ all have the states $(Q^+_{\rm AKLT})^n|\psi_0, A_{\rm S,SP}\rangle$ as eigenstates. 

Vertical lines represent families of Hamiltonians that share a common form, but with different operators and scar towers due to a distinct choice of base state.
Explicitly, the leftmost vertical line $\overline{AD}$ corresponds to Hamiltonians of the form of $H_{\rm unify}$ in Eq.~\ref{Eq:AKLTcanon} for states of the form in Eq.~\ref{Eq:defex} interpolating between $A_{\rm S,AKLT}$ and $A_{\rm S, SP}$. Similarly, the rightmost vertical line $\overline{CF}$ corresponds to Hamiltonians of the form of $H^{(C)}_{\rm sym}$ in Eq.~\ref{Eq:canonical} for said states. The middle line $\overline{BE}$ corresponds to Hamiltonians of the form $\sum_{k=0}^{S} a_m |t=2k, t^z = 0\rangle \langle t=2k, t^z = 0|$ with coefficients chosen to annihilate the interpolating states. Finally, the oblique line $\overline{AE}$ corresponds to a more general path in Hamiltonian space between $H_{\rm S,AKLT}$ and $H_{\rm S, SP}$; an example would be the deformation given in Ref.~\cite{MoudMPS} for $S=1$.

The horizontal lines between adjacent points $X$ and $Y$ interpolate between the relevant models, e.g. $\lambda X + (1-\lambda)Y$, with $\lambda$ varying from $0$ to $1$. More generally, moving along the horizontal lines corresponds to changing the specific choices of $H_A$ and $\Hc$ for the fixed base state (as well as changing the coefficient of $\Hs$ to change the energy spacing of the scar states, such as along $\overline{AB}$). 

The paths and points are color-coded according to whether they have spin-SU(2) or $Q$-SU(2) symmetry. The vertical line $\overline{BE}$ has $Q$-SU(2) symmetry, as $Q^+$ and $Q^-$ annihilate states of the form $|t=2k, t^z = 0\rangle$. The vertical line $\overline{CF}$ has $Q$-SU(2) symmetry by the careful design given in Section \ref{sec:parent}. We do not share any explicit paths in Hamiltonian space with spin-SU(2) symmetry, and it is not clear a priori that such paths exist that interpolate between the spin-$S$ AKLT tower of states and properly chosen towers of eigenstates in the singlet projector model. We wouldn't necessarily expect such paths a priori, as $Q$-SU(2) symmetry and $S$-SU(2) are on different footings because the ladder operator is a generator for $Q$-SU(2) and not $S$-SU(2). Furthermore, there are more $Q$-SU(2) than $S$-SU(2) symmetric Hamiltonians. This follows because of the reducibility of the $Q$-SU(2) representation at the single-site level, which yields a larger number of $Q$-SU(2) symmetric nearest-neighbor bonds from the larger number of two-site multiplets with definite $Q^2$ eigenvalues.

Finally, we emphasize that this schematic shows a very small selection of paths between the points. Each horizontal line lies in a hyperplane of parameters like $\omega$ and those in $h_\mathcal{{MM}}$. There are, e.g., paths between $A$ and $C$ that do not pass through $B$, and paths between $B$ and $C$ lying in the horizontal hyperplane that lack $Q$-SU(2) symmetry. There are also many vertical and oblique paths which will depend not only the choices of $\omega$ and $h_\mathcal{{MM}}$ but also on the particular path in state space interpolating between $|\psi_0, A_{\rm S,AKLT}\rangle$ and $|\psi_0, A_{\rm S,SP}\rangle$ (in this case, how each $c_m$ interpolates between $1$ and $\sqrt{(-1)^m \binom{2S}{S+m}}$). There is a very rich space of scarred Hamiltonians to explore.

\section{Conclusions and Outlook}
\label{sec:conclude}

In this work, we have presented a general framework for understanding how quantum scars emerge from parent Hamiltonians with non-Abelian (and possibly $q$-deformed) symmetries. Generators of the symmetry furnish a natural set of operators with spectrum generating commutation relations, and the parent Hamiltonians have rich structure in their eigenspectrum as a consequence of the symmetry. In particular, the spectrum of $\Hc$ is organized as degenerate multiplets (`tunnels') that transform as irreps of the symmetry. Scars emerge when perturbations generically destroy the symmetry and give a thermal spectrum, but do so in a manner that preserves a \emph{shadow} of the symmetry so that a particular multiplet of low entanglement states fails to mix with the rest of the Hilbert space. This furnishes one qualitative  `picture' for how and when one might expect scars to arise generally, something that has thus far been largely missing in the literature. 

Our framework applies to several known models with scars in the literature, but it has allowed us to also introduce several new models with exact quantum scars, significantly generalizing the types of systems known to harbor this phenomenon. These models fall into two broad classes.  In the first class, the scar states transform in a single irreducible representation of the symmetry group.  Our examples in this class include models where the symmetry is a $q$-deformation of SU(2), as well where the relevant symmetry is SU(3) rather than SU(2).  In the second class, the scar states do not belong to a single representation of the relevant symmetry group, which requires the parent Hamiltonian to have an enhanced symmetry. We have presented examples of this type not previously known in the literature, including generalizations of the AKLT model and families of scarred Hamiltonians that can be smoothly deformed into each other. It is interesting to note that prior studies have tried to explain scars in the PXP model via deformations towards integrable models~\cite{SigInt2019} and those with approximate SU(2) symmetry~\cite{AbaninPerfectScars,Papic2020}. Hence this symmetry based framework may prove key to eventually fully understanding scars in the PXP model, too. 

Our framework leaves open many important questions about the qualitative features that distinguish Hamiltonians with quantum scars from their  ETH-satisfying peers.  For example, what distinguishes models in which the symmetry is broken in a generic way from those in which a scarred subspace persists? A key general question concerns the stability of scars to perturbations, and whether scars can survive these perturbations either in an asymptotic or `prethermal' sense~\cite{SigInt2019, LinChandranMotrunich2020}. Indeed, to classify scars as a new kind of dynamical \emph{phase} of matter with `intermediate' thermalization properties - neither fully ergodic nor fully localized - requires scarred models to display some degree of stability in phase space. One important consequence of our picture is that it furnishes a family of scarred models emanating from symmetric parent Hamiltonians, and thus shows that scarred models can have at least some degree of stability to certain classes of perturbations.

{\it Acknowledgments---} The authors are grateful to Daniel Renard, Cheng-Ju Lin and Lesik Motrunich, for helpful discussions.  A.C. and V.K. are supported by the Sloan Foundation through a Sloan Research Fellowship. This work was supported by the National Science Foundation through the awards NSF DMR-1752759 (A.C.) and DMR-1928166 (F.J.B.), and by the US Department of Energy, Office of Science, Basic Energy Sciences, under Early Career Award No. DE-SC0021111 (V.K. and N.O.D). 

{\it Note Added---} While we were finalizing our manuscript, we became aware of two related recent works, Refs.~\cite{Klebanov2020} and ~\cite{Fang2020}, which also apply group theoretic considerations to scarred Hamiltonians. Our results agree where they overlap, although the scope of our work is broader than Ref.~\cite{Klebanov2020} which only considers Casimir singlets, and our results on $q$-deformed and asymmetric scar towers also lie outside the constructions of Ref.~\cite{Fang2020}.

\appendix

\section{MPO representations for $(Q^+)^n$}
\label{App:MPO}

In the text, we discuss creating scars by repeated action of some operators on a base state. We show in this appendix that for  operators $Q^+$ of the form discussed here, $(Q^+)^n$ has a simple MPO representation with a bond dimension of only $n+1$.  This implies that acting $(Q^+)^{Poly(L)}$ on some base state can only increase the entanglement entropy of the base state by at most $O(\log(L))$. Our approach is inspired by the examples given in \cite{MoudEntang}. We extend the results there and take a different style of proof.   We emphasize that our arguments apply to raising operators $Q_{\balpha}$ discussed above for general Lie groups, as well as the raising operators relevant to $q$-deformed SU(2).

Consider an operator $Q^+$ that can be expressed as a momentum-$k$ sum of operators:
\begin{equation}
    Q^+(L) = \sum_{m=1}^L e^{ikm} (\otimes_{j=1}^{m-1} \mathcal{L}_j) \otimes Q^+_m \otimes (\otimes_{j=m+1}^{L} R_j) \ .
\end{equation}
Note that each term in the sum can have a non-local ``tail", represented by the products over $\mathcal{L}_j$ and $R_j$.
We will additionally require that at the single-site level, 
\begin{equation} 
\label{Eq: lrcomm}
    [\mathcal{L}_i,R_i]=0,\, \mathcal{L}_iQ^+_i = \frac{1}{l}Q^+_i \mathcal{L}_i,\,\text{ and } R_iQ^+_i = r Q^+_i R_i
\end{equation}
for some arbitrary numbers $r$ and $l$.

As an example, we could have $Q^+_j = S^+_j$, $\mathcal{L}_j=R_j=I_j$, and $k=0$; the commutation relations are satisfied with $r,l=0$. This choice corresponds to $Q^+ = S^+$.
As another example, consider $Q^+_j = \Tilde{S}^+_j$, $\mathcal{L}_j=q^{-S^z_j}$, $R_j =q^{S^z_j}$ and $k=0$; this has $r = q$ and $l = \frac{1}{q}$. This choice corresponds to the raising operator for SU$_q$(2), as discussed in Section \ref{sec:qdef}. 

We begin by specializing to $k=0$. Define the $n+1$ by $n+1$ matrix 
\begin{equation}
\label{Eq: mform}
\boldsymbol{M}_{ \alpha, \beta}(X, Y, Z)= \frac{1}{[[\beta -\alpha]]_{rl}!} ( X^{n+1-\beta} Y^{\beta - \alpha} Z^{\alpha-1})    
\end{equation}
for $\beta \geq \alpha$ and 0 otherwise.
Here we have defined
\begin{equation}
    [[n]]_q  = \frac{1-q^n}{1-q}, \, [[n]]_q !  = \prod_{m=1}^n [[m]]_q,
\end{equation}
 for $m$ and $n$ positive integers.
From $\boldsymbol{M}$, we define the $L$ matrices \begin{equation}
M_{j} = \boldsymbol{M}(\mathcal{L}_j, Q^+_j, R_j).
\end{equation}   
 
Our claim is that
\begin{equation}
\label{Eq: lmr}
    v_l^T (\prod_{i=1}^L M_i)v_r = \frac{1}{[[n]]_{rl}!}(Q^+(L))^n,
\end{equation}
where $(v_r)_i = \delta_{i,1}$, and $(v_l)_i = \delta_{i,n+1}$.

We'll prove \ref{Eq: lmr} by way of a stronger result:
\begin{equation}
\label{Eq: lmr2}
    \prod_{i=1}^L M_i = \boldsymbol{M}_{\alpha, \beta}(\otimes_i^L \mathcal{L}_i, Q^+(L), \otimes_i^L R_i)
\end{equation}
This stronger result implies equation \ref{Eq: lmr} by the definition \ref{Eq: mform} through
\begin{equation}
\boldsymbol{M}_{1, n+1}(\otimes_i^L \mathcal{L}_i, Q^+(L), \otimes_i^L R_i) = \frac{1}{[[n]]_{rl}!} (Q^+(L))^n
\end{equation}

We will prove equation \ref{Eq: lmr2} by induction on $L$. The base case of $L=1$ follows by inspection.

Assume the form in equation \ref{Eq: lmr2} holds for $L-1$.
Then 
\begin{widetext}

\begin{align}
(\prod_{i=1}^L M_i)_{\alpha \beta} = & \sum_\gamma (\prod_{i=1}^{L-1} M_i)_{\alpha \gamma}M_{L_{\gamma \beta}} \\
=& \sum_\gamma \frac{1}{[[\gamma - \alpha]]_{rl}!} (\otimes_{i=1}^{L-1} \mathcal{L}_i)^{n+1 -\gamma}(Q^+(L-1))^{\gamma - \alpha} (\otimes_{i=1}^{L-1} R_i)^{\alpha -1} \otimes\frac{1}{[[\beta - \gamma]]_{rl}!}  \mathcal{L}_L^{n+1-\beta} (Q^+_L)^{\beta - \gamma} R_L^{\gamma-1}  \\
=& \text{$\frac{1}{[[\beta - \alpha]]_{rl}!}$} (\otimes_{i=1}^{L} \mathcal{L}_i)^{n+1-\beta} \left[ \sum_\gamma (\frac{[[\beta - \alpha]]_{rl}!}{[[\beta - \gamma]]_{rl}! [[\gamma - \alpha]]_{rl}!}   (\otimes_{i=1}^{L-1} \mathcal{L}_i \otimes Q^+_L)^{\beta - \gamma} (Q^+(L-1)\otimes R_L)^{\gamma - \alpha}\right] (\otimes_{i=1}^{L} R_i)^{\alpha -1}
\label{Eq: bracksum}
\end{align}

\end{widetext}

We can simplify the sum on $\gamma$ by noting that for two matrices $A$ and $B$ such that $BA = x AB$, 
\begin{equation}
    (A+B)^n = \sum_{p=0}^n \frac{[[n]]_x!}{[[n-p]]_x! [[p]]_x!} A^{p}B^{n-p}.
\end{equation}
See, for example, Ref.~\cite{qgroupbook}'s ``q-Calculus" chapter. We see that the bracketed sum in \ref{Eq: bracksum} is indeed of this form for $x=lr$ as follows from the commutation relations in \ref{Eq: lrcomm}. Simplifying the bracketed sum yields
\begin{widetext}
\begin{align}
(\prod_{i=1}^L M_i)_{\alpha \beta} = & \text{$\frac{1}{[[\beta - \alpha]]_{rl}!}$} (\otimes_{i=1}^{L} \mathcal{L}_i)^{n+1-\beta} (\otimes_{i=1}^{L-1} \mathcal{L}_i \otimes Q^+_L + Q^+(L-1)\otimes R_L)^{\beta - \alpha} (\otimes_{i=1}^{L} R_i)^{\alpha -1} \\
= & \pmb{M}(\otimes_{i=1}^L \mathcal{L}_i, Q^+(L), \otimes_{i=1}^L R_i)_{\alpha \beta}
\end{align}
\end{widetext}

This concludes the proof for $k=0$ of equation \ref{Eq: lmr2} which, as noted above, immediately implies our goal of \ref{Eq: lmr}.

For $k \neq 0$, note that $\mathcal{L}_{j}' = e^{ik} \mathcal{L}_j$, $(Q^+)_{j}' = e^{ik} Q^+_j$, $R_{j}' = R_j$ also satisfy the commutation relations in equation \ref{Eq: lrcomm}.
Thus, the proof above holds for these primed single-site operators.
This implies that for $(M_i')_{\alpha \beta} = \pmb{M}(\mathcal{L}_i', (Q^+)_i', R_i')$, 
\begin{align}
v_l^T \left(\prod_{i=1}^L M_{i}'\right) v_r =& \left(\sum_i^L (\otimes_{j=1}^{i-1} \mathcal{L}_{j}') \otimes (Q^+)_{i}' \otimes (\otimes_{j=i+1}^{L} R_{j}')\right)^n \\
=& \left(\sum_j^L e^{ikj} (\otimes_{i=1}^{j-1} \mathcal{L}_i) \otimes Q^+_j \otimes \left(\otimes_{i=j+1}^{L} R_i\right)\right)^n,
\end{align}
i.e. the correct form for the momentum-$k$ sum. Thus we've proven that the same MPO with $L \rightarrow e^{ik}L$ and $Q^+_j \rightarrow e^{ik}Q^+_j$ works for the momentum-$k$ version of $(Q^+)^n$.

\section{Spin chains and higher rank symmetries}
\label{App:SpinLie}

Here, we present a detailed derivation of how higher-rank Lie groups are related to raising operators of the form $(S^+)^2$ acting on spin-$s$ representations, with $s > 3/2$. We will use notation more conventional to Lie algebras, such that $E^{(i)}_{a,b }$ are raising operators, $F^{(i)}_{a,b} $ are lowering operators, and $H$ will denote operators in the CSA.

We have:
\begin{equation}
(S^+)_{i+1,i+2} = \sqrt{s (s + 1) - (i - s) (i + 1 - s)}
\end{equation}
and
\begin{align}
(S^+)^2_{i, i+2} &= (S^+)_{i,i+1} (S^+)_{i+1,i+2} 
\end{align}
Notice that 
\be
(S^+)_{i+1,i+2} = (S^+)_{2s - i ,2s - i +1}
\ee
and therefore,
\be
(S^+)^2_{i, i+2} =  (S^+)^2_{2s - i,2s - i +2}
\ee

We thus define the raising operators
\begin{align}
E^{(i)}_{a,b } =& \delta_{a,i}  \delta_{b, i+2} + \delta_{a, 2s -i} \delta_{b, 2s- i +2} \nonumber \\
F^{(i)}_{a,b} =& (E^{(i)}_{a,b })^\dag
\end{align}
where $i$ is between $1$ and $s-1/2$ for half-integer $s$ and between $1$ and $s$ for integer $s-1$. 
For integer $s$, note that if $i=s$, the two matrices in the first line are the same, and we additionally define 
\be
E^{(s)}_{a,b } =  \delta_{a,s}  \delta_{b, s+2}.
\ee 
Other raising operators, which raise the spin by $4$ or more, can be obtained by taking commutators of these $E^{(i)}$-- i.e. the roots associated with these generators will be the simple roots. 

The Cartan subalgebra is equal to the vector space of diagonal matrices $H^{(i)} = [E^{(i)}, F^{(i)}]$. For $i\neq s$,
\begin{align}\label{Eq:Hdelta}
    H^{(i)}_{a,b} &= \delta_{a,i}\delta_{b,i} + \delta_{a, 2s-i}\delta_{b, 2s-i} \\
    &- \delta_{a,i+2}\delta_{b,i+2} -  \delta_{a, 2s-i+2}\delta_{b, 2s-i+2}
\end{align}
and additionally, for integer $s$,
\begin{equation}\label{Eq:Hsdelta}
    H^{(s)}_{a,b} = \delta_{a,s}\delta_{b,s}  - \delta_{a,s+2}\delta_{b,s+2}.
\end{equation}

Finally, in this basis, we have 
\be
(S^+)^2 = \sum_i a_i (s) E^{(i)} \ .
\ee
where
$a_i(s) = \sqrt{s (s + 1) - (i - s-1) (i  - s)}$ $\times \sqrt{s (s + 1) - (i - s) (i + 1 - s)} $.

\subsection{Half-integer spin case}

If the spin is half-integer, the matrices $E^{(i)}, F^{(i)}$, and $H^{(i)} = \left[ E^{(i)}, F^{(i)}\right]$ are a linear combination of the fundamental and anti-fundamental representations of the Lie group SU($s + \frac{1}{2}$). 
Specifically, the spins separate into two sets: $(-S, -S+2, ... S-3, S-1)$ and $(-S+1, -S+3, ... S-2, S)$.  The difference between spins in the first set and those in the second set is always an odd integer, so these two sets are not connected by the action of our generators.  However, each raising operator acts simultaneously on one spin from each set, such that the matrices $E^{(i)}, F^{(i)}$, and $H^{(i)}$ are linear combinations of a matrix in the fundamental representation, and one in the anti-fundamental representation.  As a result this set of matrices is not block diagonal, and the matrix commutators behave like a single irreducible representation of $SU(s+1/2)$.

To see this, we will compute the Cartan matrix of this representation.  For a rank $R$ Lie algebra, the Cartan matrix is an $R \times R$ integer matrix defined by
\be
C_{ij} = 2 \frac{ \balpha^{(i)} \cdot \balpha^{(j)} }{ \balpha^{(i)}  \cdot \balpha^{(i)} } \ .
\ee
where $\balpha^{(i)} $ are the simple roots.

Note that the roots $\balpha^{(i)}$ must be computed using an orthonormal basis $\{\tilde{H}^{(a)}\}$ of the CSA for which the inner product $Tr[\tilde{H}^{(a)}\tilde{H}^{(b)}] = \delta_{ab}$. For the matrices $H^{(a)} = [E^{(a)}, F^{(a)}]$ of Eq.~\ref{Eq:Hdelta}, we have $Tr[H^{(a)}H^{(a)}] = 4$, $Tr[H^{(a)}H^{(a+2)}] = -2$, and $Tr[H^{(s-1/2)}H^{(s-3/2)}] = -2$, with all other inner products $0$; the ${H^{(a)}}$ basis is not orthonormal.  
Using these inner products, it is straightforward to verify that an orthonormal basis is given by 
\begin{align}
&\tilde{H}^{(1)} = N_1 H^{(1)} \ , \ \ \tilde{H}^{(2)} = N_2 H^{(2)} \ , \\
&\tilde{H}^{(j)} =N_j \left ( \frac{1}{N_{j-2}} \tilde{H}^{(j-2)} + \text{ceil}\left( \frac{j}{2} \right ) H^{(j)} \right) \ \\
&\,\,\,\,\,\,\,\,\,\,\,\,\,\,\,\,\,\text{for } 2 < j < s - 1/2, \\
&\tilde{H}^{(s-1/2)} =  N_{s-1/2} (H^{(s-1/2)}+ C_1 \tilde{H}^{(s-5/2)}+ C_2 \tilde{H}^{(s-3/2)})
\end{align}
where
\ba
C_1 =\sqrt{ \frac{2 \ \text{ceil}[(s-5/2)/2] }{\text{ceil}[ (s-1/2)/2]}  }\n
 C_2 =\sqrt{ \frac{2 \ \text{ceil}[(s-3/2)/2] }{\text{ceil}[ (s+1/2)/2]} }
\ea
Here $N_i$ are normalization constants:
\be
N_j = \frac{1}{\sqrt{2 \text{ceil}(j/2)  \text{ceil}(j/2+1)}}
\ee
for $j< s-1/2$,
and 
\be
N_{s-1/2} =\frac{1}{\sqrt{4 - C_1^2-C_2^2}}\ . 
\ee

In this orthonormal basis, the simple roots are
\be
\left[H^{(a)}, E^{(b)}\right] = \alpha^{(b)}_a E^{(b)}
\ee

To find the roots in this orthonormal basis, we exploit the fact that
\be
\left[H^{(a)}, E^{(b)} \right] = (2 \delta_{a,b} - \delta_{a, b+2} - \delta_{a, b-2})E^{(b)} 
\ee
for $a< s-1/2$, and 
\begin{align}
    & \left [H^{(s-1/2)}, E^{(b)} \right ] = (2 \delta_{b, s-1/2} - \delta_{b, s-3/2} - \delta_{ b, s-5/2}) E^{(b)}\\
    &\left [H^{(s-3/2)}, E^{(s-1/2)} \right ]  =-  E^{(s-1/2)}
\end{align}

From these, we obtain:
\be
\left[\tilde{H}^{(a)}, E^{(b)} \right] =\sqrt{  \frac{\text{ceil}[a/2 +1 ]}{2 \text{ceil}[a/2  ]}} \delta_{a,b} - \sqrt{\frac{\text{ceil}[a/2 ]}{2 \text{ceil}[a/2 +1 ]}} \delta_{a, b-2} 
\ee
for $a,b<s-1/2$. 
The remaining nonzero commutators are
\ba
\left[\tilde{H}^{(s-1/2)}, E^{(s-1/2)} \right ] =& N_{s-1/2}(2 - \frac{1}{2} (C_1^2 + C_2^2)) E^{(s-1/2)} \n
\left[\tilde{H}^{(s-3/2)}, E^{(s-1/2)} \right ] =& -\frac{1}{ 2} C_2 
E^{(s-1/2)} \n
\left[\tilde{H}^{(s-5/2)}, E^{(s-1/2)} \right ] =& -\frac{1}{ 2} C_1 E^{(s-1/2)}
\ea

Thus the nonzero components of the roots are:
\begin{align}
\alpha^{(a)}_{a} &=   \sqrt{  \frac{\text{ceil}[a/2 +1 ]}{2 \text{ceil}[a/2  ]}} \\
 \alpha^{(a)}_{a-2} &= - \sqrt{\frac{\text{ceil}[a/2-1 ]}{2 \text{ceil}[a/2 ]}}
\end{align}
for $a < s-1/2$, and
\begin{align}
\alpha^{(s-1/2)}_{s-1/2} &= N_{s-1/2}(2 - \frac{1}{2} (C_1^2 + C_2^2))  \\
&= \sqrt{1-\frac{1}{4}(C_1^2+C_2^2)} \\
 \alpha^{(s-1/2)}_{s-3/2} &= -\frac{1}{ 2} C_2 \\
  \alpha^{(s-1/2)}_{s-5/2} &= -\frac{1}{ 2} C_1
\end{align} 
where here the superscript indexes the raising operator, and the subscript is a vector index.
We can see that these roots obey
\ba
\balpha^{(a)} \cdot \balpha^{(a)} = 1  \\
\balpha^{(a)} \cdot \balpha^{(a \pm 2)} = - \frac{1}{2} \\
\balpha^{(s-1/2)} \cdot \balpha^{(s -3/ 2)} = - \frac{1}{2}
\ea
with all other root vectors orthogonal. After re-ordering the roots (taking all odd roots in increasing order, followed by all even roots in decreasing order), this gives the Cartan matrix of SU$(s+1/2)$:
\be
C= \begin{pmatrix} 2 & -1 & 0 & 0 & ... & 0 \\
-1 & 2 & -1 & 0 & ... & 0 \\
\vdots &\vdots & \vdots & \vdots &...&\vdots \\
0&  ... & 0 &-1 & 2 & -1 \\
0&  ...& 0 &0 & -1 & 2 \\
\end{pmatrix}
\ee

\subsection{Integer spin case}

For integer spin, $E^{(2i)}$ connects only states with even integer $S^z$, while $E^{(2i)+1}$ connects only states with odd integer $S^z$.  Thus the raising and lowering operators split into two commuting sets -- one acting on even $S^z$, which we refer to as the even generators, and one acting on odd $S^z$, which we refer to as the odd generators.  Each of these two independent sets constitutes a matrix representation of a Lie group, which we will show here is either $Sp(n)$ or $SO(n)$, for $n= s, s+1$. 

Again, in order to construct the Cartan matrix, we must use an orthonormal basis of our CSA. For integer $s$, we still have $Tr[H^{(a)}H^{(a+2)}] = -2$, but note that $Tr[H^{(a)}H^{(a)}] = 4$ only for $a<s-1$. Notice that Eqns. $\ref{Eq:Hdelta}$ and $\ref{Eq:Hsdelta}$ imply that $Tr[H^{(s-1)}H^{(s-1)}] =Tr[H^{(s)}H^{(s)}] = 2$. From these inner products, we construct an orthonormal basis $\{\tilde{H}^{(a)}\}$:
\be
\tilde{H}^{(2 a+1)} = N_a \left ( \sum_{b=1}^{a+1} b  H^{(2b-1)} \right )
\ee
on odd generators, and similarly
\be
\tilde{H}^{(2 a)} = N_a \left ( \sum_{b=1}^{a} b  H^{(2b)}  \right )
\ee
on even generators.  Here 
\be
N_a = \begin{cases} \frac{1}{ \sqrt{2 a(a+1)}}  & a < a_{\text{max}}\\
\frac{1}{\sqrt{ 2 a} }& a = a_{\text{max}}
\end{cases}
\ee
where $a_{\text{max}} = \text{ceil}[s/2]$ ($\text{floor}[s/2]$) for the odd (even) generators, respectively.

If $s$ is an odd integer, 
then we have $(s+1)/2$ odd raising operators $E^{(2b+1)}$, and $(s-1)/2$ even raising operators  $E^{(2b)}$.  
The commutation relations of the odd generators with our original basis for the CSA are:
\be
\left [H^{(2a-1)}, E^{(2b-1)} \right] = (2 \delta_{a,b} - \delta_{a, b+1} - \delta_{a, b-1}) E^{(2b+1)}  
\ee
for $1  \leq a \leq (s+1)/2$, and $1  \leq b \leq (s-1)/2$.
Also,
\be
\left[H^{(2a-1)}, E^{(s)} \right] = 2 ( \delta_{a,(s+1)/2} - \delta_{a, (s-1)/2} ) E^{(s)} \ .
\ee
For the operators acting on even $S^z$ states, we have
\be
\left [H^{(2a)}, E^{(2b)} \right] = (2 \delta_{a,b} - \delta_{a, b+1} - \delta_{a, b-1}) E^{(2b)}   
\ee
for $1  \leq a \leq (s-1)/2$, and $1  \leq b \leq (s-3)/2$, and  
\be
\left[H^{(2b)}, E^{(s-1)} \right] = ( \delta_{b, (s-1)/2} - \delta_{b, (s-3)/2}) E^{(s-1)} \ .
\ee
The remaining commutators between raising operators and elements of the CSA vanish.

Thus if the spin $s$ is odd, we find the following roots associated with the odd generators:
\be \label{Eq:OddOddRoots}
\alpha^{(a)}_{b} = \sqrt{\frac{a+1}{2 a }} \delta_{b,a} -\sqrt{\frac{a-1}{2 a }} \delta_{b, a-1}
\ee
for $a< (s+1)/2$, and
\be
\alpha^{((s+1)/2)}_b = -2 \sqrt{\frac{s-1}{2s+2 }} \delta_{b, (s-1)/2} + \frac{2}{ \sqrt{s+1}} \delta_{b, (s+1)/2} \ .
\ee
Thus  for $a< (s+1)/2$,
\be
\balpha^{(a)} \cdot \balpha^{(a)}  = 1  \ , \ \ 
\balpha^{(a)} \cdot \balpha^{(a-1)} = -\frac{1}{2} 
\ee
while
\be
\balpha^{( (s+1)/2)} \cdot \balpha^{((s+1)/2)}  = 2 \ , \ \ 
\balpha^{( (s+1)/2)} \cdot \balpha^{((s-1)/2)} = -1 \ .
\ee
The resulting Cartan matrix is
\be
C = \begin{pmatrix} 
2 & -1 & 0 & 0 & ... & 0 \\
-1 & 2 & -1 & 0 & ... & 0 \\
\vdots & \vdots & \vdots & \vdots & ... &\vdots  \\
 0 & ... & -1 & 2 & -1 & 0 \\
  0 & ... & 0 & -1 & 2  & -2 \\
   0 & ... & 0 & 0 & -1  & 2 \\
   \end{pmatrix}
\ee
indicating that the algebra is $Sp(S+1)$.

The even generators are associated with the roots
\be \label{Eq:OddevenRoots}
\beta^{(a)}_{b} = \sqrt{\frac{a+1}{2 a }} \delta_{b,a} -\sqrt{\frac{a-1}{2 a }} \delta_{b, a-1}
\ee
for $a< (s-1)/2$, and
\be
\beta^{((s-1)/2)}_b = - \sqrt{\frac{s-3}{2(s-1) }} \delta_{b, (s-3)/2} + \frac{1}{ \sqrt{s-1}} \delta_{b, (s-1)/2} \ .
\ee
Thus  for $a< (s-1)/2$,
\be
\bbeta^{(a)} \cdot \bbeta^{(a)}  = 1  \ , \ \ 
\bbeta^{(a)} \cdot \bbeta^{(a-1)} = -\frac{1}{2} 
\ee
while
\be
\bbeta^{( (s-1)/2)} \cdot \bbeta^{((s-1)/2)}  = \frac{1}{2} \ , \ \ 
\bbeta^{( (s-1)/2)} \cdot \bbeta^{((s-3)/2)} = -\frac{1}{2} \ .
\ee
The resulting Cartan matrix is
\be
C = \begin{pmatrix} 
2 & -1 & 0 & 0 & ... & 0 \\
-1 & 2 & -1 & 0 & ... & 0 \\
\vdots & \vdots & \vdots & \vdots & ... &\vdots  \\
 0 & ... & -1 & 2 & -1 & 0 \\
  0 & ... & 0 & -1 & 2  & -1 \\
   0 & ... & 0 & 0 & -2  & 2 \\
   \end{pmatrix}
\ee
indicating that the algebra is $SO(S)$.

If $s$ is an even integer,  
then there are $s/2$ odd raising operators $E^{(2b+1)}$, and  $s/2$ even raising operators $E^{(2b)}$.   The relevant non-vanishing commutators are:
\be
\left [H^{(2a-1)}, E^{(2b-1)} \right] = (2 \delta_{a,b} - \delta_{a, b+1} - \delta_{a, b-1}) E^{(2b-1)}  
\ee
for $1  \leq a \leq s/2$, and $1  \leq b \leq (s-2)/2$.
In addition,
\be
\left [H^{(2a-1)}, E^{(s-1)} \right] = (\delta_{a,s/2} - \delta_{a, s/2-1} ) E^{(s-1)}  \ .
\ee
For the even generators, we find
\be
\left [H^{(2a)}, E^{(2b)} \right] = (2 \delta_{a,b} - \delta_{a, b+1} - \delta_{a, b-1}) E^{(2b)}   
\ee
for $1  \leq a \leq s/2$, and $1  \leq b \leq (s-2)/2$.
Finally,
\be
\left[H^{(2b)}, E^{(s)} \right] = (2 \delta_{b, s/2} -2 \delta_{b, s/2-1}) E^{(s)} \ .
\ee

If the spin $s$ is even, Eq. (\ref{Eq:OddOddRoots}) (Eq. (\ref{Eq:OddevenRoots})) holds for $b \leq s/2$, $a< s/2$ for  odd (even) generators, and 
\ba
\alpha^{(s/2)}_b = - \sqrt{\frac{s-2}{2 s}} \delta_{b, s/2-1} +\frac{1}{ \sqrt{s}}\delta_{b, s/2} \\
 \beta^{(s/2)}_b = - 2 \sqrt{\frac{s-2}{2s}} \delta_{b, s/2-1} +\frac{2}{ \sqrt{s}}\delta_{b, s/2}\ .
\ea
where as above, $\balpha^{(i)}$ denote the roots associated with odd generators, and $\bbeta^{(i)}$ denote the roots associated with even generators.  
We therefore find the same Cartan matrices as above, but with the role of even and odd generators interchanged.  Thus we obtain $SO(s+1)$ on the even generators, and $Sp(s)$ for the odd generators.

\section{A new class of as-a-sum annihilators}
\label{app:as-a-sum}

In this appendix, we show that the alternating spin-1 AKLT Hamiltonian:
\begin{align}
    H_A = \sum_j (-1)^j P^{(2)}_{j,j+1}
    \label{Eq:AKLTalternating}
\end{align}
provides a new as-a-sum annihilator of the tower of states generated by the ladder operator $Q^+_\textrm{AKLT}$ (Eq.~\eqref{Eq:QpAKLT} with $S=1$) and the $S=1$ AKLT base state. We therefore add to the known classes of Hamiltonians~\cite{MoudMPS, MM} with the spin-1 AKLT tower of states; for example, the staggered AKLT model:
\begin{align}
    H = \sum_j c_j P^{(2)}_{j,j+1}, \quad c_j = c_{j+2}
\end{align}
has a scarred tower with energy spacing $c_1+c_2$. 

Eq.~\eqref{Eq:AKLTalternating} is a specific application of the following general theorem. Consider models $H = \sum_j h_{j,j+1}$ with the form discussed in Sec.~\ref{sec:aklt}: namely, the Hamiltonian has a scar tower generated by an operator $Q^+$ acting on a base state $|\psi_0 \rangle$, such that the corresponding sets $\mathcal{G}$ and $\mathcal{R}$ are disjoint and further such that equations $\ref{Eq:hform}$ and $\ref{Eq:qform}$ are satisfied.
Then
\begin{equation}\label{eq:HA}
H_A = \sum_{j} (-1)^j h_{j,j+1}
\end{equation}
annihilates the scar tower as-a-sum. As the generalized AKLT models introduced in Sec.~\ref{sec:aklt} satisfy Eqs.~\ref{Eq:hform} and \ref{Eq:qform}, this result allows us to identify new terms that annihilate the scar towers in the spin-$S$, the $q$-deformed spin-$S$, and the SO(2S+1) AKLT chain.

Recall that Eqs.~\ref{Eq:hform} and \ref{Eq:qform} imply that the term $A_{j,j+1} = [h_{j,j+1},q^+_{j,j+1}] - \omega q^+_{j,j+1}$ annihilates every term in the scar tower, where $Q^+ = \sum_j e^{ikj} Q^+_j$, $q^+_{j,j+1} = Q^+_j + e^{ik} Q^+_{j+1}$, and the base state has zero energy. We will show below that the above $H_{A}$ annihilates the tower in periodic chains whenever $(-1)^L e^{ikL} = 1$. Such a condition is met by periodic chains of even length with $kL$ an integer multiple of $2\pi$ or in periodic chains of odd length with $kL$ an odd-integer multiple of $\pi$. We find:
\begin{equation}
\begin{split}
    [H_A, Q^+]_{pbc} = & \sum_{j=1}^L \sum_{l=1}^L (-1)^j [h_{j,j+1}, e^{ikl} Q^+_l] \\
               = & \sum_{j=1}^L (-1)^j e^{ikj} [h_{j,j+1}, Q^+_j + e^{ik} Q^+_{j+1}] \\
               = & \sum_{j=1}^L (-1)^j e^{ikj} (\omega (Q^+_j + e^{ik} Q^+_{j+1}) + A_{j,j+1}) \\
               = & \omega \left(\sum_{j=1}^L (-1)^j e^{ikj} Q^+_j + \sum_{j=1}^L  (-1)^j e^{ik(j+1)} Q^+_{j+1}\right) \\ 
               &+  \sum_{j=1}^L (-1)^j e^{ikj} A_{j,j+1} \\
               = & -\omega e^{ik}Q^+_1(1-(-1)^L e^{ikL}) \\ &+ \sum_{j=1}^L (-1)^j e^{ikj} A_{j,j+1} \\
               = & 0 + \sum_{j=1}^L (-1)^j e^{ikj} A_{j,j+1}
\end{split}
\end{equation}
Thus,
\begin{align}
    [H_A, Q^+]_{pbc} \ket{\psi_0} = 0
\end{align}
which is of the form in Eq.~\ref{eq:MLM} with $\omega = 0$. This completes the proof that $H_A$ in Eq.~\eqref{eq:HA} annihilates the scar tower in periodic chains with $(-1)^L e^{ikL} = 1$. $H_A$ is certainly an as-a-sum (rather than bond-wise) annihilator, as without the alternating sign it would give energy to the scar states. 

In open chains, the LHS of the above expression has boundary terms proportional to $Q^+_1$ and $Q^+_L$:
\begin{equation}
\begin{split}
    [H_A, Q^+]_{\rm obc} =& -\omega(e^{ik} Q^+_1 + (-1)^L e^{ikL}Q^+_L) \\
                     +& \sum_{j=1}^{L-1} (-1)^j e^{ikj} A_{j,j+1}.
\end{split}
\end{equation}
In order for $[H_A, Q^+]_{\rm obc} \ket{\psi_0} = 0$, we need to choose a base state that is annihilated by each of the boundary terms, $Q^+_{1,L} \ket{\psi_0}=0$. For the generalized AKLT models, there is always such a base state in the ground state manifold (see Appendix \ref{appendix:OBC}, where we additionally show that such states are valid base states for building a tower of eigenstates for a Hamiltonian of the form in Eq.~\ref{Eq:hform}). With this choice of base state, $H_A$ in Eq.~\eqref{eq:HA} annihilates the scar tower.

We have thus shown that $H_A=\sum_j (-1)^j h_{j,j+1}$ annihilates the scar tower as-a-sum for all models satisfying Eqs.~\ref{Eq:hform} and \ref{Eq:qform} for suitable base states and chain lengths.

\section{Explicit forms for AKLT Projectors}\label{appendix: Projectors}

In discussing the AKLT Hamiltonians, we noted that all of them could be written as sums of two-site projectors onto manifolds with various total spin or total $q$-deformed spin eigenvalues. We did not need the explicit forms of the projectors for our discussion. In this appendix, we give forms for the AKLT projectors in terms of spin operators, which are useful for generating the AKLT Hamiltonians for exact diagonalization.

For total spin-$S$, we can project into the total-spin $t$ manifold by projecting out everything else:
\begin{equation}
    P_{12}^{(t)} = \prod_{s\neq t} \frac{(S_1 + S_2)^2 - s(s+1)}{t(t+1) - s(s+1)}
\end{equation}
will vanish when acting on a two-site state of total spin $s \neq t$, and will reduce to $1$ acting on a state of total spin $t$.

Similarly, for the $q$-deformed models, we can write 
\begin{equation}
    P_{12}^{(t)} = \Tilde{P}^{(t)} = \prod_{s\neq t} \frac{\Tilde{S}^-_{12}\Tilde{S}^+_{12} + [S^z_{12}]_q[S^z_{12}+1]_q - [s]_q[s+1]_q}{[t]_q[t+1]_q - [s]_q[s+1]_q}
\end{equation}
where we have defined
\begin{equation}
    \Tilde{S}^-_{12} = \Tilde{S}^-_1 \otimes q^{S^z_2} + q^{-S^z_1} \otimes \Tilde{S}^-_2
\end{equation}
and 
\begin{equation}
    \Tilde{S}^+_{12} = \Tilde{S}^+_1 \otimes q^{S^z_2} + q^{-S^z_1} \otimes \Tilde{S}^+_2
\end{equation}
Here we have used the fact that the Casimir for SU$_q$(2) is $\Tilde{S}^-\Tilde{S}^+ + [S^z]_q[S^z+1]_q$ with eigenvalue $[s]_q[s+1]_q$.

For the explicit form of the single-site $\Tilde{S^i}$, see equations \ref{eq: qdefz} and \ref{eq: qdefplus}.

\section{Verifying the form of $q^+$ for the $q$-deformed AKLT model}
\label{appendix:qplus1}

In this appendix, we verify that $\mathcal{G}$ and $\mathcal{R}$ are disjoint under the action of $Q^+ = \frac{1}{(2S)!}\sum_i (-1)^i (S^+_i)^2$ in the $q$-deformed AKLT model. We also show that the form for $q^+$ matches that of equation \ref{Eq:qform}. As noted in the text, these two facts will complete our proof of scars in the $q$-deformed AKLT model in periodic boundary conditions.

We'll prove these facts for a much wider set of base states than just the $q$-deformed spin-$S$ ground states. Namely, we will show that for towers built with $Q^+$ on top of the states in Eq.~\ref{Eq:MPSScarBaseSt}, repeated here for ease,
\begin{align}
    \ket{\psi_0, A} &= \sum_{\textbf{m}} \textrm{Tr}[A^{[m_1]}A^{[m_2]}...A^{[m_L]}]|m_1...m_L\rangle \\
    A^{[m]}_{ij} &= 0 \text{ for } j-i \neq m \nonumber
\end{align}
that $\mathcal{G}$ and $\mathcal{R}$ are disjoint and $q^+$ has the stated form. The ground state of the $q$-deformed spin-$S$ AKLT model is indeed a matrix product state of this form for all $q$ \cite{Motegi2010, Santos2012}.

The space of two-site bonds present in a given  translationally-invariant matrix product state of bond dimension $\chi$ will be contained within the span of the $\chi^2$ states
\begin{equation}
    |AA\rangle_{ij} = \sum_{m_1, m_2, k} A^{[m_1]}_{ik} A^{[m_2]}_{kj} |m_1 m_2\rangle.
\end{equation}
In the case at hand, this product involves a matrix with non-zero elements only on the $m_1^{th}$ diagonal and a matrix with non-zero elements only on the $m_2^{th}$ diagonal.  The result is a matrix with non-zero elements only on the $m_1+m_2^{th}$ diagonal. Thus $|AA\rangle_{ij}$ has contributions only from kets with $m_1+m_2 = j - i$.  Since $|j-i|\leq S$, it follows that $-S \leq m_1 + m_2 \leq S$ for all the states in $\mathcal{G}$.  The number of linearly independent states with magnetization $m_1+m_2$ is given by the number of choices of $i$ and $j$ with $0 \leq i,j \leq S$, and $j-i = m_1+m_2$, which is $S+1 - |m_1+m_2|$.  

To see that $\mathcal{G}$ and $\mathcal{R}$ are disjoint, note that the action of $q_j^+ = \frac{1}{(2S)!}\left( (S^+_j)^{2S} - (S^+_{j+1})^{2S} \right )$ increases the z-magnetization of $m_1 + m_2$ by $2S$.  Thus, $q^+$ takes the states in $\mathcal{G}$ with $m_1+m_2 > -S$ to those with total z-magnetization $>S$, which is outside of $\mathcal{G}$. There is only one state within $\mathcal{G}$ with magnetization $m_1+m_2 = -S$ , $|AA\rangle_{S+1, 1}$, and there is only one state in $\mathcal{G}$ with magnetization $S$, $|AA\rangle_{1, S+1}$. In order to complete the proof of disjoint $\mathcal{G}$ and $\mathcal{R}$, we have to verify that despite having the same z-magnetization as $|AA\rangle_{1, S+1}$, the state $q^+|AA\rangle_{S+1, 1}$ is orthogonal to $|AA\rangle_{1, S+1}$.

Within $|AA\rangle_{S+1, 1}$, all the $|m_1 m_2\rangle$ states are annihilated under $q^+$ except for $|-S0\rangle$ and $|0-S\rangle$, which are mapped to $|S0\rangle$ and $-|0S\rangle$ respectively. Thus, looking at the explicit form of $|AA\rangle_{ij}$, we see $q^+|AA\rangle_{S+1, 1} \propto A^{[-S]}_{S+1, 1} A^{[0]}_{11}|S0\rangle - A^{[0]}_{S+1, S+1} A^{[-S]}_{S+1, 1}|0S\rangle \propto A^{[0]}_{11}|S0\rangle - A^{[0]}_{S+1, S+1}|0S\rangle$. On the other hand, $|AA\rangle_{1, S+1} \propto A^{[0]}_{S+1,S+1}|S0\rangle + A^{[0]}_{11}|0S\rangle + \text{``terms with $m_1,m_2<S$''}$, which is orthogonal to $q^+|AA\rangle_{S+1, 1}$.  
That is,  $q^+$ maps $|AA\rangle_{S+1 1}$ to a state outside of $\mathcal{G}$, and hence all states in $\mathcal{G}$ are mapped to states outside of $\mathcal{G}$ under the action of $q^+$. We've thus shown that $\mathcal{R}$ and $\mathcal{G}$ are disjoint for towers built with $Q^+$ on top of the matrix product states described above. 

Further, noting that $(q^+)^2$ raises the total z-magnetization by $4S$ and hence annihilates all the states in $\mathcal{G}$, we see that $q^+$ annihilates the states in $\mathcal{R}$. Putting all the information together, we see that the form for $q^+$ (namely, the blocks of zeros) is indeed the one given in Eq.~\ref{Eq:qform}.

Thus, we've proven the lemma about disjoint $\mathcal{G}$ and $\mathcal{R}$ and the form of $q^+$ for the spin-$S$ $q$-deformed AKLT models, completing the proof that the tower of states are indeed eigenstates for these models. We emphasize that disjoint $\mathcal{G}$ and $\mathcal{R}$, and the form for $q^+$, were all satisfied for the large class of matrix product states in equation \ref{Eq:MPSScarBaseSt}.

\section{Verifying the form of $q^+$ for the SO(2S+1) AKLT model}
\label{appendix:qplus2}

In this appendix, we verify that $\mathcal{G}$ and $\mathcal{R}$ are disjoint under the action of $Q^+ = \frac{1}{(2S)!}\sum_i (-1)^i (S^+_i)^2$ in the SO(2S+1) AKLT model. We also show that the form for $q^+$ matches that of equation \ref{Eq:qform}. 

The Hamiltonian is $\sum_j \sum_{k=1}^{S} P^{(2k)}_{j,j+1}$, which projects onto two-site total spin $t$ even and greater than zero. Since the ground state is frustration-free, the bonds of the ground state necessarily live within either total spin odd, or zero total spin. That is, $\mathcal{G}$ is contained within the span of total spin odd states and the zero spin state.

Under the action of $q^+$, the $t$-odd states in $\mathcal{G}$ are mapped to $t$-even states, and vice versa. This is because $(S^+_1)^{2S} - (S^+_2)^{2S}$ is odd under exchange of $1$ and $2$, and states with total spin even (odd) are even (odd) under exchange of $1$ and $2$. Thus the action of $(S^+_1)^{2S} - (S^+_2)^{2S}$ acting on a total spin odd (even) state yields a state that is even (odd) under exchange, proving the stated result. This useful property of $q^+$ was noted in reference \cite{MLM}.

Under the action of $q^+$, the unique state of total (2-site) spin zero, $|t=0, t^z=0\rangle$, is annihilated. Further, no states are mapped via $q^+$ to the total spin zero state. This is because $q^+$ acting on a  state with $t^z =0$ needs to return a state with $t^z=2S$, but the only such state, $|t=2S,m=2S\rangle$, has the same total spin parity (even) as $|t=0, t^z=0\rangle$.  Similarly, the only state satisfying $t^z= -2S$ is $|t=2S, m=-2S\rangle$, which again the same total spin parity as $|t=0, m=0\rangle$, and hence cannot be mapped onto it by the action of $q^+$.

Putting this together, this means that the total spin odd states in $\mathcal{G}$ are mapped to total spin even and greater than zero, i.e. states outside of $\mathcal{G}$, while the total spin zero state in $\mathcal{G}$ is just annihilated, so $\mathcal{R}$ and $\mathcal{G}$ are disjoint.
Since $(q^+)^2$ raises $t^z$ by $4S$ and none of the bonds in $\mathcal{G}$ have $t^z = -2S$, we again see that $q^+$ must annihilate the states in $\mathcal{R}$. Putting this all together, $q^+$ must have the form given in Eq. \ref{Eq:qform}. 

This proves the lemma for the AKLT-like point of the SO(2S+1) AKLT model and completes the proof that the tower of states built off the ground state are all eigenstates of the Hamiltonian.

\section{Scar towers in generalized AKLT models with OBC}
\label{appendix:OBC}
Our discussion about the generalized AKLT models in the main text was limited to periodic boundary conditions (PBC). However, there are several key differences between PBC and open boundary conditions (OBC): in OBC, the ground state is no longer unique; instead the number of ground states grows with $S$. As a consequence, the number of linearly independent scar towers that can be formed by using one of these ground states as a base state also grows with $S$. However, not every ground state will be a good base state for a scar tower. In particular, we will show that, in OBC, for the $q$-deformed and regular spin $S$ AKLT models only $S^2$ out of $(S+1)^2$ ground states constitute the base of linearly independent scar towers; towers built on the other ground states are not eigenstates of the OBC Hamiltonian. For the SO(2S+1) models, we show that the corresponding number is $4^{S-1}$ out of $4^S$ ground states. In past work on the subject, reference \cite{MoudDisc} described one of the $S^2$ towers in the regular spin-$S$ AKLT model in OBC. 

In order to count the ground states, note that the ground states of the PBC models were frustration free, unique, and could be represented as matrix product states
\begin{equation}
|\psi_0[A]\rangle = \sum_{\textbf{m}} \textrm{Tr}[A^{[m_1]}A^{[m_2]}...A^{[m_L]}] |m_1...m_L\rangle    
\end{equation}
for some model-dependent $A$ with some model-dependent bond dimension $\chi$. In open boundary conditions, the models enjoy $\chi^2$ frustration-free ground states; i.e., . 
\begin{equation}
\label{eqn:OBC_MPS}
|\psi_0[A]\rangle_{ij} = \sum_{\textbf{m}} (A^{[m_1]}A^{[m_2]}...A^{[m_L]})_{ij} |m_1...m_L\rangle.    
\end{equation}

Such states are ground states in OBC because they are in the kernel of the projectors in the Hamiltonian: They are composed of the same bonds as in the PBC ground state save for the bond between the edge spins at $1$ and $L$. Since the bond dimension of $A$ is $S+1$ for the $q$-deformed and regular AKLT models, while the bond-dimension of $A$ is $2^S$ for the SO(2S+1) models, there are $(S+1)^2$ ground states of the $q$-deformed and regular AKLT models, while there are $4^S$ ground states of the SO(2S+1) models. A small subtlety is that the SO(2S+1) models need to have long enough chain lengths $L$ for all the ground states found this way to be linearly independent; we will assume that is the case.\footnote{Certainly we need $(2S+1)^L$, the number of states in the chain, larger than $4^S$, the ground state degeneracy; numerically we find a larger threshold value of $L$, which increases with $S$.}

The main change for the proof of scars in the above models in open boundary conditions is that the missing $h_{L,1}$ in $H_{\textrm{obc}}$ changes the commutator of $[H, Q^+]$. We had before in equations \ref{eq: Hpbc} and \ref{eq: Hobc} that
\begin{equation}
  [H_{\textrm{pbc}}, Q^+] = 2\omega Q^+ + \sum_{j=1}^{L}e^{ikj} A_{j,j+1}  
\end{equation}
where $A_{j,j+1}$ annihilated the $j, j+1$ bond in all the states of the tower. Now we'll have 
\begin{equation}
    [H_{\textrm{obc}}, Q^+] =2\omega Q^+ +\sum_{j=1}^{L-1}e^{ikj}A_{j,j+1} - \omega e^{ik} Q^+_1 - \omega e^{ikL}Q^+_L
\end{equation}
where, in the cases discussed here, $k=\pi$, $\omega=1$ and $O$ is proportional to $(S^+)^{2S}$.

The ground states of the OBC Hamiltonian, given in equation \ref{eqn:OBC_MPS}, and towers of states built on top of them by $Q^+$ will be annihilated by each $A_{j,j+1}$ in $\sum_{j}^{L-1}e^{ikj}A_{j,j+1}$. This follows because the set of bonds in the states and the towers built on top of the states are the same as in $PBC$; i.e. $\mathcal{G}$, $\mathcal{R}$, and $\mathcal{M}$ are independent of boundary conditions, the towers contain only bonds in $\mathcal{G}$ and $\mathcal{R}$, and $A_{j,j+1}$ annihilates $\mathcal{G}$ and $\mathcal{R}$. (Of course, the ground states for $OBC$ may have bonds between $L$ and $1$ that are not in $\mathcal{G}$ or $\mathcal{R}$, but $A_{L,1}$ is also not in the sum $\sum_{j=1}^{L-1}A_{j,j+1}$.) However,  the $\chi^2$ OBC ground states $|\psi_0[A]\rangle_{ij}$ will not generically be annihilated by $Q^+_1$ and $Q^+_L$.  

There is a simple sufficient condition for a scar tower with OBC to be annihilated by $Q^+_1$ and $Q^+_L$: its base state must be a sum of $S^z$ product states such that in each product state, the edge spins are annihilated by $Q^+_1$ and $Q^+_L$. Clearly this condition ensures that the base state is annihilated by $Q^+_1$ and $Q^+_L$. Furthermore, it guarantees that on each product state within the base state, the action of $Q^+$ cannot change the edge spins.  Hence each state in the tower will remain a sum of product states whose edge spins are annihilated by $Q^+_1$ and $Q^+_L$. Thus the whole tower of states will be annihilated by $Q^+_1$ and $Q^+_L$.  

In our cases, $Q^+_1$ and $Q^+_L$ are proportional to $(S^+)^{2S}$, so satisfying the above condition simply means that the edge spins in the product states comprising the base state can't have $|S^z=-S\rangle$.

For the $q$-deformed and regular AKLT models, $S^2$ out of the $(S+1)^2$ OBC ground states satisfy this condition, and hence host towers of eigenstates. This follows from the explicit form of the ground states $|\psi_0[ A]\rangle_{ij}$ in equation \ref{eqn:OBC_MPS}. $A^{[m]}$ is an $S+1$ by $S+1$ dimensional matrix that lives only on the $m$th diagonal (i.e. $A^{[m]}$ is nonzero only for $j-i=m$). The possible values of $S^z_1$ in $|\psi_0[ A]\rangle_{ij}$ are determined by the leftmost matrix $A^{[m_1]}_{ik}$, where $k$ is summed over in the matrix product with $A^{[m_2]}$. Since the eigenvalue of $S^z_1$ is $k-i$, the eigenvalues of $S^z_1$ will range from $1-i$ to $ S+1 -i$ on the different product states in $|\psi_0[ A]\rangle_{ij}$ depending on the value of $k \in \{1,2,...,S+1\}$. Similarly,  $-S-1+j \leq S^z_L \leq j-1$. Thus for the $S^2$ states with both $i<S+1$ and $j>1$, the edge spins are not $|-S\rangle$ and hence are annihilated by $Q^+_1$ and $Q^+_L$. That is, for $i<S+1$ and $j>1$, the towers built on top of $|\psi_0[A]\rangle$ are all eigenstates. 

The $q$-deformed AKLT PBC ground state includes contributions from $|\psi_0[A]\rangle_{11}$ and $|\psi_0[A]\rangle_{(S+1)(S+1)}$, which contain some product states with $|-S\rangle$ at the right and left edges respectively. This means that the PBC ground state, despite also being one of the ground states of the OBC Hamiltonian, does not satisfy the above sufficient condition of $Q^+_1$ and $Q^+_L$ annihilating all the edge states.

We now show that the condition that our base state be a sum of product states, each of which is annihilated by both  $Q^+_1$ and $Q^+_L$, is in fact necessary to produce a scar tower in these models. 
To show this, it is sufficient to show that if $Q^+_1 + e^{ik(L-1)}Q^+_L$ annihilates an AKLT ground state $|\psi[A] \rangle$, then {\it both} $Q^+_1$ and $Q^+_L$ 
must annihilate $|\psi[A] \rangle$.  

First, let us consider one of the ground states $|\psi_0[A]\rangle_{ij}$ of the $q$-deformed or regular spin-$S$ AKLT models.  In general,  $|\psi_0[A]\rangle_{ij}$ has $1-i \leq S^z_1 \leq S+1 -i$, and   $-S-1+j \leq S^z_L \leq j-1$.  We have already shown that the ground states with both $i<S+1$ and $j>1$ are annihilated by $Q^+_1$ and $Q^+_L$ individually.  If $i= S+1$ we have $-S \leq S_1^z\leq 0$, such that for any $j$, $|\psi_0[A]\rangle_{S+1,j}$ generically contains product states with $S^z_1 =-S$ and $S^z_L =0$.  Similarly, for $j=1$, $-S \leq S_L^z\leq 0$, and $|\psi_0[A]\rangle_{i,1}$ generically contains product states with $S^z_1 =0$ and $S^z_L =-S$. 
For $L>2$, each of these ground states with  $i= S+1$ and/or $j=1$ will therefore contain at least one product state with one of the two configurations $(S^z_1 =0, S^z_L =-S)$ or $(S^z_1 =-S, S^z_L =0)$.  These terms are annihilated by $Q^+_1$ ($Q^+_L$); under $Q^+_L$ ($Q^+_1$), they are sent to terms of the form  $S^z_1 =0$, $S^z_L =+S$ (or vice versa).  However, since there is no state $|\psi\rangle$ such that $Q^+_1 |\psi \rangle$ has $S^z_1 =0$ ($Q^+_L |\psi \rangle$ has $S^z_L =0$), these terms cannot be cancelled by terms obtained from $Q^+_1|\psi_0[A]\rangle_{ij}$ ($Q^+_L |\psi \rangle$), i.e. $Q^+_1+e^{ik(L-1)}Q^+_L$ cannot annihilate these ground states.

Next, we show that all superpositions of the OBC ground states $|\psi_0[A]\rangle_{ij}$ with either $i=S+1$ or $j=1$ necessarily contain at least one of these problematic product states; i.e. we cannot construct a superposition in which the problematic terms cancel.  Because the terms to be cancelled are product states in the $S^z_i$ basis, only terms with the same eigenvalue under $S^z_{tot}$ can cancel. The $|\psi_0[A]\rangle_{ij}$ are eigenstates of $S^z_{tot}$ with eigenvalue $j-i$. There is only a single ground state with both $i=S+1$ and $j=1$, which is the unique ground state with $S^z_{tot}=-S$; for this state, there are no others which can be used to cancel out its problematic product states. Now consider the possible choices of $i$ and $j$ with a given eigenvalue $j-i > -S$ of $S^z_{tot}$: a single one of them has $i=S+1$ and a single one has $j=1$. The corresponding state with $i=S+1$ will contain some product states that begin with $|-S\rangle$ and end in $|0\rangle$ and the state with $j=1$ will contain product states that begin with $|0\rangle$ and end with $|-S\rangle$, but \textit{not} vice-versa. As they contain different problematic product states, these states cannot be superposed to eliminate the problematic product states. We have thus shown that the existence of problematic product states ensures that the sufficient condition that the base state is a sum of product states that are each annihilated by $Q^+_1$ and $Q^+_L$ is in fact a necessary condition.

We can make a similar set of arguments for the spin-$S$ SO(2S+1) AKLT model to show that $4^{S-1}$ of the $4^{S}$ OBC ground states host towers of exact eigenstates generated by $Q^+$. For these models, $A^{[0]} = -\otimes_{i=1}^S \sigma^z_i$ and for $m>0$, $A^{[\pm m]} = (\pm 1)^{m} \sqrt{2}(\otimes_{i=1}^{S-m} \sigma^z_i) \otimes \sigma^{\pm}_{S+1-m} \otimes (\otimes_{j=S+2-m}^S \sigma^0_j)$. This form of the MPS, though not quite explicitly given in Ref.~\cite{Tu2008} for general $S$, follows from that reference up to similarity transformations of the $A$. This means $A^{[-S]} = (-1)^{S} \sqrt{2} \sigma^{-}_{1} (\otimes_{j=2}^S \sigma^0_j)$, so the only non-zero values in $A^{[-S]}$ fall on the $-2^{S-1}$th diagonal. Then $|\psi_0[A]\rangle_{ij}$ doesn't have $|-S\rangle$ edge spins when $i\leq2^{S-1}$ and $j>2^{S-1}$; all these $4^{S-1}$ states will thus host towers of exact eigenstates generated by $Q^+$ as they satisfy the sufficient condition.

We believe that, similar to the case of the $q$-deformed AKLT model, 
it is in fact necessary that both $Q^+_1$ and $Q^+_L$ annihilate the base state for the SO(2S+1) AKLT model. 
This is because that each ground states in equation \ref{eqn:OBC_MPS} that fails to satisfy $i\leq2^{S-1}$ or $j>2^{S-1}$ will contain problematic product states\footnote{That is, states that have $S^z_1=-S$ and $S^z_L \neq S$ or vice-versa.} for large enough $L$.  However, we will not prove necessity in this case, since here it is more challenging to prove that every superposition of these ground states contains the problematic product states.

Finally, note that the discussion of the $q$-deformed AKLT model in OBC used only the property that $A^{[m]}$ is only nonzero on its $m$th diagonal, so the proofs there extend to all the states in Eq.~\ref{Eq:MPSScarBaseSt}. 

\section{Verifying the strict form of $q^+$ for the generalized AKLT models} \label{appendix:qplus3}

In this appendix, we verify that the ``strict" form of $q^+$ given in \ref{Eq:qform2} holds for the base states in Eq.~\ref{Eq:MPSScarBaseSt} (which include the regular and $q$-deformed spin-$S$ AKLT models) as well as for the base states of the SO(2S+1) AKLT models.

First, the fact that $\mathcal{R}$ is annihilated under $q^+$ was already shown in Appendices \ref{appendix:qplus1} and \ref{appendix:qplus2}. Additionally, that $q^+$ maps $\mathcal{G}$ only to $\mathcal{R}$ the definition of $\mathcal{R}$ and those same appendices. It follows that $q^+_\mathcal{\Tilde{M}R}, q^+_\mathcal{RR}, q^+_\mathcal{GR}, q^+_\mathcal{LR}$ and $q^+_\mathcal{\Tilde{M}G},q^+_\mathcal{GG},q^+_\mathcal{LG}$ are all zero.  

That $q^+$ cannot send $\mathcal{L}$ to $\mathcal{L}$ nor $\Tilde{\mathcal{M}}$ to $\mathcal{L}$ follows from the fact that $\mathcal{L}$ is not in the image of $q^+$. To see this, first note that none of the base states contain the fully polarized bond $|S,S\rangle$: as noted in Appendix \ref{appendix:qplus1}, the base states in Eq.~\ref{Eq:MPSScarBaseSt} only contain bonds with total z-magnetization between $-S$ and $S$, while Appendix \ref{appendix:qplus2} noted that the SO(2S+1) base states do not contain any bonds with even total spin above 0. This means that the states in $\mathcal{L}$ will have total z-magnetization below $0$, as $q^-$ lowers the z-magnetization of its image by $2S$. In turn, this means that $\mathcal{L}$ cannot be in the image of $q^+$, as the states in the image of $q^+$ all have total z-magnetization greater than or equal to $0$. Hence, we have $q^+_\mathcal{L\Tilde{M}}, q^+_\mathcal{LL} = 0$.

Because $\Tilde{\mathcal{M}}$ is disjoint from $\mathcal{L}$ by definition, it follows that the image of $q^+$ on $\Tilde{\mathcal{M}}$ has null intersection with $\mathcal{G}$. If it didn't, then the image of $q^-$ on $\mathcal{G}$ would have non-trivial intersection with $\Tilde{\mathcal{M}}$, a contradiction. Hence, we have $q^+_\mathcal{G\Tilde{M}} = 0$.

Finally, we must now show that $q^+_\mathcal{\Tilde{M}\Tilde{M}}$  and $q^+_\mathcal{RL}$ are all zero; namely that we cannot map $\Tilde{\mathcal{M}}$ to $\Tilde{\mathcal{M}}$; nor can we map $\mathcal{L}$ to $\mathcal{R}$. We will show these facts separately for the base states in Eq.~\ref{Eq:MPSScarBaseSt} and the base states of the SO(2S+1) models.

We showed in Appendix \ref{appendix:qplus1} that $\mathcal{R}$ for the base states in Eq.~\ref{Eq:MPSScarBaseSt} contains $A_{11}|S0\rangle - A_{(S+1)(S+1)}|0S\rangle$. Note that this is the only total z-magnetization $t^z = S$ bond in $\mathcal{R}$, as it is in the image of the only $t^z = -S$ state in $\mathcal{G}$. Analogously, $A_{(S+1)(S+1)}|-S0\rangle - A_{11}|0-S\rangle$ is the only $t^z = -S$ state in $\mathcal{L}$. Under the action of $q^+$, $A_{(S+1)(S+1)}|-S0\rangle - A_{11}|0-S\rangle$ is mapped to $A_{(S+1)(S+1)}|S0\rangle + A_{11}|0S\rangle$, which is orthogonal to $A_{11}|S0\rangle - A_{(S+1)(S+1)}|0S\rangle$. Since no other states in $\mathcal{L}$ and $\mathcal{R}$ have a $t^z$ that is $2S$ apart, it follows that $q^+_\mathcal{RL} = 0$. 

In showing that $q^+_\mathcal{\Tilde{M}\Tilde{M}}=0$, we will use the fact that $\Tilde{\mathcal{M}}$ is disjoint from $\mathcal{L}$, $\mathcal{R}$, and $\mathcal{G}$ and hence must only contain states orthogonal to the states in $\mathcal{L}$, $\mathcal{R}$, and $\mathcal{G}$.  Quite generically, the states in $\mathcal{L}$ with z-magnetization $S^z=m<-S$ will span the two-state space $|-S, m+S\rangle, |m+S, -S\rangle$, as said states have at least two states in $\mathcal{G}$ in their preimage\footnote{If many of the generically non-zero entries of the $m$th diagonal of $A^{[m]}$ are fine-tuned or set to zero, there could be pathological cases where $\mathcal{L}$ does not contain $|-S, m\rangle, |m, -S\rangle$. For these measure-zero cases, in designing the operator $\ref{Eq:HZ22}$, the projector $P_{j,j+1}^{(\mathcal{L})}$ should project onto the union of $\mathcal{L}$ and the states $|-S,m\rangle$ for all $m<0$, and analogously for $P_{j,j+1}^{(\mathcal{R})}$.}. Furthermore, from the above paragraph, the union of $\mathcal{L}$ and $\mathcal{G}$ generically contain both $|-S0\rangle$ and $|0-S\rangle$.  Accordingly, the bonds in $\Tilde{\mathcal{M}}$ with $S^z \leq -S$ will not contain any factors of $|-S\rangle$ and hence will be annihilated under the action of $q^+$. A similar argument shows that the bonds in $\Tilde{\mathcal{M}}$ with $S^z \geq S$ do not contain any factor of $|S\rangle$ and hence have an empty preimage under $q^+$, so the bonds in $\Tilde{\mathcal{M}}$ with $t^z \geq -S$ (i.e. those that are mapped to bonds with $t^z \geq S$) are mapped to bonds orthogonal to $\Tilde{\mathcal{M}}$. Thus $q^+_\mathcal{\Tilde{M}\Tilde{M}} = 0$.

Now we turn to the base states of the SO(2S+1) AKLT model. Note that $\mathcal{G}$ corresponds exactly to those bonds with odd total spin or total spin zero, and so it follows by the disjoint nature of $\mathcal{G},\Tilde{\mathcal{M}},\mathcal{R}$ that the states in $\tilde{\mathcal{M}}$ and $\tilde{\mathcal{R}}$ have total spin even and greater than zero. As noted in Appendix \ref{appendix:qplus2}, $q^+$ changes total spin even to total spin odd and vice-versa, so $q^+_\mathcal{\Tilde{M}\Tilde{M}}$ and $q^+_\mathcal{R\Tilde{M}}$ are immediately zero.

This concludes the proof of the form of $q^+$ for all these base states.

\bibliography{TunnelTower.bib}

%Control: production of eprint (-1) disabled
\begin{thebibliography}{65}%
\makeatletter
\providecommand \@ifxundefined [1]{%
 \@ifx{#1\undefined}
}%
\providecommand \@ifnum [1]{%
 \ifnum #1\expandafter \@firstoftwo
 \else \expandafter \@secondoftwo
 \fi
}%
\providecommand \@ifx [1]{%
 \ifx #1\expandafter \@firstoftwo
 \else \expandafter \@secondoftwo
 \fi
}%
\providecommand \natexlab [1]{#1}%
\providecommand \enquote  [1]{``#1''}%
\providecommand \bibnamefont  [1]{#1}%
\providecommand \bibfnamefont [1]{#1}%
\providecommand \citenamefont [1]{#1}%
\providecommand \href@noop [0]{\@secondoftwo}%
\providecommand \href [0]{\begingroup \@sanitize@url \@href}%
\providecommand \@href[1]{\@@startlink{#1}\@@href}%
\providecommand \@@href[1]{\endgroup#1\@@endlink}%
\providecommand \@sanitize@url [0]{\catcode `\\12\catcode `\$12\catcode
  `\&12\catcode `\#12\catcode `\^12\catcode `\_12\catcode `\%12\relax}%
\providecommand \@@startlink[1]{}%
\providecommand \@@endlink[0]{}%
\providecommand \url  [0]{\begingroup\@sanitize@url \@url }%
\providecommand \@url [1]{\endgroup\@href {#1}{\urlprefix }}%
\providecommand \urlprefix  [0]{URL }%
\providecommand \Eprint [0]{\href }%
\providecommand \doibase [0]{http://dx.doi.org/}%
\providecommand \selectlanguage [0]{\@gobble}%
\providecommand \bibinfo  [0]{\@secondoftwo}%
\providecommand \bibfield  [0]{\@secondoftwo}%
\providecommand \translation [1]{[#1]}%
\providecommand \BibitemOpen [0]{}%
\providecommand \bibitemStop [0]{}%
\providecommand \bibitemNoStop [0]{.\EOS\space}%
\providecommand \EOS [0]{\spacefactor3000\relax}%
\providecommand \BibitemShut  [1]{\csname bibitem#1\endcsname}%
\let\auto@bib@innerbib\@empty
%</preamble>
\bibitem [{\citenamefont {Jensen}\ and\ \citenamefont
  {Shankar}(1985)}]{Jensen:1985aa}%
  \BibitemOpen
  \bibfield  {author} {\bibinfo {author} {\bibfnamefont {R.~V.}\ \bibnamefont
  {Jensen}}\ and\ \bibinfo {author} {\bibfnamefont {R.}~\bibnamefont
  {Shankar}},\ }\bibfield  {title} {\enquote {\bibinfo {title} {Statistical
  behavior in deterministic quantum systems with few degrees of freedom},}\
  }\href@noop {} {\bibfield  {journal} {\bibinfo  {journal} {Phys. Rev. Lett.}\
  }\textbf {\bibinfo {volume} {54}},\ \bibinfo {pages} {1879--1882} (\bibinfo
  {year} {1985})}\BibitemShut {NoStop}%
\bibitem [{\citenamefont {Deutsch}(1991)}]{Deutsch1}%
  \BibitemOpen
  \bibfield  {author} {\bibinfo {author} {\bibfnamefont {J.~M.}\ \bibnamefont
  {Deutsch}},\ }\bibfield  {title} {\enquote {\bibinfo {title} {Quantum
  statistical mechanics in a closed system},}\ }\href {\doibase
  10.1103/PhysRevA.43.2046} {\bibfield  {journal} {\bibinfo  {journal} {Phys.
  Rev. A}\ }\textbf {\bibinfo {volume} {43}},\ \bibinfo {pages} {2046--2049}
  (\bibinfo {year} {1991})}\BibitemShut {NoStop}%
\bibitem [{\citenamefont {Srednicki}(1994)}]{Sred1}%
  \BibitemOpen
  \bibfield  {author} {\bibinfo {author} {\bibfnamefont {Mark}\ \bibnamefont
  {Srednicki}},\ }\bibfield  {title} {\enquote {\bibinfo {title} {Chaos and
  quantum thermalization},}\ }\href {\doibase 10.1103/PhysRevE.50.888}
  {\bibfield  {journal} {\bibinfo  {journal} {Phys. Rev. E}\ }\textbf {\bibinfo
  {volume} {50}},\ \bibinfo {pages} {888--901} (\bibinfo {year}
  {1994})}\BibitemShut {NoStop}%
\bibitem [{\citenamefont {Rigol}\ \emph {et~al.}(2008)\citenamefont {Rigol},
  \citenamefont {Dunjko},\ and\ \citenamefont {Olshanii}}]{Rigol:2008bh}%
  \BibitemOpen
  \bibfield  {author} {\bibinfo {author} {\bibfnamefont {Marcos}\ \bibnamefont
  {Rigol}}, \bibinfo {author} {\bibfnamefont {Vanja}\ \bibnamefont {Dunjko}}, \
  and\ \bibinfo {author} {\bibfnamefont {Maxim}\ \bibnamefont {Olshanii}},\
  }\bibfield  {title} {\enquote {\bibinfo {title} {Thermalization and its
  mechanism for generic isolated quantum systems},}\ }\href@noop {} {\bibfield
  {journal} {\bibinfo  {journal} {Nature}\ }\textbf {\bibinfo {volume} {452}},\
  \bibinfo {pages} {854--858} (\bibinfo {year} {2008})}\BibitemShut {NoStop}%
\bibitem [{\citenamefont {D'Alessio}\ \emph {et~al.}(2016)\citenamefont
  {D'Alessio}, \citenamefont {Kafri}, \citenamefont {Polkovnikov},\ and\
  \citenamefont {Rigol}}]{DAlessio:2016aa}%
  \BibitemOpen
  \bibfield  {author} {\bibinfo {author} {\bibfnamefont {Luca}\ \bibnamefont
  {D'Alessio}}, \bibinfo {author} {\bibfnamefont {Yariv}\ \bibnamefont
  {Kafri}}, \bibinfo {author} {\bibfnamefont {Anatoli}\ \bibnamefont
  {Polkovnikov}}, \ and\ \bibinfo {author} {\bibfnamefont {Marcos}\
  \bibnamefont {Rigol}},\ }\bibfield  {title} {\enquote {\bibinfo {title} {From
  quantum chaos and eigenstate thermalization to statistical mechanics and
  thermodynamics},}\ }\href@noop {} {\bibfield  {journal} {\bibinfo  {journal}
  {Advances in Physics}\ }\textbf {\bibinfo {volume} {65}},\ \bibinfo {pages}
  {239--362} (\bibinfo {year} {2016})}\BibitemShut {NoStop}%
\bibitem [{\citenamefont {{Kim}}\ \emph {et~al.}(2014)\citenamefont {{Kim}},
  \citenamefont {{Ikeda}},\ and\ \citenamefont {{Huse}}}]{KimIkedaHuse}%
  \BibitemOpen
  \bibfield  {author} {\bibinfo {author} {\bibfnamefont {Hyungwon}\
  \bibnamefont {{Kim}}}, \bibinfo {author} {\bibfnamefont {Tatsuhiko~N.}\
  \bibnamefont {{Ikeda}}}, \ and\ \bibinfo {author} {\bibfnamefont {David~A.}\
  \bibnamefont {{Huse}}},\ }\bibfield  {title} {\enquote {\bibinfo {title}
  {{Testing whether all eigenstates obey the eigenstate thermalization
  hypothesis}},}\ }\href {\doibase 10.1103/PhysRevE.90.052105} {\bibfield
  {journal} {\bibinfo  {journal} {Phys. Rev. E}\ }\textbf {\bibinfo {volume}
  {90}},\ \bibinfo {eid} {052105} (\bibinfo {year} {2014})}\BibitemShut
  {NoStop}%
\bibitem [{\citenamefont {Basko}\ \emph {et~al.}(2006)\citenamefont {Basko},
  \citenamefont {Aleiner},\ and\ \citenamefont {Altshuler}}]{Basko2006}%
  \BibitemOpen
  \bibfield  {author} {\bibinfo {author} {\bibfnamefont {D.~M.}\ \bibnamefont
  {Basko}}, \bibinfo {author} {\bibfnamefont {I.~L.}\ \bibnamefont {Aleiner}},
  \ and\ \bibinfo {author} {\bibfnamefont {B.~L.}\ \bibnamefont {Altshuler}},\
  }\bibfield  {title} {\enquote {\bibinfo {title} {Metal-insulator transition
  in a weakly interacting many-electron system with localized single-particle
  states},}\ }\href {\doibase 10.1016/j.aop.2005.11.014} {\bibfield  {journal}
  {\bibinfo  {journal} {Ann. Phys. (Amsterdam)}\ }\textbf {\bibinfo {volume}
  {321}},\ \bibinfo {pages} {1126--1205} (\bibinfo {year} {2006})}\BibitemShut
  {NoStop}%
\bibitem [{\citenamefont {Nandkishore}\ and\ \citenamefont
  {Huse}(2015)}]{Nandkishore2015}%
  \BibitemOpen
  \bibfield  {author} {\bibinfo {author} {\bibfnamefont {R.}~\bibnamefont
  {Nandkishore}}\ and\ \bibinfo {author} {\bibfnamefont {D.~A.}\ \bibnamefont
  {Huse}},\ }\bibfield  {title} {\enquote {\bibinfo {title} {Many-body
  localization and thermalization in quantum statistical mechanics},}\ }\href
  {\doibase 10.1146/annurev-conmatphys-031214-014726} {\bibfield  {journal}
  {\bibinfo  {journal} {Annu. Rev. Condens. Matter Phys.}\ }\textbf {\bibinfo
  {volume} {6}},\ \bibinfo {pages} {15--38} (\bibinfo {year}
  {2015})}\BibitemShut {NoStop}%
\bibitem [{\citenamefont {Abanin}\ \emph {et~al.}(2019)\citenamefont {Abanin},
  \citenamefont {Altman}, \citenamefont {Bloch},\ and\ \citenamefont
  {Serbyn}}]{AbaninRMP2019}%
  \BibitemOpen
  \bibfield  {author} {\bibinfo {author} {\bibfnamefont {D.~A.}\ \bibnamefont
  {Abanin}}, \bibinfo {author} {\bibfnamefont {E.}~\bibnamefont {Altman}},
  \bibinfo {author} {\bibfnamefont {I.}~\bibnamefont {Bloch}}, \ and\ \bibinfo
  {author} {\bibfnamefont {M.}~\bibnamefont {Serbyn}},\ }\bibfield  {title}
  {\enquote {\bibinfo {title} {Colloquium: Many-body localization,
  thermalization, and entanglement},}\ }\href {\doibase
  10.1103/RevModPhys.91.021001} {\bibfield  {journal} {\bibinfo  {journal}
  {Rev. Mod. Phys.}\ }\textbf {\bibinfo {volume} {91}},\ \bibinfo {pages}
  {021001} (\bibinfo {year} {2019})}\BibitemShut {NoStop}%
\bibitem [{\citenamefont {Shiraishi}\ and\ \citenamefont
  {Mori}(2017)}]{ShiraishiMori}%
  \BibitemOpen
  \bibfield  {author} {\bibinfo {author} {\bibfnamefont {Naoto}\ \bibnamefont
  {Shiraishi}}\ and\ \bibinfo {author} {\bibfnamefont {Takashi}\ \bibnamefont
  {Mori}},\ }\bibfield  {title} {\enquote {\bibinfo {title} {Systematic
  construction of counterexamples to the eigenstate thermalization
  hypothesis},}\ }\href {\doibase 10.1103/PhysRevLett.119.030601} {\bibfield
  {journal} {\bibinfo  {journal} {Phys. Rev. Lett.}\ }\textbf {\bibinfo
  {volume} {119}},\ \bibinfo {pages} {030601} (\bibinfo {year}
  {2017})}\BibitemShut {NoStop}%
\bibitem [{\citenamefont {Moudgalya}\ \emph
  {et~al.}(2018{\natexlab{a}})\citenamefont {Moudgalya}, \citenamefont
  {Rachel}, \citenamefont {Bernevig},\ and\ \citenamefont
  {Regnault}}]{MoudDisc}%
  \BibitemOpen
  \bibfield  {author} {\bibinfo {author} {\bibfnamefont {Sanjay}\ \bibnamefont
  {Moudgalya}}, \bibinfo {author} {\bibfnamefont {Stephan}\ \bibnamefont
  {Rachel}}, \bibinfo {author} {\bibfnamefont {B.~Andrei}\ \bibnamefont
  {Bernevig}}, \ and\ \bibinfo {author} {\bibfnamefont {Nicolas}\ \bibnamefont
  {Regnault}},\ }\bibfield  {title} {\enquote {\bibinfo {title} {Exact excited
  states of nonintegrable models},}\ }\href {\doibase
  10.1103/PhysRevB.98.235155} {\bibfield  {journal} {\bibinfo  {journal} {Phys.
  Rev. B}\ }\textbf {\bibinfo {volume} {98}},\ \bibinfo {pages} {235155}
  (\bibinfo {year} {2018}{\natexlab{a}})}\BibitemShut {NoStop}%
\bibitem [{\citenamefont {Bernien}\ \emph {et~al.}(2017)\citenamefont
  {Bernien}, \citenamefont {Schwartz}, \citenamefont {Keesling}, \citenamefont
  {Levine}, \citenamefont {Omran}, \citenamefont {Pichler}, \citenamefont
  {Choi}, \citenamefont {Zibrov}, \citenamefont {Endres}, \citenamefont
  {Greiner}, \citenamefont {Vuleti{\'{c}}},\ and\ \citenamefont
  {Lukin}}]{Bernien2017}%
  \BibitemOpen
  \bibfield  {author} {\bibinfo {author} {\bibfnamefont {Hannes}\ \bibnamefont
  {Bernien}}, \bibinfo {author} {\bibfnamefont {Sylvain}\ \bibnamefont
  {Schwartz}}, \bibinfo {author} {\bibfnamefont {Alexander}\ \bibnamefont
  {Keesling}}, \bibinfo {author} {\bibfnamefont {Harry}\ \bibnamefont
  {Levine}}, \bibinfo {author} {\bibfnamefont {Ahmed}\ \bibnamefont {Omran}},
  \bibinfo {author} {\bibfnamefont {Hannes}\ \bibnamefont {Pichler}}, \bibinfo
  {author} {\bibfnamefont {Soonwon}\ \bibnamefont {Choi}}, \bibinfo {author}
  {\bibfnamefont {Alexander~S.}\ \bibnamefont {Zibrov}}, \bibinfo {author}
  {\bibfnamefont {Manuel}\ \bibnamefont {Endres}}, \bibinfo {author}
  {\bibfnamefont {Markus}\ \bibnamefont {Greiner}}, \bibinfo {author}
  {\bibfnamefont {Vladan}\ \bibnamefont {Vuleti{\'{c}}}}, \ and\ \bibinfo
  {author} {\bibfnamefont {Mikhail~D.}\ \bibnamefont {Lukin}},\ }\bibfield
  {title} {\enquote {\bibinfo {title} {Probing many-body dynamics on a 51-atom
  quantum simulator},}\ }\href {\doibase 10.1038/nature24622} {\bibfield
  {journal} {\bibinfo  {journal} {Nature}\ }\textbf {\bibinfo {volume} {551}},\
  \bibinfo {pages} {579--584} (\bibinfo {year} {2017})}\BibitemShut {NoStop}%
\bibitem [{\citenamefont {Turner}\ \emph
  {et~al.}(2018{\natexlab{a}})\citenamefont {Turner}, \citenamefont
  {Michailidis}, \citenamefont {Abanin}, \citenamefont {Serbyn},\ and\
  \citenamefont {Papi{\'{c}}}}]{TurnerWEB2018}%
  \BibitemOpen
  \bibfield  {author} {\bibinfo {author} {\bibfnamefont {C.~J.}\ \bibnamefont
  {Turner}}, \bibinfo {author} {\bibfnamefont {A.~A.}\ \bibnamefont
  {Michailidis}}, \bibinfo {author} {\bibfnamefont {D.~A.}\ \bibnamefont
  {Abanin}}, \bibinfo {author} {\bibfnamefont {M.}~\bibnamefont {Serbyn}}, \
  and\ \bibinfo {author} {\bibfnamefont {Z.}~\bibnamefont {Papi{\'{c}}}},\
  }\bibfield  {title} {\enquote {\bibinfo {title} {Weak ergodicity breaking
  from quantum many-body scars},}\ }\href {\doibase 10.1038/s41567-018-0137-5}
  {\bibfield  {journal} {\bibinfo  {journal} {Nature Physics}\ }\textbf
  {\bibinfo {volume} {14}},\ \bibinfo {pages} {745--749} (\bibinfo {year}
  {2018}{\natexlab{a}})}\BibitemShut {NoStop}%
\bibitem [{\citenamefont {Biroli}\ \emph {et~al.}(2010)\citenamefont {Biroli},
  \citenamefont {Kollath},\ and\ \citenamefont {L\"auchli}}]{LauchliWeakETH}%
  \BibitemOpen
  \bibfield  {author} {\bibinfo {author} {\bibfnamefont {Giulio}\ \bibnamefont
  {Biroli}}, \bibinfo {author} {\bibfnamefont {Corinna}\ \bibnamefont
  {Kollath}}, \ and\ \bibinfo {author} {\bibfnamefont {Andreas~M.}\
  \bibnamefont {L\"auchli}},\ }\bibfield  {title} {\enquote {\bibinfo {title}
  {Effect of rare fluctuations on the thermalization of isolated quantum
  systems},}\ }\href {\doibase 10.1103/PhysRevLett.105.250401} {\bibfield
  {journal} {\bibinfo  {journal} {Phys. Rev. Lett.}\ }\textbf {\bibinfo
  {volume} {105}},\ \bibinfo {pages} {250401} (\bibinfo {year}
  {2010})}\BibitemShut {NoStop}%
\bibitem [{\citenamefont {{Kao}}\ \emph {et~al.}(2020)\citenamefont {{Kao}},
  \citenamefont {{Li}}, \citenamefont {{Lin}}, \citenamefont
  {{Gopalakrishnan}},\ and\ \citenamefont {{Lev}}}]{Lev2020}%
  \BibitemOpen
  \bibfield  {author} {\bibinfo {author} {\bibfnamefont {Wil}\ \bibnamefont
  {{Kao}}}, \bibinfo {author} {\bibfnamefont {Kuan-Yu}\ \bibnamefont {{Li}}},
  \bibinfo {author} {\bibfnamefont {Kuan-Yu}\ \bibnamefont {{Lin}}}, \bibinfo
  {author} {\bibfnamefont {Sarang}\ \bibnamefont {{Gopalakrishnan}}}, \ and\
  \bibinfo {author} {\bibfnamefont {Benjamin~L.}\ \bibnamefont {{Lev}}},\
  }\bibfield  {title} {\enquote {\bibinfo {title} {{Creating quantum many-body
  scars through topological pumping of a 1D dipolar gas}},}\ }\href@noop {}
  {\bibfield  {journal} {\bibinfo  {journal} {arXiv e-prints}\ ,\ \bibinfo
  {eid} {arXiv:2002.10475}} (\bibinfo {year} {2020})}\BibitemShut {NoStop}%
\bibitem [{\citenamefont {Turner}\ \emph
  {et~al.}(2018{\natexlab{b}})\citenamefont {Turner}, \citenamefont
  {Michailidis}, \citenamefont {Abanin}, \citenamefont {Serbyn},\ and\
  \citenamefont {Papi\ifmmode~\acute{c}\else \'{c}\fi{}}}]{TurnerRyd2018}%
  \BibitemOpen
  \bibfield  {author} {\bibinfo {author} {\bibfnamefont {C.~J.}\ \bibnamefont
  {Turner}}, \bibinfo {author} {\bibfnamefont {A.~A.}\ \bibnamefont
  {Michailidis}}, \bibinfo {author} {\bibfnamefont {D.~A.}\ \bibnamefont
  {Abanin}}, \bibinfo {author} {\bibfnamefont {M.}~\bibnamefont {Serbyn}}, \
  and\ \bibinfo {author} {\bibfnamefont {Z.}~\bibnamefont
  {Papi\ifmmode~\acute{c}\else \'{c}\fi{}}},\ }\bibfield  {title} {\enquote
  {\bibinfo {title} {Quantum scarred eigenstates in a rydberg atom chain:
  Entanglement, breakdown of thermalization, and stability to perturbations},}\
  }\href {\doibase 10.1103/PhysRevB.98.155134} {\bibfield  {journal} {\bibinfo
  {journal} {Phys. Rev. B}\ }\textbf {\bibinfo {volume} {98}},\ \bibinfo
  {pages} {155134} (\bibinfo {year} {2018}{\natexlab{b}})}\BibitemShut
  {NoStop}%
\bibitem [{\citenamefont {Khemani}\ \emph {et~al.}(2019)\citenamefont
  {Khemani}, \citenamefont {Laumann},\ and\ \citenamefont
  {Chandran}}]{SigInt2019}%
  \BibitemOpen
  \bibfield  {author} {\bibinfo {author} {\bibfnamefont {Vedika}\ \bibnamefont
  {Khemani}}, \bibinfo {author} {\bibfnamefont {Chris~R.}\ \bibnamefont
  {Laumann}}, \ and\ \bibinfo {author} {\bibfnamefont {Anushya}\ \bibnamefont
  {Chandran}},\ }\bibfield  {title} {\enquote {\bibinfo {title} {Signatures of
  integrability in the dynamics of rydberg-blockaded chains},}\ }\href
  {\doibase 10.1103/PhysRevB.99.161101} {\bibfield  {journal} {\bibinfo
  {journal} {Phys. Rev. B}\ }\textbf {\bibinfo {volume} {99}},\ \bibinfo
  {pages} {161101} (\bibinfo {year} {2019})}\BibitemShut {NoStop}%
\bibitem [{\citenamefont {Choi}\ \emph {et~al.}(2019)\citenamefont {Choi},
  \citenamefont {Turner}, \citenamefont {Pichler}, \citenamefont {Ho},
  \citenamefont {Michailidis}, \citenamefont {Papi\ifmmode~\acute{c}\else
  \'{c}\fi{}}, \citenamefont {Serbyn}, \citenamefont {Lukin},\ and\
  \citenamefont {Abanin}}]{AbaninPerfectScars}%
  \BibitemOpen
  \bibfield  {author} {\bibinfo {author} {\bibfnamefont {Soonwon}\ \bibnamefont
  {Choi}}, \bibinfo {author} {\bibfnamefont {Christopher~J.}\ \bibnamefont
  {Turner}}, \bibinfo {author} {\bibfnamefont {Hannes}\ \bibnamefont
  {Pichler}}, \bibinfo {author} {\bibfnamefont {Wen~Wei}\ \bibnamefont {Ho}},
  \bibinfo {author} {\bibfnamefont {Alexios~A.}\ \bibnamefont {Michailidis}},
  \bibinfo {author} {\bibfnamefont {Zlatko}\ \bibnamefont
  {Papi\ifmmode~\acute{c}\else \'{c}\fi{}}}, \bibinfo {author} {\bibfnamefont
  {Maksym}\ \bibnamefont {Serbyn}}, \bibinfo {author} {\bibfnamefont
  {Mikhail~D.}\ \bibnamefont {Lukin}}, \ and\ \bibinfo {author} {\bibfnamefont
  {Dmitry~A.}\ \bibnamefont {Abanin}},\ }\bibfield  {title} {\enquote {\bibinfo
  {title} {Emergent su(2) dynamics and perfect quantum many-body scars},}\
  }\href {\doibase 10.1103/PhysRevLett.122.220603} {\bibfield  {journal}
  {\bibinfo  {journal} {Phys. Rev. Lett.}\ }\textbf {\bibinfo {volume} {122}},\
  \bibinfo {pages} {220603} (\bibinfo {year} {2019})}\BibitemShut {NoStop}%
\bibitem [{\citenamefont {Ho}\ \emph {et~al.}(2019)\citenamefont {Ho},
  \citenamefont {Choi}, \citenamefont {Pichler},\ and\ \citenamefont
  {Lukin}}]{Ho2019}%
  \BibitemOpen
  \bibfield  {author} {\bibinfo {author} {\bibfnamefont {Wen~Wei}\ \bibnamefont
  {Ho}}, \bibinfo {author} {\bibfnamefont {Soonwon}\ \bibnamefont {Choi}},
  \bibinfo {author} {\bibfnamefont {Hannes}\ \bibnamefont {Pichler}}, \ and\
  \bibinfo {author} {\bibfnamefont {Mikhail~D.}\ \bibnamefont {Lukin}},\
  }\bibfield  {title} {\enquote {\bibinfo {title} {Periodic orbits,
  entanglement, and quantum many-body scars in constrained models: Matrix
  product state approach},}\ }\href {\doibase 10.1103/PhysRevLett.122.040603}
  {\bibfield  {journal} {\bibinfo  {journal} {Phys. Rev. Lett.}\ }\textbf
  {\bibinfo {volume} {122}},\ \bibinfo {pages} {040603} (\bibinfo {year}
  {2019})}\BibitemShut {NoStop}%
\bibitem [{\citenamefont {Lin}\ and\ \citenamefont {Motrunich}(2019)}]{LM2019}%
  \BibitemOpen
  \bibfield  {author} {\bibinfo {author} {\bibfnamefont {Cheng-Ju}\
  \bibnamefont {Lin}}\ and\ \bibinfo {author} {\bibfnamefont {Olexei~I.}\
  \bibnamefont {Motrunich}},\ }\bibfield  {title} {\enquote {\bibinfo {title}
  {Exact quantum many-body scar states in the rydberg-blockaded atom chain},}\
  }\href {\doibase 10.1103/PhysRevLett.122.173401} {\bibfield  {journal}
  {\bibinfo  {journal} {Phys. Rev. Lett.}\ }\textbf {\bibinfo {volume} {122}},\
  \bibinfo {pages} {173401} (\bibinfo {year} {2019})}\BibitemShut {NoStop}%
\bibitem [{\citenamefont {Iadecola}\ and\ \citenamefont
  {Schecter}(2020)}]{ISDWC}%
  \BibitemOpen
  \bibfield  {author} {\bibinfo {author} {\bibfnamefont {Thomas}\ \bibnamefont
  {Iadecola}}\ and\ \bibinfo {author} {\bibfnamefont {Michael}\ \bibnamefont
  {Schecter}},\ }\bibfield  {title} {\enquote {\bibinfo {title} {Quantum
  many-body scar states with emergent kinetic constraints and
  finite-entanglement revivals},}\ }\href {\doibase
  10.1103/PhysRevB.101.024306} {\bibfield  {journal} {\bibinfo  {journal}
  {Phys. Rev. B}\ }\textbf {\bibinfo {volume} {101}},\ \bibinfo {pages}
  {024306} (\bibinfo {year} {2020})}\BibitemShut {NoStop}%
\bibitem [{\citenamefont {Iadecola}\ \emph {et~al.}(2019)\citenamefont
  {Iadecola}, \citenamefont {Schecter},\ and\ \citenamefont
  {Xu}}]{IadecolaMagnon2019}%
  \BibitemOpen
  \bibfield  {author} {\bibinfo {author} {\bibfnamefont {Thomas}\ \bibnamefont
  {Iadecola}}, \bibinfo {author} {\bibfnamefont {Michael}\ \bibnamefont
  {Schecter}}, \ and\ \bibinfo {author} {\bibfnamefont {Shenglong}\
  \bibnamefont {Xu}},\ }\bibfield  {title} {\enquote {\bibinfo {title} {Quantum
  many-body scars from magnon condensation},}\ }\href {\doibase
  10.1103/PhysRevB.100.184312} {\bibfield  {journal} {\bibinfo  {journal}
  {Phys. Rev. B}\ }\textbf {\bibinfo {volume} {100}},\ \bibinfo {pages}
  {184312} (\bibinfo {year} {2019})}\BibitemShut {NoStop}%
\bibitem [{\citenamefont {James}\ \emph {et~al.}(2019)\citenamefont {James},
  \citenamefont {Konik},\ and\ \citenamefont {Robinson}}]{James2019}%
  \BibitemOpen
  \bibfield  {author} {\bibinfo {author} {\bibfnamefont {Andrew J.~A.}\
  \bibnamefont {James}}, \bibinfo {author} {\bibfnamefont {Robert~M.}\
  \bibnamefont {Konik}}, \ and\ \bibinfo {author} {\bibfnamefont {Neil~J.}\
  \bibnamefont {Robinson}},\ }\bibfield  {title} {\enquote {\bibinfo {title}
  {Nonthermal states arising from confinement in one and two dimensions},}\
  }\href {\doibase 10.1103/PhysRevLett.122.130603} {\bibfield  {journal}
  {\bibinfo  {journal} {Phys. Rev. Lett.}\ }\textbf {\bibinfo {volume} {122}},\
  \bibinfo {pages} {130603} (\bibinfo {year} {2019})}\BibitemShut {NoStop}%
\bibitem [{\citenamefont {Michailidis}\ \emph {et~al.}(2020)\citenamefont
  {Michailidis}, \citenamefont {Turner}, \citenamefont
  {Papi\ifmmode~\acute{c}\else \'{c}\fi{}}, \citenamefont {Abanin},\ and\
  \citenamefont {Serbyn}}]{PhysRevX.10.011055}%
  \BibitemOpen
  \bibfield  {author} {\bibinfo {author} {\bibfnamefont {A.~A.}\ \bibnamefont
  {Michailidis}}, \bibinfo {author} {\bibfnamefont {C.~J.}\ \bibnamefont
  {Turner}}, \bibinfo {author} {\bibfnamefont {Z.}~\bibnamefont
  {Papi\ifmmode~\acute{c}\else \'{c}\fi{}}}, \bibinfo {author} {\bibfnamefont
  {D.~A.}\ \bibnamefont {Abanin}}, \ and\ \bibinfo {author} {\bibfnamefont
  {M.}~\bibnamefont {Serbyn}},\ }\bibfield  {title} {\enquote {\bibinfo {title}
  {Slow quantum thermalization and many-body revivals from mixed phase
  space},}\ }\href {\doibase 10.1103/PhysRevX.10.011055} {\bibfield  {journal}
  {\bibinfo  {journal} {Phys. Rev. X}\ }\textbf {\bibinfo {volume} {10}},\
  \bibinfo {pages} {011055} (\bibinfo {year} {2020})}\BibitemShut {NoStop}%
\bibitem [{\citenamefont {Surace}\ \emph {et~al.}(2020)\citenamefont {Surace},
  \citenamefont {Mazza}, \citenamefont {Giudici}, \citenamefont {Lerose},
  \citenamefont {Gambassi},\ and\ \citenamefont {Dalmonte}}]{Surace2020}%
  \BibitemOpen
  \bibfield  {author} {\bibinfo {author} {\bibfnamefont {Federica~M.}\
  \bibnamefont {Surace}}, \bibinfo {author} {\bibfnamefont {Paolo~P.}\
  \bibnamefont {Mazza}}, \bibinfo {author} {\bibfnamefont {Giuliano}\
  \bibnamefont {Giudici}}, \bibinfo {author} {\bibfnamefont {Alessio}\
  \bibnamefont {Lerose}}, \bibinfo {author} {\bibfnamefont {Andrea}\
  \bibnamefont {Gambassi}}, \ and\ \bibinfo {author} {\bibfnamefont {Marcello}\
  \bibnamefont {Dalmonte}},\ }\bibfield  {title} {\enquote {\bibinfo {title}
  {Lattice gauge theories and string dynamics in rydberg atom quantum
  simulators},}\ }\href {\doibase 10.1103/PhysRevX.10.021041} {\bibfield
  {journal} {\bibinfo  {journal} {Phys. Rev. X}\ }\textbf {\bibinfo {volume}
  {10}},\ \bibinfo {pages} {021041} (\bibinfo {year} {2020})}\BibitemShut
  {NoStop}%
\bibitem [{\citenamefont {Bull}\ \emph {et~al.}(2020)\citenamefont {Bull},
  \citenamefont {Desaules},\ and\ \citenamefont {Papi\ifmmode~\acute{c}\else
  \'{c}\fi{}}}]{Papic2020}%
  \BibitemOpen
  \bibfield  {author} {\bibinfo {author} {\bibfnamefont {Kieran}\ \bibnamefont
  {Bull}}, \bibinfo {author} {\bibfnamefont {Jean-Yves}\ \bibnamefont
  {Desaules}}, \ and\ \bibinfo {author} {\bibfnamefont {Zlatko}\ \bibnamefont
  {Papi\ifmmode~\acute{c}\else \'{c}\fi{}}},\ }\bibfield  {title} {\enquote
  {\bibinfo {title} {Quantum scars as embeddings of weakly broken lie algebra
  representations},}\ }\href {\doibase 10.1103/PhysRevB.101.165139} {\bibfield
  {journal} {\bibinfo  {journal} {Phys. Rev. B}\ }\textbf {\bibinfo {volume}
  {101}},\ \bibinfo {pages} {165139} (\bibinfo {year} {2020})}\BibitemShut
  {NoStop}%
\bibitem [{\citenamefont {Moudgalya}\ \emph
  {et~al.}(2018{\natexlab{b}})\citenamefont {Moudgalya}, \citenamefont
  {Regnault},\ and\ \citenamefont {Bernevig}}]{MoudEntang}%
  \BibitemOpen
  \bibfield  {author} {\bibinfo {author} {\bibfnamefont {Sanjay}\ \bibnamefont
  {Moudgalya}}, \bibinfo {author} {\bibfnamefont {Nicolas}\ \bibnamefont
  {Regnault}}, \ and\ \bibinfo {author} {\bibfnamefont {B.~Andrei}\
  \bibnamefont {Bernevig}},\ }\bibfield  {title} {\enquote {\bibinfo {title}
  {Entanglement of exact excited states of affleck-kennedy-lieb-tasaki models:
  Exact results, many-body scars, and violation of the strong eigenstate
  thermalization hypothesis},}\ }\href {\doibase 10.1103/PhysRevB.98.235156}
  {\bibfield  {journal} {\bibinfo  {journal} {Phys. Rev. B}\ }\textbf {\bibinfo
  {volume} {98}},\ \bibinfo {pages} {235156} (\bibinfo {year}
  {2018}{\natexlab{b}})}\BibitemShut {NoStop}%
\bibitem [{\citenamefont {Schecter}\ and\ \citenamefont
  {Iadecola}(2019)}]{ISXY}%
  \BibitemOpen
  \bibfield  {author} {\bibinfo {author} {\bibfnamefont {Michael}\ \bibnamefont
  {Schecter}}\ and\ \bibinfo {author} {\bibfnamefont {Thomas}\ \bibnamefont
  {Iadecola}},\ }\bibfield  {title} {\enquote {\bibinfo {title} {Weak
  ergodicity breaking and quantum many-body scars in spin-1 $xy$ magnets},}\
  }\href {\doibase 10.1103/PhysRevLett.123.147201} {\bibfield  {journal}
  {\bibinfo  {journal} {Phys. Rev. Lett.}\ }\textbf {\bibinfo {volume} {123}},\
  \bibinfo {pages} {147201} (\bibinfo {year} {2019})}\BibitemShut {NoStop}%
\bibitem [{\citenamefont {Chattopadhyay}\ \emph {et~al.}(2020)\citenamefont
  {Chattopadhyay}, \citenamefont {Pichler}, \citenamefont {Lukin},\ and\
  \citenamefont {Ho}}]{Chattopadhyay}%
  \BibitemOpen
  \bibfield  {author} {\bibinfo {author} {\bibfnamefont {Sambuddha}\
  \bibnamefont {Chattopadhyay}}, \bibinfo {author} {\bibfnamefont {Hannes}\
  \bibnamefont {Pichler}}, \bibinfo {author} {\bibfnamefont {Mikhail~D.}\
  \bibnamefont {Lukin}}, \ and\ \bibinfo {author} {\bibfnamefont {Wen~Wei}\
  \bibnamefont {Ho}},\ }\bibfield  {title} {\enquote {\bibinfo {title} {Quantum
  many-body scars from virtual entangled pairs},}\ }\href {\doibase
  10.1103/PhysRevB.101.174308} {\bibfield  {journal} {\bibinfo  {journal}
  {Phys. Rev. B}\ }\textbf {\bibinfo {volume} {101}},\ \bibinfo {pages}
  {174308} (\bibinfo {year} {2020})}\BibitemShut {NoStop}%
\bibitem [{\citenamefont {Sala}\ \emph {et~al.}(2020)\citenamefont {Sala},
  \citenamefont {Rakovszky}, \citenamefont {Verresen}, \citenamefont {Knap},\
  and\ \citenamefont {Pollmann}}]{Sala2020}%
  \BibitemOpen
  \bibfield  {author} {\bibinfo {author} {\bibfnamefont {Pablo}\ \bibnamefont
  {Sala}}, \bibinfo {author} {\bibfnamefont {Tibor}\ \bibnamefont {Rakovszky}},
  \bibinfo {author} {\bibfnamefont {Ruben}\ \bibnamefont {Verresen}}, \bibinfo
  {author} {\bibfnamefont {Michael}\ \bibnamefont {Knap}}, \ and\ \bibinfo
  {author} {\bibfnamefont {Frank}\ \bibnamefont {Pollmann}},\ }\bibfield
  {title} {\enquote {\bibinfo {title} {Ergodicity breaking arising from hilbert
  space fragmentation in dipole-conserving hamiltonians},}\ }\href {\doibase
  10.1103/PhysRevX.10.011047} {\bibfield  {journal} {\bibinfo  {journal} {Phys.
  Rev. X}\ }\textbf {\bibinfo {volume} {10}},\ \bibinfo {pages} {011047}
  (\bibinfo {year} {2020})}\BibitemShut {NoStop}%
\bibitem [{\citenamefont {Khemani}\ \emph {et~al.}(2020)\citenamefont
  {Khemani}, \citenamefont {Hermele},\ and\ \citenamefont
  {Nandkishore}}]{Shattering2020}%
  \BibitemOpen
  \bibfield  {author} {\bibinfo {author} {\bibfnamefont {Vedika}\ \bibnamefont
  {Khemani}}, \bibinfo {author} {\bibfnamefont {Michael}\ \bibnamefont
  {Hermele}}, \ and\ \bibinfo {author} {\bibfnamefont {Rahul}\ \bibnamefont
  {Nandkishore}},\ }\bibfield  {title} {\enquote {\bibinfo {title}
  {Localization from hilbert space shattering: From theory to physical
  realizations},}\ }\href {\doibase 10.1103/PhysRevB.101.174204} {\bibfield
  {journal} {\bibinfo  {journal} {Phys. Rev. B}\ }\textbf {\bibinfo {volume}
  {101}},\ \bibinfo {pages} {174204} (\bibinfo {year} {2020})}\BibitemShut
  {NoStop}%
\bibitem [{\citenamefont {Mark}\ \emph {et~al.}(2020)\citenamefont {Mark},
  \citenamefont {Lin},\ and\ \citenamefont {Motrunich}}]{MLM}%
  \BibitemOpen
  \bibfield  {author} {\bibinfo {author} {\bibfnamefont {Daniel~K.}\
  \bibnamefont {Mark}}, \bibinfo {author} {\bibfnamefont {Cheng-Ju}\
  \bibnamefont {Lin}}, \ and\ \bibinfo {author} {\bibfnamefont {Olexei~I.}\
  \bibnamefont {Motrunich}},\ }\bibfield  {title} {\enquote {\bibinfo {title}
  {Unified structure for exact towers of scar states in the
  affleck-kennedy-lieb-tasaki and other models},}\ }\href {\doibase
  10.1103/PhysRevB.101.195131} {\bibfield  {journal} {\bibinfo  {journal}
  {Phys. Rev. B}\ }\textbf {\bibinfo {volume} {101}},\ \bibinfo {pages}
  {195131} (\bibinfo {year} {2020})}\BibitemShut {NoStop}%
\bibitem [{\citenamefont {{Moudgalya}}\ \emph
  {et~al.}(2020{\natexlab{a}})\citenamefont {{Moudgalya}}, \citenamefont
  {{O'Brien}}, \citenamefont {{Bernevig}}, \citenamefont {{Fendley}},\ and\
  \citenamefont {{Regnault}}}]{MoudMPS}%
  \BibitemOpen
  \bibfield  {author} {\bibinfo {author} {\bibfnamefont {Sanjay}\ \bibnamefont
  {{Moudgalya}}}, \bibinfo {author} {\bibfnamefont {Edward}\ \bibnamefont
  {{O'Brien}}}, \bibinfo {author} {\bibfnamefont {B.~Andrei}\ \bibnamefont
  {{Bernevig}}}, \bibinfo {author} {\bibfnamefont {Paul}\ \bibnamefont
  {{Fendley}}}, \ and\ \bibinfo {author} {\bibfnamefont {Nicolas}\ \bibnamefont
  {{Regnault}}},\ }\bibfield  {title} {\enquote {\bibinfo {title} {{Large
  Classes of Quantum Scarred Hamiltonians from Matrix Product States}},}\
  }\href@noop {} {\bibfield  {journal} {\bibinfo  {journal} {arXiv e-prints}\
  ,\ \bibinfo {eid} {arXiv:2002.11725}} (\bibinfo {year}
  {2020}{\natexlab{a}})}\BibitemShut {NoStop}%
\bibitem [{\citenamefont {{Moudgalya}}\ \emph
  {et~al.}(2020{\natexlab{b}})\citenamefont {{Moudgalya}}, \citenamefont
  {{Regnault}},\ and\ \citenamefont {{Bernevig}}}]{MoudHubb}%
  \BibitemOpen
  \bibfield  {author} {\bibinfo {author} {\bibfnamefont {Sanjay}\ \bibnamefont
  {{Moudgalya}}}, \bibinfo {author} {\bibfnamefont {Nicolas}\ \bibnamefont
  {{Regnault}}}, \ and\ \bibinfo {author} {\bibfnamefont {B.~Andrei}\
  \bibnamefont {{Bernevig}}},\ }\bibfield  {title} {\enquote {\bibinfo {title}
  {{Eta-Pairing in Hubbard Models: From Spectrum Generating Algebras to Quantum
  Many-Body Scars}},}\ }\href@noop {} {\bibfield  {journal} {\bibinfo
  {journal} {arXiv e-prints}\ ,\ \bibinfo {eid} {arXiv:2004.13727}} (\bibinfo
  {year} {2020}{\natexlab{b}})}\BibitemShut {NoStop}%
\bibitem [{\citenamefont {{Mark}}\ and\ \citenamefont
  {{Motrunich}}(2020)}]{MM}%
  \BibitemOpen
  \bibfield  {author} {\bibinfo {author} {\bibfnamefont {Daniel~K.}\
  \bibnamefont {{Mark}}}\ and\ \bibinfo {author} {\bibfnamefont {Olexei~I.}\
  \bibnamefont {{Motrunich}}},\ }\bibfield  {title} {\enquote {\bibinfo {title}
  {{Eta-pairing states as true scars in an extended Hubbard Model}},}\
  }\href@noop {} {\bibfield  {journal} {\bibinfo  {journal} {arXiv e-prints}\
  ,\ \bibinfo {eid} {arXiv:2004.13800}} (\bibinfo {year} {2020})}\BibitemShut
  {NoStop}%
\bibitem [{\citenamefont {Ok}\ \emph {et~al.}(2019)\citenamefont {Ok},
  \citenamefont {Choo}, \citenamefont {Mudry}, \citenamefont {Castelnovo},
  \citenamefont {Chamon},\ and\ \citenamefont {Neupert}}]{Ok2019}%
  \BibitemOpen
  \bibfield  {author} {\bibinfo {author} {\bibfnamefont {Seulgi}\ \bibnamefont
  {Ok}}, \bibinfo {author} {\bibfnamefont {Kenny}\ \bibnamefont {Choo}},
  \bibinfo {author} {\bibfnamefont {Christopher}\ \bibnamefont {Mudry}},
  \bibinfo {author} {\bibfnamefont {Claudio}\ \bibnamefont {Castelnovo}},
  \bibinfo {author} {\bibfnamefont {Claudio}\ \bibnamefont {Chamon}}, \ and\
  \bibinfo {author} {\bibfnamefont {Titus}\ \bibnamefont {Neupert}},\
  }\bibfield  {title} {\enquote {\bibinfo {title} {Topological many-body scar
  states in dimensions one, two, and three},}\ }\href {\doibase
  10.1103/PhysRevResearch.1.033144} {\bibfield  {journal} {\bibinfo  {journal}
  {Phys. Rev. Research}\ }\textbf {\bibinfo {volume} {1}},\ \bibinfo {pages}
  {033144} (\bibinfo {year} {2019})}\BibitemShut {NoStop}%
\bibitem [{\citenamefont {Pancotti}\ \emph {et~al.}(2020)\citenamefont
  {Pancotti}, \citenamefont {Giudice}, \citenamefont {Cirac}, \citenamefont
  {Garrahan},\ and\ \citenamefont {Ba\~nuls}}]{Pancotti2020}%
  \BibitemOpen
  \bibfield  {author} {\bibinfo {author} {\bibfnamefont {Nicola}\ \bibnamefont
  {Pancotti}}, \bibinfo {author} {\bibfnamefont {Giacomo}\ \bibnamefont
  {Giudice}}, \bibinfo {author} {\bibfnamefont {J.~Ignacio}\ \bibnamefont
  {Cirac}}, \bibinfo {author} {\bibfnamefont {Juan~P.}\ \bibnamefont
  {Garrahan}}, \ and\ \bibinfo {author} {\bibfnamefont {Mari~Carmen}\
  \bibnamefont {Ba\~nuls}},\ }\bibfield  {title} {\enquote {\bibinfo {title}
  {Quantum east model: Localization, nonthermal eigenstates, and slow
  dynamics},}\ }\href {\doibase 10.1103/PhysRevX.10.021051} {\bibfield
  {journal} {\bibinfo  {journal} {Phys. Rev. X}\ }\textbf {\bibinfo {volume}
  {10}},\ \bibinfo {pages} {021051} (\bibinfo {year} {2020})}\BibitemShut
  {NoStop}%
\bibitem [{\citenamefont {Serbyn}\ \emph {et~al.}(2013)\citenamefont {Serbyn},
  \citenamefont {Papi\ifmmode~\acute{c}\else \'{c}\fi{}},\ and\ \citenamefont
  {Abanin}}]{Abanin2013CL}%
  \BibitemOpen
  \bibfield  {author} {\bibinfo {author} {\bibfnamefont {Maksym}\ \bibnamefont
  {Serbyn}}, \bibinfo {author} {\bibfnamefont {Z.}~\bibnamefont
  {Papi\ifmmode~\acute{c}\else \'{c}\fi{}}}, \ and\ \bibinfo {author}
  {\bibfnamefont {Dmitry~A.}\ \bibnamefont {Abanin}},\ }\bibfield  {title}
  {\enquote {\bibinfo {title} {Local conservation laws and the structure of the
  many-body localized states},}\ }\href {\doibase
  10.1103/PhysRevLett.111.127201} {\bibfield  {journal} {\bibinfo  {journal}
  {Phys. Rev. Lett.}\ }\textbf {\bibinfo {volume} {111}},\ \bibinfo {pages}
  {127201} (\bibinfo {year} {2013})}\BibitemShut {NoStop}%
\bibitem [{\citenamefont {Huse}\ \emph {et~al.}(2014)\citenamefont {Huse},
  \citenamefont {Nandkishore},\ and\ \citenamefont {Oganesyan}}]{Huse:2014ac}%
  \BibitemOpen
  \bibfield  {author} {\bibinfo {author} {\bibfnamefont {David~A.}\
  \bibnamefont {Huse}}, \bibinfo {author} {\bibfnamefont {Rahul}\ \bibnamefont
  {Nandkishore}}, \ and\ \bibinfo {author} {\bibfnamefont {Vadim}\ \bibnamefont
  {Oganesyan}},\ }\bibfield  {title} {\enquote {\bibinfo {title} {Phenomenology
  of fully many-body-localized systems},}\ }\href@noop {} {\bibfield  {journal}
  {\bibinfo  {journal} {Physical Review B}\ }\textbf {\bibinfo {volume} {90}},\
  \bibinfo {pages} {174202} (\bibinfo {year} {2014})}\BibitemShut {NoStop}%
\bibitem [{\citenamefont {Shibata}\ \emph {et~al.}(2020)\citenamefont
  {Shibata}, \citenamefont {Yoshioka},\ and\ \citenamefont
  {Katsura}}]{Shibata}%
  \BibitemOpen
  \bibfield  {author} {\bibinfo {author} {\bibfnamefont {Naoyuki}\ \bibnamefont
  {Shibata}}, \bibinfo {author} {\bibfnamefont {Nobuyuki}\ \bibnamefont
  {Yoshioka}}, \ and\ \bibinfo {author} {\bibfnamefont {Hosho}\ \bibnamefont
  {Katsura}},\ }\bibfield  {title} {\enquote {\bibinfo {title} {Onsager's scars
  in disordered spin chains},}\ }\href {\doibase
  10.1103/PhysRevLett.124.180604} {\bibfield  {journal} {\bibinfo  {journal}
  {Phys. Rev. Lett.}\ }\textbf {\bibinfo {volume} {124}},\ \bibinfo {pages}
  {180604} (\bibinfo {year} {2020})}\BibitemShut {NoStop}%
\bibitem [{\citenamefont {{Medenjak}}\ \emph {et~al.}(2020)\citenamefont
  {{Medenjak}}, \citenamefont {{Bu{\v{c}}a}},\ and\ \citenamefont
  {{Jaksch}}}]{Buca2020}%
  \BibitemOpen
  \bibfield  {author} {\bibinfo {author} {\bibfnamefont {Marko}\ \bibnamefont
  {{Medenjak}}}, \bibinfo {author} {\bibfnamefont {Berislav}\ \bibnamefont
  {{Bu{\v{c}}a}}}, \ and\ \bibinfo {author} {\bibfnamefont {Dieter}\
  \bibnamefont {{Jaksch}}},\ }\bibfield  {title} {\enquote {\bibinfo {title}
  {{Isolated Heisenberg magnet as a quantum time crystal}},}\ }\href {\doibase
  10.1103/PhysRevB.102.041117} {\bibfield  {journal} {\bibinfo  {journal}
  {Phys. Rev. B}\ }\textbf {\bibinfo {volume} {102}},\ \bibinfo {eid} {041117}
  (\bibinfo {year} {2020})}\BibitemShut {NoStop}%
\bibitem [{\citenamefont {{Qi}}\ and\ \citenamefont
  {{Ranard}}(2017)}]{XLQ2017}%
  \BibitemOpen
  \bibfield  {author} {\bibinfo {author} {\bibfnamefont {Xiao-Liang}\
  \bibnamefont {{Qi}}}\ and\ \bibinfo {author} {\bibfnamefont {Daniel}\
  \bibnamefont {{Ranard}}},\ }\bibfield  {title} {\enquote {\bibinfo {title}
  {{Determining a local Hamiltonian from a single eigenstate}},}\ }\href@noop
  {} {\bibfield  {journal} {\bibinfo  {journal} {arXiv e-prints}\ ,\ \bibinfo
  {eid} {arXiv:1712.01850}} (\bibinfo {year} {2017})}\BibitemShut {NoStop}%
\bibitem [{\citenamefont {Yang}(1989)}]{Yangeta}%
  \BibitemOpen
  \bibfield  {author} {\bibinfo {author} {\bibfnamefont {Chen~Ning}\
  \bibnamefont {Yang}},\ }\bibfield  {title} {\enquote {\bibinfo {title}
  {\ensuremath{\eta} pairing and off-diagonal long-range order in a hubbard
  model},}\ }\href {\doibase 10.1103/PhysRevLett.63.2144} {\bibfield  {journal}
  {\bibinfo  {journal} {Phys. Rev. Lett.}\ }\textbf {\bibinfo {volume} {63}},\
  \bibinfo {pages} {2144--2147} (\bibinfo {year} {1989})}\BibitemShut {NoStop}%
\bibitem [{\citenamefont {Vafek}\ \emph {et~al.}(2017)\citenamefont {Vafek},
  \citenamefont {Regnault},\ and\ \citenamefont {Bernevig}}]{Vafek2017}%
  \BibitemOpen
  \bibfield  {author} {\bibinfo {author} {\bibfnamefont {Oskar}\ \bibnamefont
  {Vafek}}, \bibinfo {author} {\bibfnamefont {Nicolas}\ \bibnamefont
  {Regnault}}, \ and\ \bibinfo {author} {\bibfnamefont {B.~Andrei}\
  \bibnamefont {Bernevig}},\ }\bibfield  {title} {\enquote {\bibinfo {title}
  {{Entanglement of Exact Excited Eigenstates of the Hubbard Model in Arbitrary
  Dimension}},}\ }\href {\doibase 10.21468/SciPostPhys.3.6.043} {\bibfield
  {journal} {\bibinfo  {journal} {SciPost Phys.}\ }\textbf {\bibinfo {volume}
  {3}},\ \bibinfo {pages} {043} (\bibinfo {year} {2017})}\BibitemShut {NoStop}%
\bibitem [{\citenamefont {Vernier}\ \emph {et~al.}(2019)\citenamefont
  {Vernier}, \citenamefont {O'Brien},\ and\ \citenamefont
  {Fendley}}]{Vernier2019}%
  \BibitemOpen
  \bibfield  {author} {\bibinfo {author} {\bibfnamefont {Eric}\ \bibnamefont
  {Vernier}}, \bibinfo {author} {\bibfnamefont {Edward}\ \bibnamefont
  {O'Brien}}, \ and\ \bibinfo {author} {\bibfnamefont {Paul}\ \bibnamefont
  {Fendley}},\ }\bibfield  {title} {\enquote {\bibinfo {title} {Onsager
  symmetries in {\textdollar}u(1){\textdollar} -invariant clock models},}\
  }\href {\doibase 10.1088/1742-5468/ab11c0} {\bibfield  {journal} {\bibinfo
  {journal} {Journal of Statistical Mechanics: Theory and Experiment}\ }\textbf
  {\bibinfo {volume} {2019}},\ \bibinfo {pages} {043107} (\bibinfo {year}
  {2019})}\BibitemShut {NoStop}%
\bibitem [{\citenamefont {Zadnik}\ \emph {et~al.}(2016)\citenamefont {Zadnik},
  \citenamefont {Medenjak},\ and\ \citenamefont {Prosen}}]{ProsenXXZ2016}%
  \BibitemOpen
  \bibfield  {author} {\bibinfo {author} {\bibfnamefont {Lenart}\ \bibnamefont
  {Zadnik}}, \bibinfo {author} {\bibfnamefont {Marko}\ \bibnamefont
  {Medenjak}}, \ and\ \bibinfo {author} {\bibfnamefont {Tomaž}\ \bibnamefont
  {Prosen}},\ }\bibfield  {title} {\enquote {\bibinfo {title} {Quasilocal
  conservation laws from semicyclic irreducible representations of uq(sl2) in
  xxz spin-1/2 chains},}\ }\href {\doibase
  https://doi.org/10.1016/j.nuclphysb.2015.11.023} {\bibfield  {journal}
  {\bibinfo  {journal} {Nuclear Physics B}\ }\textbf {\bibinfo {volume}
  {902}},\ \bibinfo {pages} {339 -- 353} (\bibinfo {year} {2016})}\BibitemShut
  {NoStop}%
\bibitem [{\citenamefont {{Majid}}(1990)}]{Majid1990}%
  \BibitemOpen
  \bibfield  {author} {\bibinfo {author} {\bibfnamefont {Shahn}\ \bibnamefont
  {{Majid}}},\ }\bibfield  {title} {\enquote {\bibinfo {title}
  {{Quasitriangular Hopf Algebras and Yang-Baxter Equations}},}\ }\href
  {\doibase 10.1142/S0217751X90000027} {\bibfield  {journal} {\bibinfo
  {journal} {International Journal of Modern Physics A}\ }\textbf {\bibinfo
  {volume} {5}},\ \bibinfo {pages} {1--91} (\bibinfo {year}
  {1990})}\BibitemShut {NoStop}%
\bibitem [{\citenamefont {Lerda}\ and\ \citenamefont
  {Sciuto}(1993)}]{lerda1993anyons}%
  \BibitemOpen
  \bibfield  {author} {\bibinfo {author} {\bibfnamefont {Alberto}\ \bibnamefont
  {Lerda}}\ and\ \bibinfo {author} {\bibfnamefont {Stefano}\ \bibnamefont
  {Sciuto}},\ }\bibfield  {title} {\enquote {\bibinfo {title} {Anyons and
  quantum groups},}\ }\href@noop {} {\bibfield  {journal} {\bibinfo  {journal}
  {Nuclear Physics B}\ }\textbf {\bibinfo {volume} {401}},\ \bibinfo {pages}
  {613--643} (\bibinfo {year} {1993})}\BibitemShut {NoStop}%
\bibitem [{\citenamefont {Bonatsos}\ and\ \citenamefont
  {Daskaloyannis}(1999)}]{bonatsos1999quantum}%
  \BibitemOpen
  \bibfield  {author} {\bibinfo {author} {\bibfnamefont {Dennis}\ \bibnamefont
  {Bonatsos}}\ and\ \bibinfo {author} {\bibfnamefont {C}~\bibnamefont
  {Daskaloyannis}},\ }\bibfield  {title} {\enquote {\bibinfo {title} {Quantum
  groups and their applications in nuclear physics},}\ }\href@noop {}
  {\bibfield  {journal} {\bibinfo  {journal} {Progress in particle and nuclear
  physics}\ }\textbf {\bibinfo {volume} {43}},\ \bibinfo {pages} {537--618}
  (\bibinfo {year} {1999})}\BibitemShut {NoStop}%
\bibitem [{\citenamefont {Klimyk}\ and\ \citenamefont
  {Schm{\"u}dgen}(2012)}]{qgroupbook}%
  \BibitemOpen
  \bibfield  {author} {\bibinfo {author} {\bibfnamefont {Anatoli}\ \bibnamefont
  {Klimyk}}\ and\ \bibinfo {author} {\bibfnamefont {Konrad}\ \bibnamefont
  {Schm{\"u}dgen}},\ }\href@noop {} {\emph {\bibinfo {title} {Quantum groups
  and their representations}}}\ (\bibinfo  {publisher} {Springer Science \&
  Business Media},\ \bibinfo {year} {2012})\BibitemShut {NoStop}%
\bibitem [{\citenamefont {{Motegi}}(2010)}]{Motegi2010}%
  \BibitemOpen
  \bibfield  {author} {\bibinfo {author} {\bibfnamefont {Kohei}\ \bibnamefont
  {{Motegi}}},\ }\bibfield  {title} {\enquote {\bibinfo {title} {{The matrix
  product representation for the q-VBS state of one-dimensional higher integer
  spin model}},}\ }\href {\doibase 10.1016/j.physleta.2010.05.055} {\bibfield
  {journal} {\bibinfo  {journal} {Physics Letters A}\ }\textbf {\bibinfo
  {volume} {374}},\ \bibinfo {pages} {3112--3115} (\bibinfo {year}
  {2010})}\BibitemShut {NoStop}%
\bibitem [{\citenamefont {Santos}\ \emph {et~al.}(2012)\citenamefont {Santos},
  \citenamefont {Paraan}, \citenamefont {Korepin},\ and\ \citenamefont
  {Klümper}}]{Santos2012}%
  \BibitemOpen
  \bibfield  {author} {\bibinfo {author} {\bibfnamefont {Raul~A}\ \bibnamefont
  {Santos}}, \bibinfo {author} {\bibfnamefont {Francis N~C}\ \bibnamefont
  {Paraan}}, \bibinfo {author} {\bibfnamefont {Vladimir~E}\ \bibnamefont
  {Korepin}}, \ and\ \bibinfo {author} {\bibfnamefont {Andreas}\ \bibnamefont
  {Klümper}},\ }\bibfield  {title} {\enquote {\bibinfo {title} {Entanglement
  spectra ofq-deformed higher spin {VBS} states},}\ }\href {\doibase
  10.1088/1751-8113/45/17/175303} {\bibfield  {journal} {\bibinfo  {journal}
  {Journal of Physics A: Mathematical and Theoretical}\ }\textbf {\bibinfo
  {volume} {45}},\ \bibinfo {pages} {175303} (\bibinfo {year}
  {2012})}\BibitemShut {NoStop}%
\bibitem [{\citenamefont {Tu}\ \emph {et~al.}(2008)\citenamefont {Tu},
  \citenamefont {Zhang},\ and\ \citenamefont {Xiang}}]{Tu2008}%
  \BibitemOpen
  \bibfield  {author} {\bibinfo {author} {\bibfnamefont {Hong-Hao}\
  \bibnamefont {Tu}}, \bibinfo {author} {\bibfnamefont {Guang-Ming}\
  \bibnamefont {Zhang}}, \ and\ \bibinfo {author} {\bibfnamefont {Tao}\
  \bibnamefont {Xiang}},\ }\bibfield  {title} {\enquote {\bibinfo {title}
  {Class of exactly solvable $so(n)$ symmetric spin chains with matrix product
  ground states},}\ }\href {\doibase 10.1103/PhysRevB.78.094404} {\bibfield
  {journal} {\bibinfo  {journal} {Phys. Rev. B}\ }\textbf {\bibinfo {volume}
  {78}},\ \bibinfo {pages} {094404} (\bibinfo {year} {2008})}\BibitemShut
  {NoStop}%
\bibitem [{\citenamefont {Affleck}\ \emph {et~al.}(1987)\citenamefont
  {Affleck}, \citenamefont {Kennedy}, \citenamefont {Lieb},\ and\ \citenamefont
  {Tasaki}}]{AKLT}%
  \BibitemOpen
  \bibfield  {author} {\bibinfo {author} {\bibfnamefont {Ian}\ \bibnamefont
  {Affleck}}, \bibinfo {author} {\bibfnamefont {Tom}\ \bibnamefont {Kennedy}},
  \bibinfo {author} {\bibfnamefont {Elliott~H.}\ \bibnamefont {Lieb}}, \ and\
  \bibinfo {author} {\bibfnamefont {Hal}\ \bibnamefont {Tasaki}},\ }\bibfield
  {title} {\enquote {\bibinfo {title} {Rigorous results on valence-bond ground
  states in antiferromagnets},}\ }\href {\doibase 10.1103/PhysRevLett.59.799}
  {\bibfield  {journal} {\bibinfo  {journal} {Phys. Rev. Lett.}\ }\textbf
  {\bibinfo {volume} {59}},\ \bibinfo {pages} {799--802} (\bibinfo {year}
  {1987})}\BibitemShut {NoStop}%
\bibitem [{\citenamefont {Kennedy}\ and\ \citenamefont
  {Tasaki}(1992)}]{KennedyTasaki}%
  \BibitemOpen
  \bibfield  {author} {\bibinfo {author} {\bibfnamefont {Tom}\ \bibnamefont
  {Kennedy}}\ and\ \bibinfo {author} {\bibfnamefont {Hal}\ \bibnamefont
  {Tasaki}},\ }\bibfield  {title} {\enquote {\bibinfo {title} {Hidden symmetry
  breaking and the haldane phase ins=1 quantum spin chains},}\ }\href {\doibase
  10.1007/BF02097239} {\bibfield  {journal} {\bibinfo  {journal}
  {Communications in Mathematical Physics}\ }\textbf {\bibinfo {volume}
  {147}},\ \bibinfo {pages} {431--484} (\bibinfo {year} {1992})}\BibitemShut
  {NoStop}%
\bibitem [{\citenamefont {Pollmann}\ \emph {et~al.}(2012)\citenamefont
  {Pollmann}, \citenamefont {Berg}, \citenamefont {Turner},\ and\ \citenamefont
  {Oshikawa}}]{PollmannSPT2012}%
  \BibitemOpen
  \bibfield  {author} {\bibinfo {author} {\bibfnamefont {Frank}\ \bibnamefont
  {Pollmann}}, \bibinfo {author} {\bibfnamefont {Erez}\ \bibnamefont {Berg}},
  \bibinfo {author} {\bibfnamefont {Ari~M.}\ \bibnamefont {Turner}}, \ and\
  \bibinfo {author} {\bibfnamefont {Masaki}\ \bibnamefont {Oshikawa}},\
  }\bibfield  {title} {\enquote {\bibinfo {title} {Symmetry protection of
  topological phases in one-dimensional quantum spin systems},}\ }\href
  {\doibase 10.1103/PhysRevB.85.075125} {\bibfield  {journal} {\bibinfo
  {journal} {Phys. Rev. B}\ }\textbf {\bibinfo {volume} {85}},\ \bibinfo
  {pages} {075125} (\bibinfo {year} {2012})}\BibitemShut {NoStop}%
\bibitem [{\citenamefont {Batchelor}\ \emph {et~al.}(1990)\citenamefont
  {Batchelor}, \citenamefont {Mezincescu}, \citenamefont {Nepomechie},\ and\
  \citenamefont {Rittenberg}}]{Batchelorqdef}%
  \BibitemOpen
  \bibfield  {author} {\bibinfo {author} {\bibfnamefont {M~T}\ \bibnamefont
  {Batchelor}}, \bibinfo {author} {\bibfnamefont {L}~\bibnamefont
  {Mezincescu}}, \bibinfo {author} {\bibfnamefont {R~I}\ \bibnamefont
  {Nepomechie}}, \ and\ \bibinfo {author} {\bibfnamefont {V}~\bibnamefont
  {Rittenberg}},\ }\bibfield  {title} {\enquote {\bibinfo {title}
  {q-deformations of the o(3) symmetric spin-1 heisenberg chain},}\ }\href
  {\doibase 10.1088/0305-4470/23/4/003} {\bibfield  {journal} {\bibinfo
  {journal} {Journal of Physics A: Mathematical and General}\ }\textbf
  {\bibinfo {volume} {23}},\ \bibinfo {pages} {L141--L144} (\bibinfo {year}
  {1990})}\BibitemShut {NoStop}%
\bibitem [{\citenamefont {Kl{\"u}mper}\ \emph {et~al.}(1992)\citenamefont
  {Kl{\"u}mper}, \citenamefont {Schadschneider},\ and\ \citenamefont
  {Zittartz}}]{Klumperqdef}%
  \BibitemOpen
  \bibfield  {author} {\bibinfo {author} {\bibfnamefont {A.}~\bibnamefont
  {Kl{\"u}mper}}, \bibinfo {author} {\bibfnamefont {A.}~\bibnamefont
  {Schadschneider}}, \ and\ \bibinfo {author} {\bibfnamefont {J.}~\bibnamefont
  {Zittartz}},\ }\bibfield  {title} {\enquote {\bibinfo {title} {Groundstate
  properties of a generalized vbs-model},}\ }\href {\doibase
  10.1007/BF01309281} {\bibfield  {journal} {\bibinfo  {journal} {Zeitschrift
  f{\"u}r Physik B Condensed Matter}\ }\textbf {\bibinfo {volume} {87}},\
  \bibinfo {pages} {281--287} (\bibinfo {year} {1992})}\BibitemShut {NoStop}%
\bibitem [{\citenamefont {Totsuka}\ and\ \citenamefont
  {Suzuki}(1994)}]{Totsuka1994}%
  \BibitemOpen
  \bibfield  {author} {\bibinfo {author} {\bibfnamefont {K}~\bibnamefont
  {Totsuka}}\ and\ \bibinfo {author} {\bibfnamefont {M}~\bibnamefont
  {Suzuki}},\ }\bibfield  {title} {\enquote {\bibinfo {title} {Hidden symmetry
  breaking in a generalized valence-bond solid model},}\ }\href {\doibase
  10.1088/0305-4470/27/19/017} {\bibfield  {journal} {\bibinfo  {journal}
  {Journal of Physics A: Mathematical and General}\ }\textbf {\bibinfo {volume}
  {27}},\ \bibinfo {pages} {6443--6456} (\bibinfo {year} {1994})}\BibitemShut
  {NoStop}%
\bibitem [{\citenamefont {{Quella}}(2020)}]{Quella}%
  \BibitemOpen
  \bibfield  {author} {\bibinfo {author} {\bibfnamefont {Thomas}\ \bibnamefont
  {{Quella}}},\ }\bibfield  {title} {\enquote {\bibinfo {title} {{Symmetry
  protected topological phases beyond groups: The q-deformed AKLT model}},}\
  }\href@noop {} {\bibfield  {journal} {\bibinfo  {journal} {arXiv e-prints}\
  ,\ \bibinfo {eid} {arXiv:2005.09072}} (\bibinfo {year} {2020})}\BibitemShut
  {NoStop}%
\bibitem [{\citenamefont {Batchelor}\ and\ \citenamefont
  {Barber}(1990)}]{Batchelor_1990}%
  \BibitemOpen
  \bibfield  {author} {\bibinfo {author} {\bibfnamefont {M~T}\ \bibnamefont
  {Batchelor}}\ and\ \bibinfo {author} {\bibfnamefont {M~N}\ \bibnamefont
  {Barber}},\ }\bibfield  {title} {\enquote {\bibinfo {title} {Spin-s quantum
  chains and temperley-lieb algebras},}\ }\href {\doibase
  10.1088/0305-4470/23/1/004} {\bibfield  {journal} {\bibinfo  {journal}
  {Journal of Physics A: Mathematical and General}\ }\textbf {\bibinfo {volume}
  {23}},\ \bibinfo {pages} {L15--L21} (\bibinfo {year} {1990})}\BibitemShut
  {NoStop}%
\bibitem [{\citenamefont {Klumper}(1990)}]{Klumper_1990}%
  \BibitemOpen
  \bibfield  {author} {\bibinfo {author} {\bibfnamefont {A}~\bibnamefont
  {Klumper}},\ }\bibfield  {title} {\enquote {\bibinfo {title} {The spectra of
  q-state vertex models and related antiferromagnetic quantum spin chains},}\
  }\href {\doibase 10.1088/0305-4470/23/5/023} {\bibfield  {journal} {\bibinfo
  {journal} {Journal of Physics A: Mathematical and General}\ }\textbf
  {\bibinfo {volume} {23}},\ \bibinfo {pages} {809--823} (\bibinfo {year}
  {1990})}\BibitemShut {NoStop}%
\bibitem [{\citenamefont {Lin}\ \emph {et~al.}(2020)\citenamefont {Lin},
  \citenamefont {Chandran},\ and\ \citenamefont
  {Motrunich}}]{LinChandranMotrunich2020}%
  \BibitemOpen
  \bibfield  {author} {\bibinfo {author} {\bibfnamefont {Cheng-Ju}\
  \bibnamefont {Lin}}, \bibinfo {author} {\bibfnamefont {Anushya}\ \bibnamefont
  {Chandran}}, \ and\ \bibinfo {author} {\bibfnamefont {Olexei~I.}\
  \bibnamefont {Motrunich}},\ }\bibfield  {title} {\enquote {\bibinfo {title}
  {Slow thermalization of exact quantum many-body scar states under
  perturbations},}\ }\href {\doibase 10.1103/PhysRevResearch.2.033044}
  {\bibfield  {journal} {\bibinfo  {journal} {Phys. Rev. Research}\ }\textbf
  {\bibinfo {volume} {2}},\ \bibinfo {pages} {033044} (\bibinfo {year}
  {2020})}\BibitemShut {NoStop}%
\bibitem [{\citenamefont {{Pakrouski}}\ \emph {et~al.}(2020)\citenamefont
  {{Pakrouski}}, \citenamefont {{Pallegar}}, \citenamefont {{Popov}},\ and\
  \citenamefont {{Klebanov}}}]{Klebanov2020}%
  \BibitemOpen
  \bibfield  {author} {\bibinfo {author} {\bibfnamefont {Kiryl}\ \bibnamefont
  {{Pakrouski}}}, \bibinfo {author} {\bibfnamefont {Preethi~N.}\ \bibnamefont
  {{Pallegar}}}, \bibinfo {author} {\bibfnamefont {Fedor~K.}\ \bibnamefont
  {{Popov}}}, \ and\ \bibinfo {author} {\bibfnamefont {Igor~R.}\ \bibnamefont
  {{Klebanov}}},\ }\bibfield  {title} {\enquote {\bibinfo {title} {{Many Body
  Scars as a Group Invariant Sector of Hilbert Space}},}\ }\href@noop {}
  {\bibfield  {journal} {\bibinfo  {journal} {arXiv e-prints}\ ,\ \bibinfo
  {eid} {arXiv:2007.00845}} (\bibinfo {year} {2020})}\BibitemShut {NoStop}%
\bibitem [{\citenamefont {{Ren}}\ \emph {et~al.}(2020)\citenamefont {{Ren}},
  \citenamefont {{Liang}},\ and\ \citenamefont {{Fang}}}]{Fang2020}%
  \BibitemOpen
  \bibfield  {author} {\bibinfo {author} {\bibfnamefont {Jie}\ \bibnamefont
  {{Ren}}}, \bibinfo {author} {\bibfnamefont {Chenguang}\ \bibnamefont
  {{Liang}}}, \ and\ \bibinfo {author} {\bibfnamefont {Chen}\ \bibnamefont
  {{Fang}}},\ }\bibfield  {title} {\enquote {\bibinfo {title} {{Quasi-symmetry
  groups and many-body scar dynamics}},}\ }\href@noop {} {\bibfield  {journal}
  {\bibinfo  {journal} {arXiv e-prints}\ ,\ \bibinfo {eid} {arXiv:2007.10380}}
  (\bibinfo {year} {2020})}\BibitemShut {NoStop}%
\end{thebibliography}%

\end{document}